\documentclass[preprint,12pt]{elsarticle}

\usepackage{amssymb}
\usepackage{amsmath}

\usepackage{tabularx}
\usepackage{algorithm}
\usepackage{booktabs}
\usepackage{algpseudocode}
\usepackage{subcaption}

\usepackage{xcolor}
\usepackage{booktabs}
\usepackage{multirow}
\usepackage{siunitx}
\usepackage{graphicx} 

\begin{document}

\begin{frontmatter}
	
	
	
	\title{TUC-PPO: Team Utility-Constrained Proximal Policy Optimization for Spatial Public Goods Games}
	
	
	\author[1,2]{Zhaoqilin Yang}
	
	
	\ead{zqlyang@gzu.edu.cn}
	
	\tnotetext[1]{https://github.com/geek12138/TUC-PPO}
	
	\affiliation[1]{organization={State Key Laboratory of Public Big Data, College of Computer Science and Technology},
		addressline={Guizhou University}, 
		city={Guiyang},
		postcode={550025}, 
		state={Guizhou},
		country={China}}

	\affiliation[2]{organization={Institute of Cryptography and Data Security},
		addressline={Guizhou University}, 
		city={Guiyang},
		postcode={550025}, 
		state={Guizhou},
		country={China}}
		
		\author[3]{Xin Wang}  
		
		\ead{xinwang2@bjtu.edu.cn}
		
		\affiliation[3]{organization={School of Mathematics and Statistics},
			addressline={Beijing Jiaotong University}, 
			city={Beijing},
			postcode={100044}, 
			state={Beijing},
			country={China}}
			
		\author[4]{Ruichen Zhang}
		
		\ead{ruichen.zhang@ntu.edu.sg}
		
		\affiliation[4]{organization={College of Computing and Data Science},
			addressline={Nanyang Technological University}, 
			city={Singapore},
			postcode={639798}, 
			country={Singapore}}

		\author[5]{Chanchan Li}		

		\ead{ccli@gzu.edu.cn}
		
		\affiliation[5]{organization={State Key Laboratory of Public Big Data, College of Mathematics and Statistics},
			addressline={Guizhou University}, 
			city={Guiyang},
			postcode={550025}, 
			state={Guizhou},
			country={China}}

		\author[6,2]{Youliang Tian\corref{cor1}}
		\ead{yltian@gzu.edu.cn}
		\cortext[cor1]{Corresponding author}
		
		\affiliation[6]{organization={State Key Laboratory of Public Big Data, College of Big Data and Information Engineering},
			addressline={Guizhou University}, 
			city={Guiyang},
			postcode={550025}, 
			state={Guizhou},
			country={China}}

	\begin{abstract}
		We introduce Team Utility-Constrained Proximal Policy Optimization (TUC-PPO), a new deep reinforcement learning framework. 
		It extends Proximal Policy Optimization (PPO) by integrating team welfare objectives specifically for spatial public goods games.
		Unlike conventional approaches where cooperation emerges indirectly from individual rewards, TUC-PPO instead optimizes a bi-level objective integrating policy gradients and team utility constraints. 
		Consequently, all policy updates explicitly incorporate collective payoff thresholds.
		The framework preserves PPO's policy gradient core while incorporating constrained optimization through adaptive Lagrangian multipliers. 
		Therefore, decentralized agents dynamically balance selfish and cooperative incentives.
		The comparative analysis demonstrates superior performance of this constrained deep reinforcement learning approach compared to unmodified PPO and evolutionary game theory baselines.
		It achieves faster convergence to cooperative equilibria and greater stability against invasion by defectors.
		The framework formally integrates team objectives into policy updates. 
		This work advances multi-agent deep reinforcement learning for social dilemmas while providing new computational tools for evolutionary game theory research.
	\end{abstract}
	
	
	
	
\end{frontmatter}

	
	
	\section{Introduction}
	\label{sec1}
	
	Cooperative behavior serves as a cornerstone of human civilization, playing a pivotal role in societal development and sustainability across historical epochs \cite{dawes_1988_anomalies,perc_2016_phase,perc_2017_statistical}.
	 This fundamental mechanism is evident in agricultural communities, where collective irrigation systems demonstrate how coordinated efforts yield benefits that surpass the capabilities of individual farmers. 
	 Early Mesopotamian societies thrived precisely because farmers collaboratively maintained complex water distribution networks that no single individual could sustain alone. 
	 The evolutionary persistence of such cooperative systems underscores their profound significance in enabling human progress and adaptation to environmental challenges.
	
	  The evolutionary persistence of cooperative behavior presents a fundamental paradox in complex adaptive systems, where individual optimization frequently opposes collective welfare \cite{pennisi_2005_did,kennedy_2005_don}. The public goods game framework provides a mathematical formulation of this tension. It captures the essential conflict between private incentives and group benefits through rigorous payoff structures \cite{nowak_1993_spatial,macy_2002_learning,wang_2015_universal}. International climate agreements exemplify this tension. Participating states confront dichotomous options: cooperative emission reduction versus defection through free-riding. The 2015 Paris Agreement illustrates this dynamic, as countries voluntarily set emission targets while facing economic incentives to minimize their costly contributions. Evolutionary game theory delivers a comprehensive framework for dilemma analysis. It specifically elucidates how strategic interactions and population configurations mold cooperation patterns at varying scales \cite{nowak_1992_evolutionary, hauert_2005_game,szabo_2007_evolutionary}. The agreement's design incorporates key theoretical insights, including reciprocal commitments and transparency mechanisms that mimic the reputation systems studied in evolutionary models. Network reciprocity mechanisms effectively resolve trust dilemmas in structured populations \cite{Chica_2018_TEVC}. In contrast, Lim et al. \cite{Lim_2024_TEVC} propose an asymmetric N-player trust game where investors retain agency to abstain from interactions, resolving the critical flaw of `investor extinction'. However, conventional models inadequately characterize the dynamic interactions among adaptive agents \cite{Waibel_2009_TEVC}.
	
	Research has identified some primary mechanisms that sustain cooperation against free-riding in social systems, each with distinct real-world manifestations. Positive reinforcement mechanisms \cite{chen_2015_first,dos_2015_evolution,okada_2015_effect} operate through incentive structures, as seen in modern carbon credit markets where companies receive tradable credits for emission reductions. Reputation systems \cite{Ren_2024_TEVC, Li_2017_TEVC, quan_2020_information} function similarly to credit scoring, where businesses maintaining strong environmental records gain preferential access to green financing. Negative constraints \cite{helbing_2010_punish,chen_2014_probabilistic,chen_2015_competition,liu_2018_synergy} appear in various sanctioning forms, exemplified by international trade penalties imposed on nations violating environmental agreements. Exclusion practices \cite{Wei_2024_TEVC, liu_2017_competitions,szolnoki_2017_alliance} manifest in professional networks blacklisting unethical members, paralleling evolutionary models of ostracism. Institutional innovations \cite{griffin_2017_cyclic,wang_2021_tax,lee_2024_supporting} include progressive carbon taxation systems that scale levies with emission levels. Separately, differential investment rules \cite{cao_2010_evolutionary} resemble tiered membership structures in sustainability certification programs. These mechanisms collectively demonstrate how theoretical frameworks translate into practical governance tools for maintaining cooperative systems across economic and environmental domains.
	
	 The integration of reinforcement learning (RL) paradigms with evolutionary game theory has significantly advanced our understanding of strategic decision-making processes in social dilemmas. Traditional analytical frameworks based on Fermi update rules and replicator dynamics effectively capture immediate payoff effects and local interactions. However, they neglect the complexity of individual learning processes \cite{szabo_1998_evolutionary,schuster_1983_replicator}, exemplified by businesses in a trade association adjusting their sustainability investments based on peers' performance. RL algorithms implement sophisticated closed-loop learning systems that optimize strategies through iterative state-action-reward cycles and experience accumulation \cite{sutton_1998_reinforcement,izquierdo_2007_transient,lipowski_2009_statistical}. These systems provide more accurate simulations of human decision-making by accounting for both immediate gains and long-term strategic considerations \cite{jia_2021_local,wang_2022_levy,song_2022_reinforcement}.

	Among these RL approaches, Q-learning's robust value iteration mechanism has proven effective at maintaining cooperative equilibria despite strong free-riding incentives in various game settings \cite{watkins_1992_q,han_2021_evolutionary,shi_2022_analysis}. Its applications demonstrate consistent performance across different network topologies in spatial games, with the algorithm's temporal difference learning enabling effective strategy adaptation \cite{szolnoki_2009_topology,szolnoki_2010_impact,szabo_2012_selfishness,szabo_2013_coexistence}. Recent methodological innovations have expanded Q-learning's applicability through novel combinations with periodic strategy updates and adaptive punishment mechanisms \cite{yan_2024_periodic}. Moreover, Shen et al. \cite{shen_2024_learning} significantly enhance agents' cooperative inclinations through fused Q-learning/Fermi dynamics.

	While traditional RL methods like Q-learning have shown promising results, deep RL approaches offer superior capabilities in handling complex, high-dimensional strategy spaces. Modern deep RL methods like Proximal Policy Optimization (PPO) \cite{John_2017_arxiv} address critical dimensionality challenges through their neural network architectures. They overcome limitations of tabular methods in large-scale problems. A compelling economic parallel exists in central bank monetary policy committees. There, policymakers employ PPO-like reasoning by continuously adjusting interest rates (policy parameters) based on complex economic indicators (high-dimensional state space). They simultaneously maintain stability through constrained adjustments (clipped policy updates). The architecture enables direct policy optimization with stable training. Stability leverages clipped objectives and adaptive learning rates \cite{sun_2024_intuitionistic,yu_2022_surprising}. This parallels central banks' policy-stability balancing. Yang et al. \cite{YANG_2025_116762} were the first to introduce PPO into spatial public goods games (SPGG), developing a two-stage curriculum learning framework that enhances agents' cooperative tendencies. However, significant theoretical challenges remain in understanding how these modern machine learning approaches interact with population structures and evolutionary dynamics in complex social systems. These open questions illuminate frontier research integrating evolutionary game theory with multi-agent RL. Applications encompass socio-technical system design, notably modeling regulatory-financial co-dynamics.

	 We propose a Team Utility-Constrained Proximal Policy Optimization (TUC-PPO) framework that establishes a new paradigm for studying cooperation evolution in SPGG. By integrating constrained deep RL with evolutionary dynamics, this approach enables adaptive policy optimization under explicit team utility requirements. The architecture's dual-objective design bridges individual strategic adaptation with collective welfare preservation, achieving robust equilibrium maintenance in non-stationary multi-agent environments. Systematic game-based evaluations verify the sustainability of cooperation under strong free-riding incentives. Quantitative comparisons further confirm the framework's superior convergence efficiency and behavioral stability versus baselines.
	﻿
	Our research makes three fundamental contributions:
	\begin{itemize}
		\item We establish the first integration of team utility constraints into policy gradient optimization for evolutionary games. This achieves a controlled balance between individual rationality and collective welfare. Furthermore, the Lagrangian dual-ascent mechanism provides new mathematical foundations for cooperation sustainability.
		﻿
		\item We design a self-adjusting constraint mechanism that dynamically adapts penalty coefficients through batch-wise violation evaluation. This ensures team utility thresholds while maintaining policy update stability via constrained gradient updates.
		﻿
		\item The PPO-based framework uniquely addresses SPGG challenges through clipped policy updates and advantage estimation. This enables stable strategy adaptation in high-dimensional non-stationary environments where traditional evolutionary methods exhibit oscillation and convergence failures.
	\end{itemize}
	
	The rest of this paper is organized as follows. Section~\ref{sec:model} establishes the theoretical framework of SPGG. It further provides rigorous derivation of the TUC-PPO algorithm integrating constrained policy gradients with evolutionary strategy updates. Analysis in \ref{sec:exp} systematically examines cooperation evolution under varying initial conditions through three complementary dimensions: (1) sensitivity analysis of hyperparameter configurations, (2) comparative performance against conventional algorithms, and (3) Evolution under different initialization strategies. Section \ref{sec:con} concludes the study.
	
	\section{Model}
	\label{sec:model}
	
	We consider an SPGG on an $L\times L$ periodic lattice with von Neumann neighborhood ($k=4$). Each agent participates in $G=5$ game groups centered on itself and its four neighbors. The strategy space is $\mathcal{S}=\{C,D\}$, where cooperators ($C$) contribute 1 unit to the public pool while defectors ($D$) contribute nothing. The payoff for agent $i$ in group $g$ is:
	
	\begin{equation}
		\Pi(s_i^g) = 
		\begin{cases}
			\frac{rN_C^g}{k+1} - 1, & s_i^g = C \\
			\frac{rN_C^g}{k+1}, & s_i^g = D
		\end{cases},
	\end{equation}
	
	where $N_C^g = \sum_{j \in g} \mathbb{I}(s_j^g=C)$ counts cooperators in group $g$, and $r>1$ is the enhancement factor. The total payoff aggregates over all groups:
	
	\begin{equation}
		\Pi_i = \sum_{g\in\mathcal{G}_i} \Pi(s_i^g).
	\end{equation}
	
	This SPGG framework provides the foundation for adapting proximal PPO to evolutionary game dynamics. The key innovation of our TUC-PPO lies in transforming conventional individual reward maximization into a team-optimization paradigm. In this paradigm, agents learn strategies that balance personal gains with collective welfare requirements.
	Unlike standard PPO, TUC-PPO introduces team utility as a fundamental constraint on policy updates, requiring each agent's actions to maintain minimum contribution thresholds for their local game groups. Unlike standard PPO, TUC-PPO introduces team utility as a fundamental constraint on policy updates. This requires each agent's actions to maintain minimum contribution thresholds for their local game groups.
	
	\subsection{TUC-PPO}
	
	The proposed method extends PPO by introducing team utility constraints and adaptive reward balancing. The core formulation begins with a constrained optimization problem. This formulation aims to maximize the expected cumulative composite reward while ensuring the average team utility meets a minimum threshold $\tau$. This translates mathematically to:
	
	\begin{equation}\label{eq:main_problem}
		\begin{cases}
			&\max_\theta \mathbb{E}_{\tau \sim \pi_\theta}\left[\sum_{t=0}^T (1-w_t)r_t^{\text{ind}} + w_t r_t^{\text{team}}\right] \\
			&\text{s.t.} \quad \mathbb{E}_{\tau \sim \pi_\theta}\left[\frac{1}{T}\sum_{t=1}^T r_t^{\text{team}}\right] \geq \tau
		\end{cases}
	\end{equation}
	where $\theta\in\mathbb{R}^d$ denotes the $d$-dimensional trainable parameters of the policy network, $\mathbb{E}_{\tau \sim \pi_\theta}$ represents the expectation over trajectories $\tau = (\mathbf{S}_0,a_0,r_0,\ldots)$ generated by policy $\pi_\theta$. $\mathbf{S}_t\in \mathbb{R}^{L\times L\times 3}$ is the encoded state tensor in combining strategy matrix, neighborhood cooperation counts, and global cooperation rate. $a_t$ is the joint action matrix in $\{0,1\}^{L\times L}$ sampled from the policy $\pi_\theta$. $r_t\in\mathbb{R}^{L\times L}$ is the immediate reward in combining individual and team components through adaptive weighting. $T$ is the episode horizon, $w_t \in [0,1]$ is the adaptive weight, and $\tau$ is the minimum team performance threshold. The threshold $\tau=0.5$ corresponds to a minimum $50\%$ neighbors cooperating in a Prisoner's Dilemma and is Equivalent to the Nash equilibrium payoff in a 2-player game.
	The individual reward $r_t^{\text{ind}}$ for agent $i$ is:
	
	\begin{equation}\label{eq:rewards_ind}
		r_t^{\text{ind}}(i) = \sum_{g\in\mathcal{G}_i} \Pi(s_i^g).
	\end{equation}
	
	The team utility $r_t^{\text{team}}$ for agent $i$ is:
	
	\begin{equation}\label{eq:rewards_team}
		r_t^{\text{team}}(i) = 
		\begin{cases} 
			\frac{r}{k+1}\left(N_C^{\text{neigh}}(i) + \mathbb{I}(s_i=C)\right) - 1 & \text{if } s_i = C \\
			0 & \text{if } s_i = D
		\end{cases}
	\end{equation}
	where $N_C^{\text{neigh}}(i)$ counts cooperating neighbors in agent $i$'s von Neumann neighborhood (4 adjacent cells). $\mathbb{I}(s_i=C)$ is an indicator function that equals 1 when agent $i$ cooperates ($s_i=C$) and 0 otherwise. $\mathcal{G}_i$ represents the 5 game groups agent $i$ participates in (centered on itself and its 4 neighbors). The division by 5 normalizes the payoff across all participating groups.
	
	The constrained optimization problem is transformed through Lagrangian relaxation, yielding the primal-dual formulation:
	\begin{equation}\label{eq:dual_formulation}
		\max_{\theta}\min_{\eta\geq 0} \mathcal{L}(\theta,\eta) = \mathbb{E}_\tau\left[\sum_{t=0}^T \gamma^t R_t\right] + \eta\left(\mathbb{E}_\tau\left[\frac{1}{T}\sum_{t=1}^T r_t^{\text{team}}\right] - \tau\right)
	\end{equation}
	
	The inequality constraint is converted using the ReLU \cite{Glorot_2011_relu} function:
	
	\begin{equation}\label{eq:max}
		\max\left(0, \tau - \mathbb{E}_{\tau \sim \pi_\theta}\left[\frac{1}{T}\sum_{t=1}^T r_t^{\text{team}}\right]\right) = 0
	\end{equation}
	
	In implementation, we approximate the expectation using mini-batch averages:
	
	\begin{equation}\label{eq:averages}
		\mathbb{E}_{\tau \sim \pi_\theta}\left[\frac{1}{T}\sum_{t=1}^T r_t^{\text{team}}\right] \approx \frac{1}{B}\sum_{i=1}^B r_i^{\text{team}}
	\end{equation}
	where $B$  is the batch size.
	
	This leads to the penalty term with the dual variable $\eta \geq 0$:
	
	\begin{equation}\label{eq:L_CV}
		L^{\text{CV}}(\eta)=\eta \cdot \underbrace{\left(\tau - \frac{1}{B}\sum_{i=1}^B r_i^{\text{team}}\right)_+}_{\text{Constraint violation}}
	\end{equation}
	where $\max(0,x)$ ensures one-sided penalty. Here, $\eta$ represents the Lagrange multiplier that dynamically scales the magnitude of the constraint violation.

	The dual variable update rule is:
	\begin{equation}\label{eq:dual}
		\eta \leftarrow \eta + \zeta \cdot L^{\text{CV}}(\eta)
	\end{equation}
	with $\zeta$ denoting the dual learning rate that controls the adjustment speed of $\eta$. This update follows the dual gradient ascent principle.
	
	Applying Lagrangian relaxation transforms this into an unconstrained optimization problem with dual variable $\eta$:
	
	\begin{equation}\label{eq:L_TUC}
		L^{\text{TUC}}(\theta,\eta) = L^{\text{CLIP}}(\theta) + \delta L^{\text{VF}}(\theta) - \rho L^{\text{ENT}}(\theta) + \eta L^{\text{CV}}(\eta)
	\end{equation}
	where $L^{\text{CLIP}}(\theta)$ is the clipped policy objective that prevents excessively large policy updates. $\delta$ is the coefficient balancing policy and value function updates. $L^{\text{VF}}(\theta)$ is the value function loss that minimizes the squared error between predicted and actual returns. $\rho$ is the entropy coefficient that controls the strength of the exploration incentive. $L^{\text{ENT}}(\theta)$ is the policy entropy term that encourages exploration by penalizing low-entropy policies.
		
	The clipped policy objective $L^{\text{CLIP}}(\theta)$ prevents drastic policy changes by constraining the update ratio $\pi_\theta/\pi_{\theta_{old}}$ to $[1-\epsilon,1+\epsilon]$. The policy ratio $r_t(\theta) = \pi_\theta(a_t|s_t)/\pi_{\theta_{\text{old}}}(a_t|s_t)$ is constrained within $[1-\epsilon, 1+\epsilon]$ where $\epsilon$ controls the maximum policy deviation:
	
	\begin{equation}\label{eq:L_CLIP}
		L^{\text{CLIP}}(\theta) = -\mathbb{E}_t\left[\min\left( r_t(\theta)\hat{A}_t, \text{clip}(r_t(\theta),1-\epsilon,1+\epsilon)\hat{A}_t \right)\right]
	\end{equation}
	where the advantage estimates $\hat{A}_t$ are computed using a weighted combination of individual and team rewards across the trajectory. This generalized advantage estimation balances immediate and long-term returns through discount factors $\gamma$ (reward decay) and $\lambda$ (bias-variance trade-off):
	
	\begin{equation}\label{eq:A_t}
		\hat{A}_t = \sum_{l=0}^{T-t}(\gamma\lambda)^l\left[(1-w_t)r_{t+l}^{\text{ind}} + w_t r_{t+l}^{\text{team}} + \gamma V(s_{t+l+1}) - V(s_{t+l})\right]
	\end{equation}
	
	The value function loss $L^{\text{VF}}(\theta)$ trains the critic network by minimizing the prediction error:
	
	\begin{equation}\label{eq:L_VF}
		L^{\text{VF}}(\theta) = \mathbb{E}_t\left[(V_\theta(s_t) - R_t)^2\right]
	\end{equation}
	where the target return $R_t$ is computed through backward recursion:
	
	\begin{equation}\label{eq:R_t}
		R_t = \gamma \left[ (1-w_t)r_t^{\text{ind}} + w_t r_t^{\text{team}} + (1 - \mathbb{I}_t^{\text{done}}) R_{t+1} \right]
	\end{equation}
	with $\mathbb{I}_t^{\text{done}}$ indicating episode termination. This recursive form is mathematically equivalent to the discounted sum but more computationally efficient. The adaptive weight $w_t$ dynamically balances individual and team rewards based on their historical ratio:
	
	\begin{equation}\label{eq:w_t}
		w_t = \sigma\left( \cdot \frac{\sum_{i=1}^t r_i^{\text{team}}}{\sum_{i=1}^t r_i^{\text{ind}} + 10^{-8}}\right)
	\end{equation}
	
	To encourage exploration, entropy regularization $L^{\text{ENT}}(\theta)$ adds a bonus proportional to the policy's information entropy:
	
	\begin{equation}\label{eq::_ENT}
		L^{\text{ENT}}(\theta) = \mathbb{E}_t\left[-\sum_{a\in\mathcal{A}}\pi_\theta(a|s_t)\log\pi_\theta(a|s_t)\right]
	\end{equation}

	\subsection{Actor-Critic Network Architecture}
\begin{figure}[htbp!]
		\centering
		\includegraphics[width=0.3\linewidth]{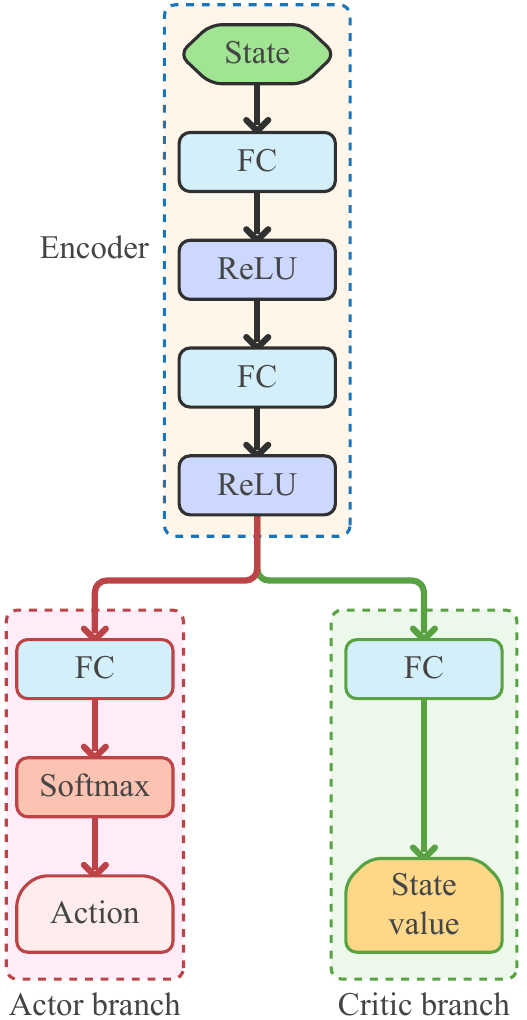}
		\caption{The architecture of actor-critic network.}
		\label{fig:tuc-ppo_AC}
	\end{figure}

	The Actor-Critic network processes spatial game states through the following mathematical operations. Let $s_t^i = [x_t^i, n_t^i, g_t] \in \mathbb{R}^3$ denote the encoded state for agent $i$ at time $t$, where $x_t^i \in \{0,1\}$ represents the current strategy (0: defection, 1: cooperation). $n_t^i \in \mathbb{N}$ counts cooperating neighbors within the von Neumann neighborhood. $g_t \in [0,1]$ indicates the global cooperation rate. Fig. \ref{fig:tuc-ppo_AC} shows the actor-critic network architecture.

	The Encoder transforms inputs through:
	
	\begin{equation}\label{eq:encoder}
		h_t^i = \sigma(W_2\sigma(W_1s_t^i + b_1) + b_2)
	\end{equation}
	where $\sigma(\cdot)$ denotes the ReLU activation function, $W_1 \in \mathbb{R}^{64\times3}$ and $W_2 \in \mathbb{R}^{64\times64}$ are weight matrices, $b_1,b_2 \in \mathbb{R}^{64}$ are bias terms. The Actor branch computes action probabilities via:
	
	\begin{equation}\label{eq:actor_branch}
		\pi_\theta(a_t^i|s_t^i) = \text{softmax}(W_ah_t^i + b_a)
	\end{equation}
	with $W_a \in \mathbb{R}^{2\times64}$ and $b_a \in \mathbb{R}^2$ producing a categorical distribution over actions (cooperate/defect). The Critic branch estimates state values as:
	
	\begin{equation}\label{eq:critic_branch}
		V_\theta(s_t^i) = w_v^Th_t^i + b_v
	\end{equation}
	where $w_v \in \mathbb{R}^{64}$ and $b_v \in \mathbb{R}$. For an $L\times L$ grid, the joint policy factorizes as $\pi_\theta(a_t|s_t) = \prod_{i=1}^{L^2}\pi_\theta(a_t^i|s_t^i)$.
	
	The proposed TUC-PPO framework is described as Algorithm \ref{alg:ppo_tuc}, which integrates team utility constraints with adaptive reward balancing in a spatially structured environment. The algorithm is as follows:
	
	\begin{algorithm}[H]
		\caption{PPO-TUC Framework for SPGG}
		\label{alg:ppo_tuc}
		\begin{algorithmic}[1]
			\State \textbf{Initialize:}
			\State \quad Policy network $\pi_\theta$ and value network $V_\phi$ with shared features
			\State \quad Team utility threshold $\tau$, dual variable $\eta$
			\State \quad Hyperparameters: $\delta$, $\rho$, $\zeta$, clip range $\epsilon$
			
			\State
			\For{each epoch $t = 1$ \textbf{to} $T$}
			\For{each agent $i$ in grid}
			\State Sample action $a_t^i \sim \pi_\theta(s_t^i)$ (Eq.~\ref{eq:actor_branch})
			\State Compute composite reward $r_t^i = (1-w_t^i)r_t^{\text{ind},i} + w_t^i r_t^{\text{team},i}$
			\EndFor
			
			\State Compute constraint violation $L^{\text{CV}}$  (Eq.~\ref{eq:L_CV})
			\State Compute GAE advantage $\hat{A}_t$ (Eq.~\ref{eq:A_t})
			\State Update dual variable $\eta$ (Eq.~\ref{eq:dual})
			\State Optimize surrogate objective:
			\State \quad $L^{\text{TUC}} = L^{\text{CLIP}} + \delta L^{\text{VF}} - \rho L^{\text{ENT}} + \eta L^{\text{CV}}$ (Eq.~\ref{eq:L_TUC})
			\State \quad Update $\theta$ via gradient descent on $L^{\text{TUC}}$
			\State Clear experience buffer
			\EndFor
		\end{algorithmic}
	\end{algorithm}

	\section{Experimental results}
	\label{sec:exp}
	\subsection{Experimental setup}
	
	The default configuration employs a $200\times200$ spatial grid with parameter settings. Learning rate $\alpha=1\times10^{-4}$, discount factor $\gamma=0.99$, generalized advantage estimation parameter $\lambda=0.95$, PPO clip threshold $\epsilon=0.2$, critic loss weight $\delta=0.5$, and entropy regularization coefficient $\rho=0.01$. The team utility mechanism operates with threshold $\tau=0.5$, and dual learning rate $\zeta=0.01$ for constrained optimization. For constraint evaluation in TUC-PPO, the effective batch size B is set to encompass the entire experience buffer. This substantial B value provides a comprehensive trajectory evaluation while maintaining reasonable computational requirements. Parameter updates utilize the Adam optimizer \cite{Diederik_2015_ICLR} with StepLR scheduling that halves the learning rate every 1,000 iterations.
	
	All experiments were executed on mobile workstation hardware featuring an AMD Ryzen 9 5900HX CPU and NVIDIA GeForce RTX 3080 Laptop GPU (16GB GDDR6) under Ubuntu 22.04.5 LTS, with PyTorch 2.2.1 leveraging CUDA 12.8 acceleration. The spatial visualization protocol encodes strategic choices using binary cell states: defection represented by black and cooperation by white RGB.

	\subsection{Comparative analysis of algorithms}
	\label{exp:Compare}
	
\begin{figure*}[htbp!]
		\begin{minipage}{\linewidth}
			\begin{minipage}{0.24\linewidth}
				\centering
				\includegraphics[width=\linewidth]{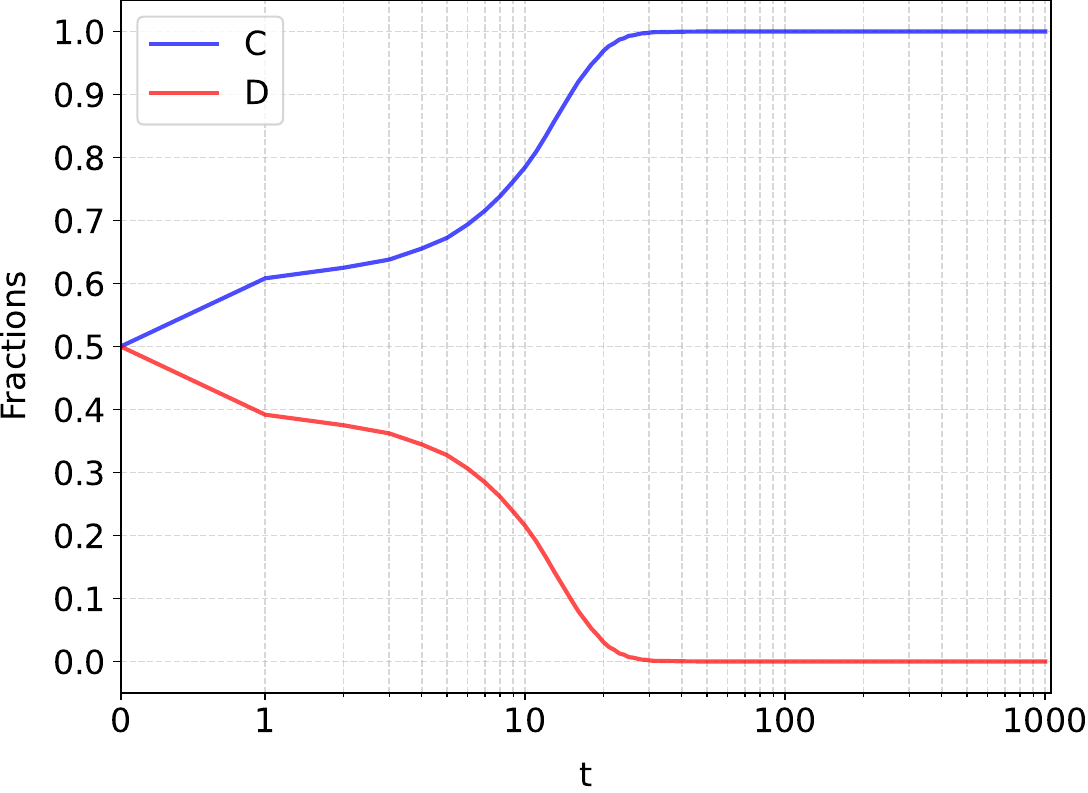}\\
			\end{minipage}
			\begin{minipage}{0.14\linewidth}
				\centering
				\fbox{\includegraphics[width=\linewidth]{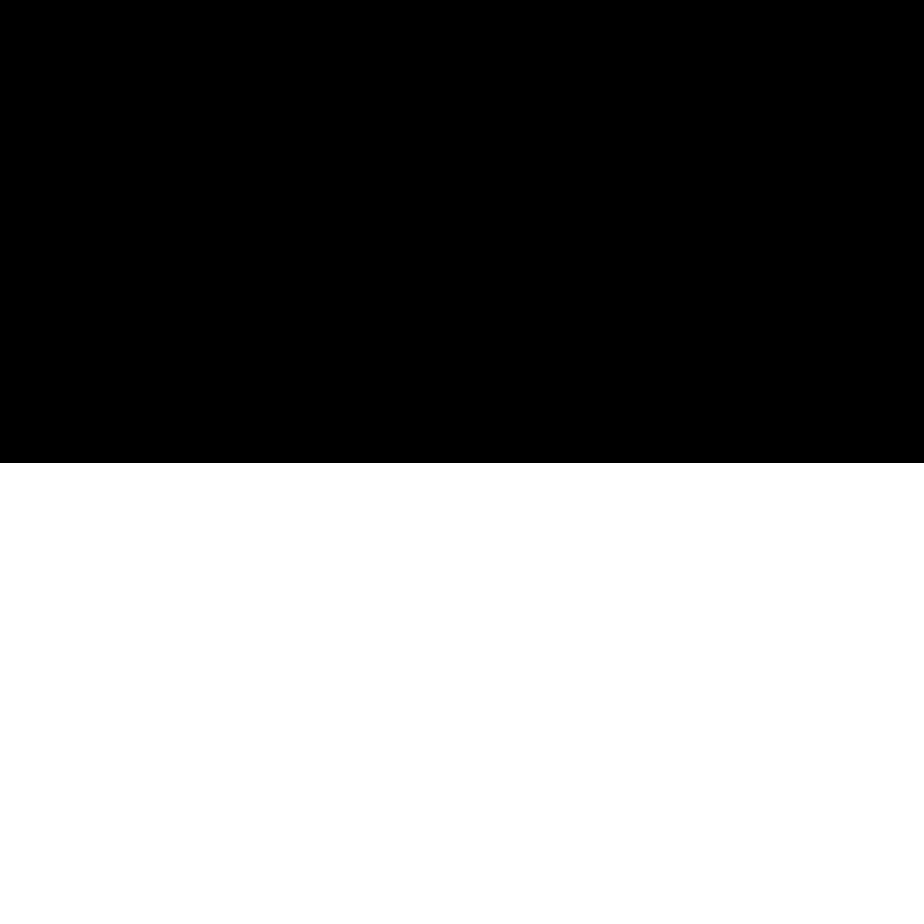}}\\
				\vspace{-2mm}
				{\footnotesize t=0}
			\end{minipage}
			\begin{minipage}{0.14\linewidth}
				\centering
				\fbox{\includegraphics[width=\linewidth]{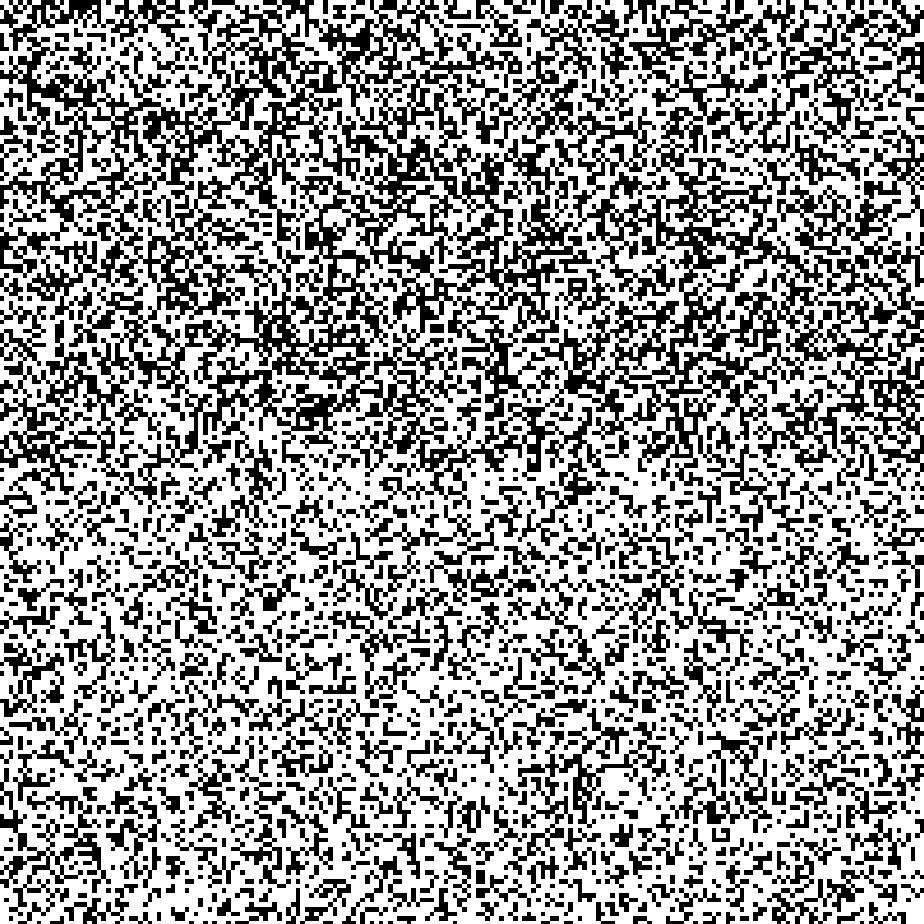}}\\
				\vspace{-2mm}
				{\footnotesize t=1}
			\end{minipage}
			\begin{minipage}{0.14\linewidth}
				\centering
				\fbox{\includegraphics[width=\linewidth]{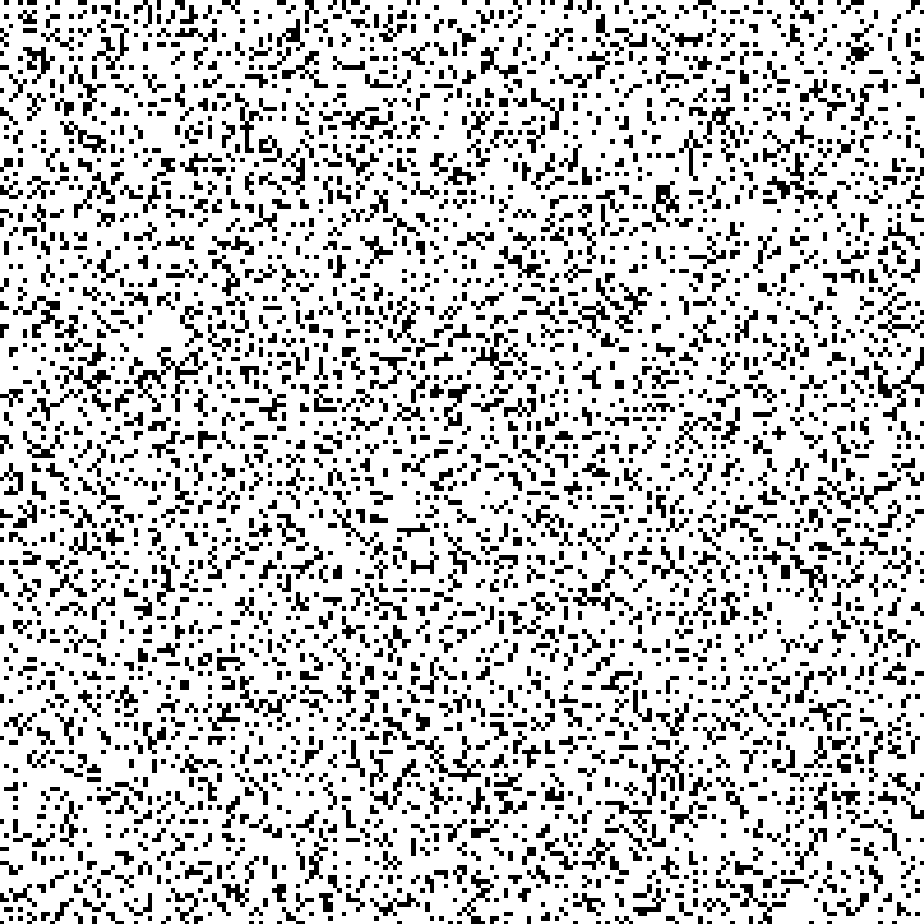}}\\
				\vspace{-2mm}
				{\footnotesize t=10}
			\end{minipage}
			\begin{minipage}{0.14\linewidth}
				\centering
				\fbox{\includegraphics[width=\linewidth]{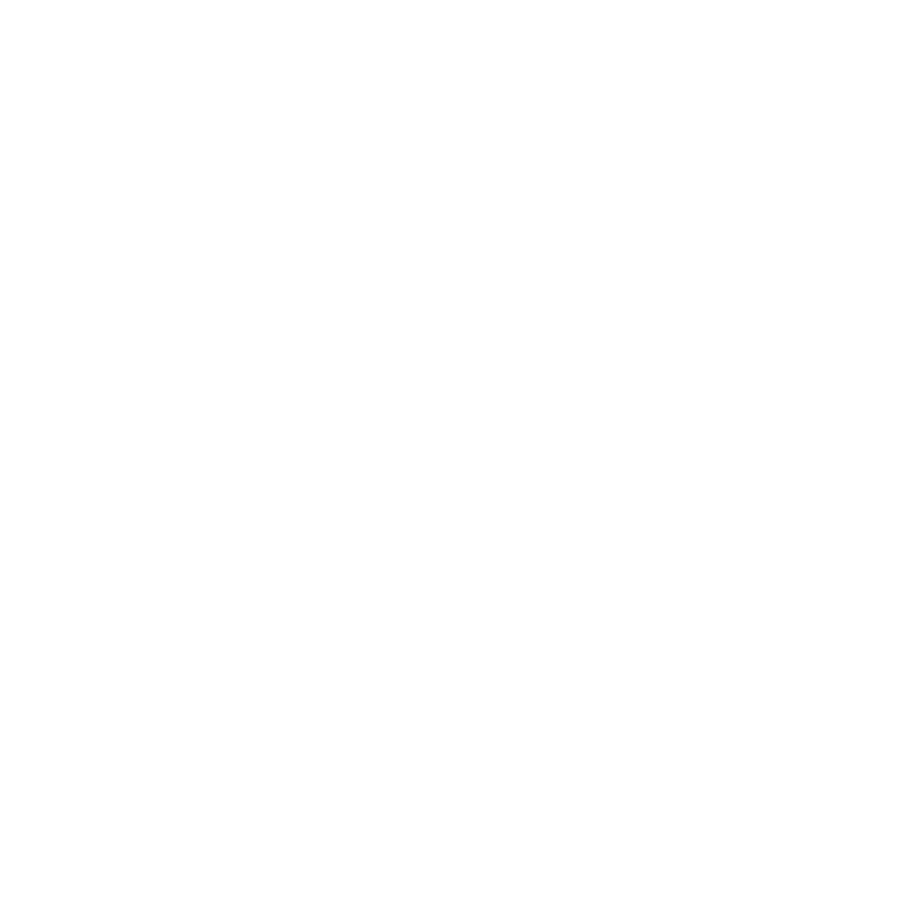}}\\
				\vspace{-2mm}
				{\footnotesize t=100}
			\end{minipage}
			\begin{minipage}{0.14\linewidth}
				\centering
				\fbox{\includegraphics[width=\linewidth]{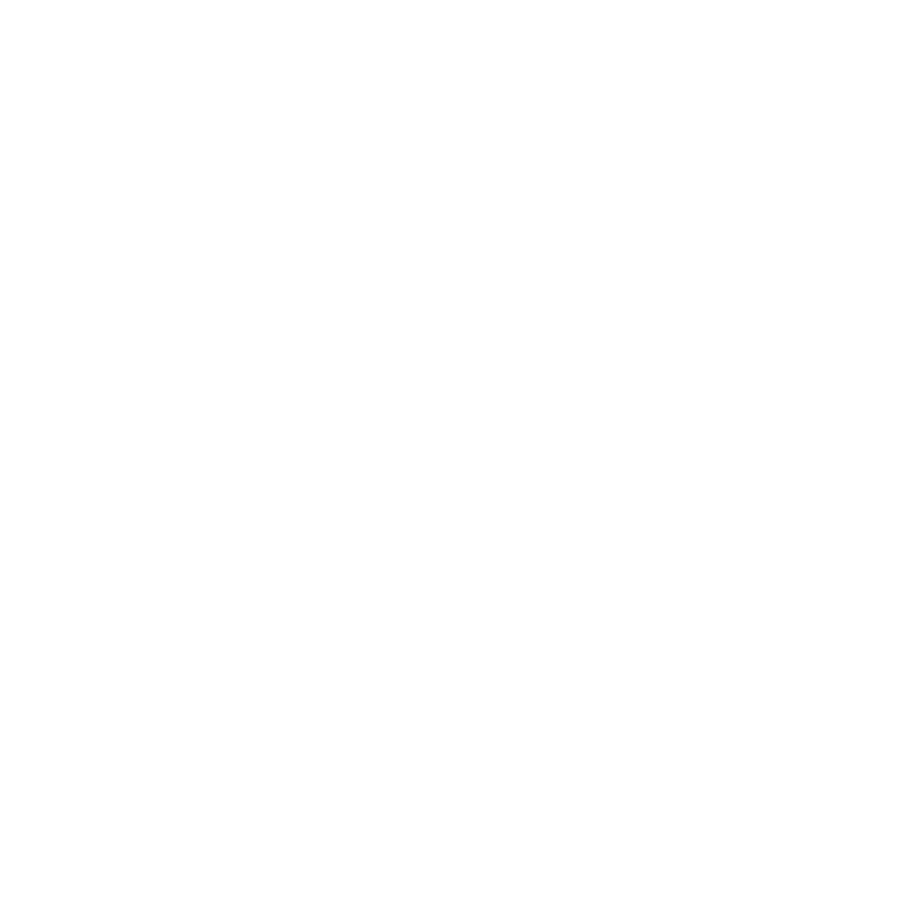}}\\
				\vspace{-2mm}
				{\footnotesize t=1000}
			\end{minipage}
			\vspace{-3mm}
			\caption*{\footnotesize (a) TUC-PPO}
		\end{minipage}
		\\[2mm]
		\begin{minipage}{\linewidth}
			\begin{minipage}{0.24\linewidth}
				\centering
				\includegraphics[width=\linewidth]{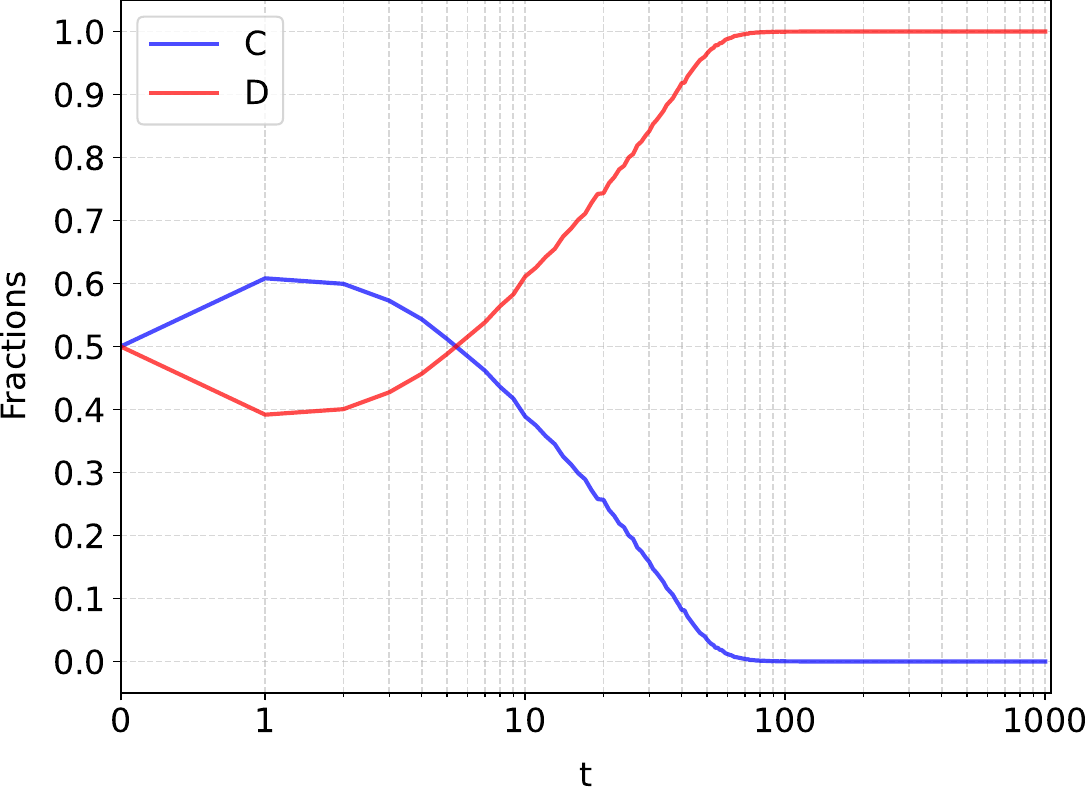}\\
			\end{minipage}
			\begin{minipage}{0.14\linewidth}
				\centering
				\fbox{\includegraphics[width=\linewidth]{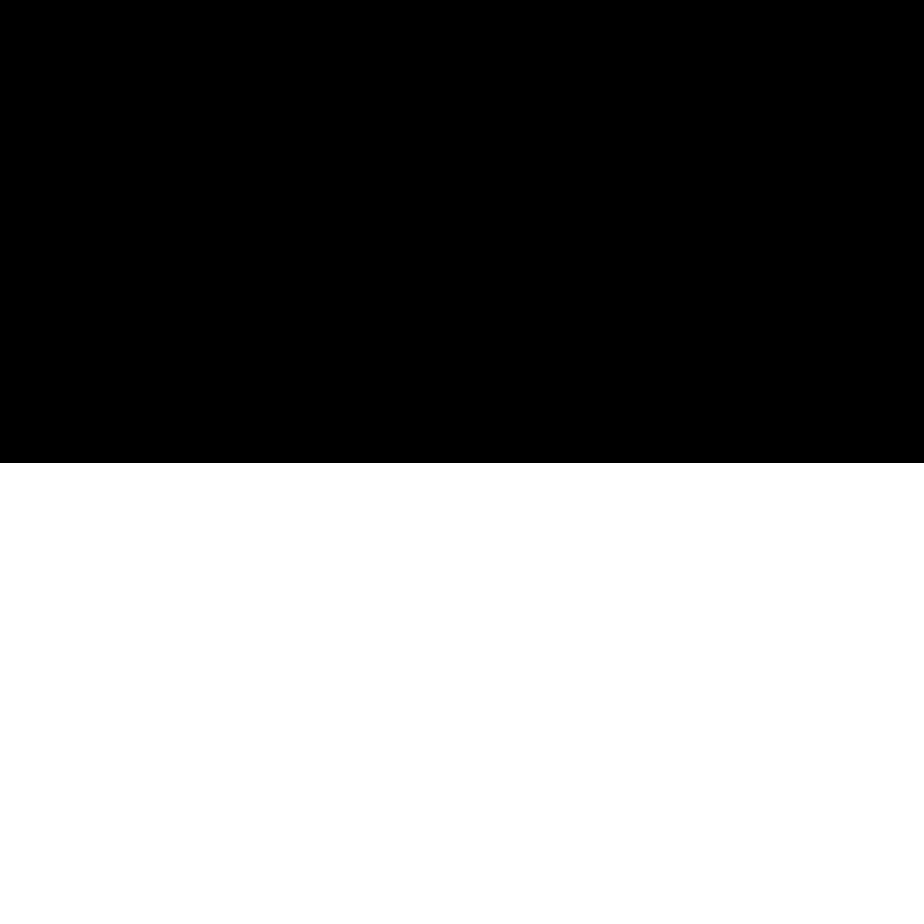}}\\
				\vspace{-2mm}
				{\footnotesize t=0}
			\end{minipage}
			\begin{minipage}{0.14\linewidth}
				\centering
				\fbox{\includegraphics[width=\linewidth]{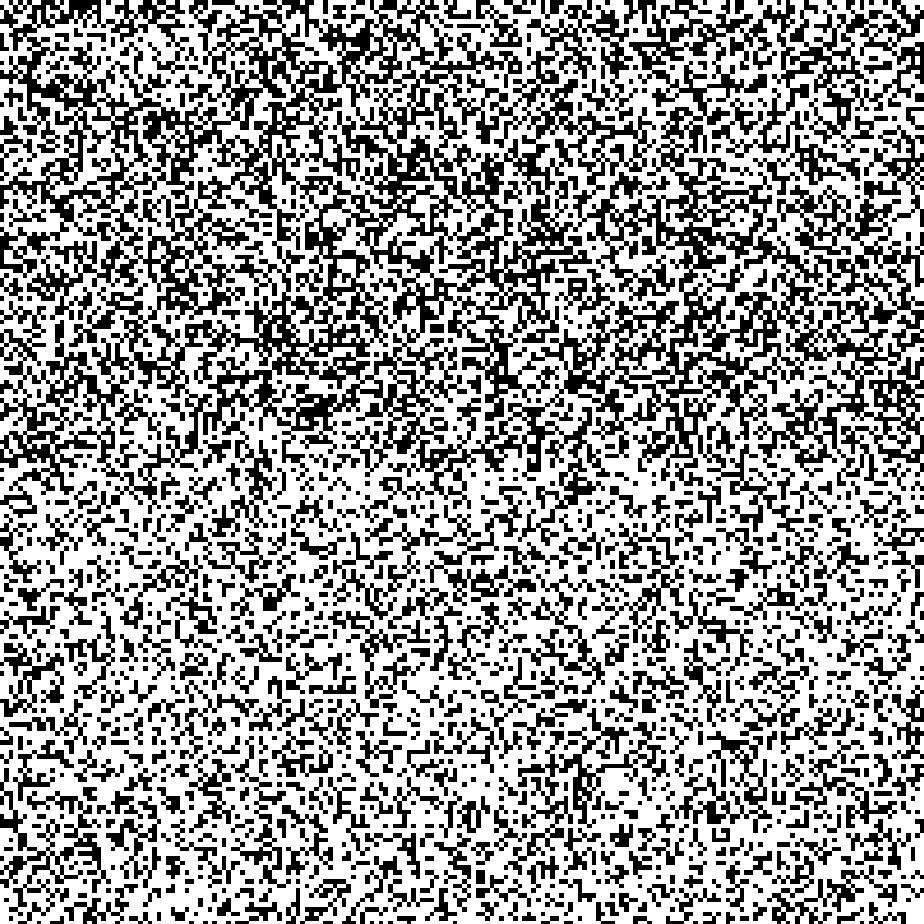}}\\
				\vspace{-2mm}
				{\footnotesize t=1}
			\end{minipage}
			\begin{minipage}{0.14\linewidth}
				\centering
				\fbox{\includegraphics[width=\linewidth]{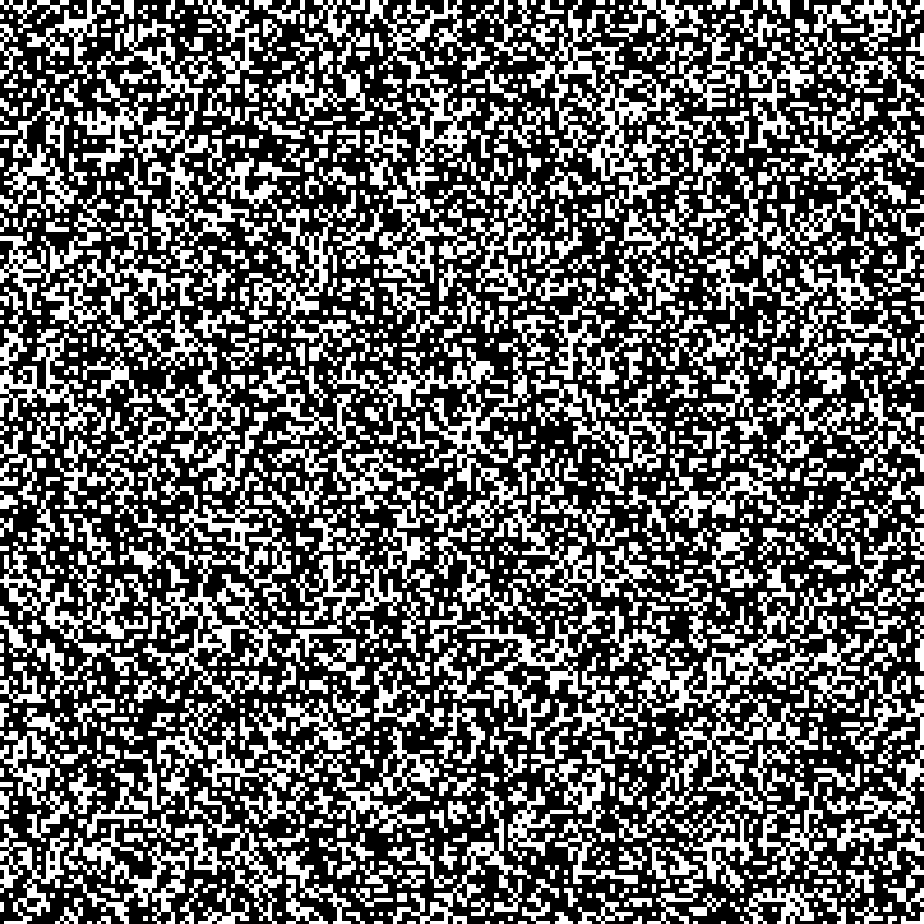}}\\
				\vspace{-2mm}
				{\footnotesize t=10}
			\end{minipage}
			\begin{minipage}{0.14\linewidth}
				\centering
				\fbox{\includegraphics[width=\linewidth]{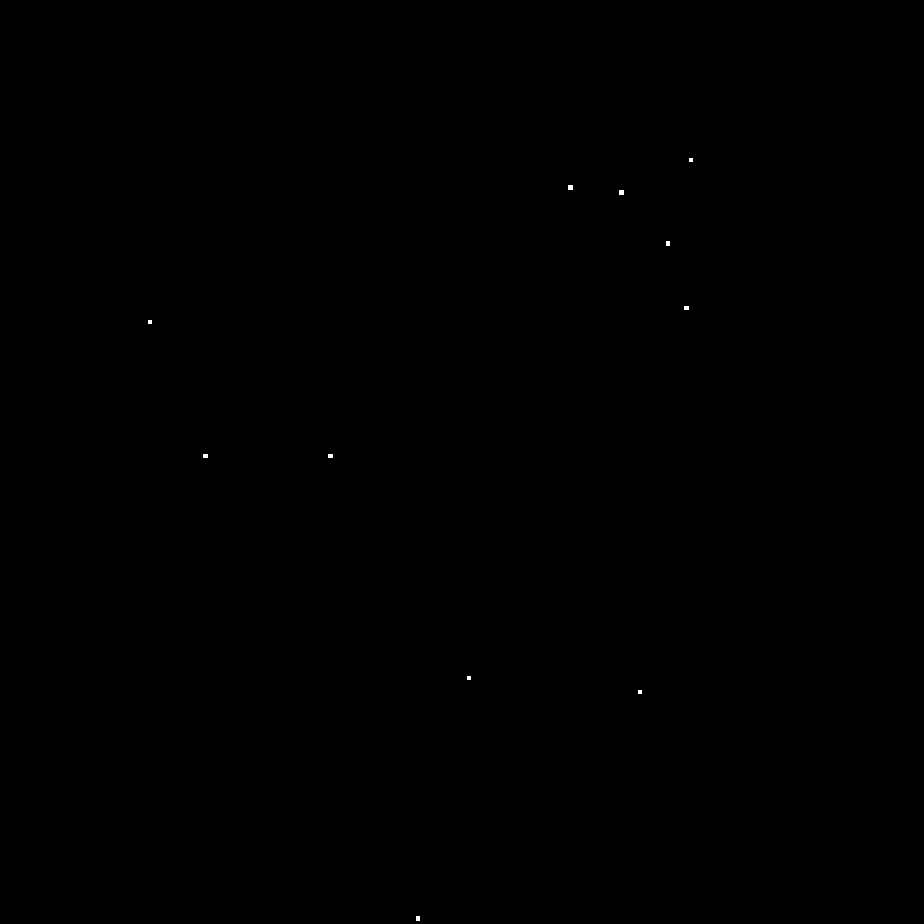}}\\
				\vspace{-2mm}
				{\footnotesize t=100}
			\end{minipage}
			\begin{minipage}{0.14\linewidth}
				\centering
				\fbox{\includegraphics[width=\linewidth]{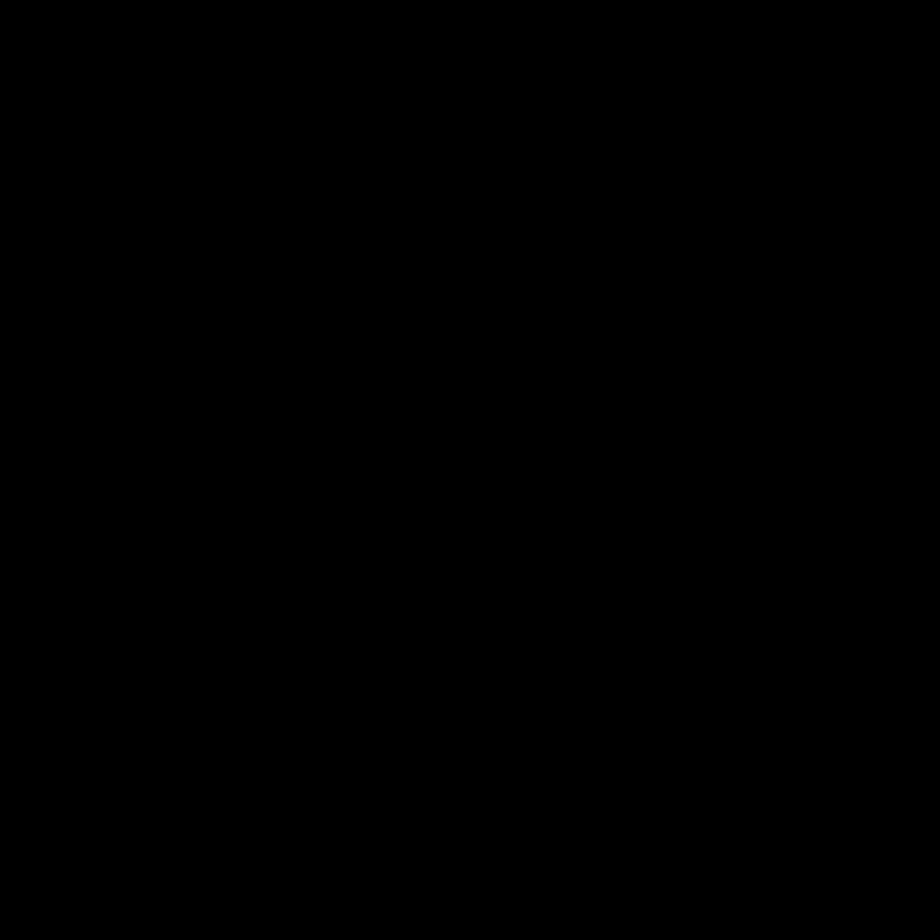}}\\
				\vspace{-2mm}
				{\footnotesize t=1000}
			\end{minipage}
			\vspace{-3mm}
			\caption*{\footnotesize (b) PPO}
		\end{minipage}
		\\[2mm]
		\begin{minipage}{\linewidth}
			\begin{minipage}{0.24\linewidth}
				\centering
				\includegraphics[width=\linewidth]{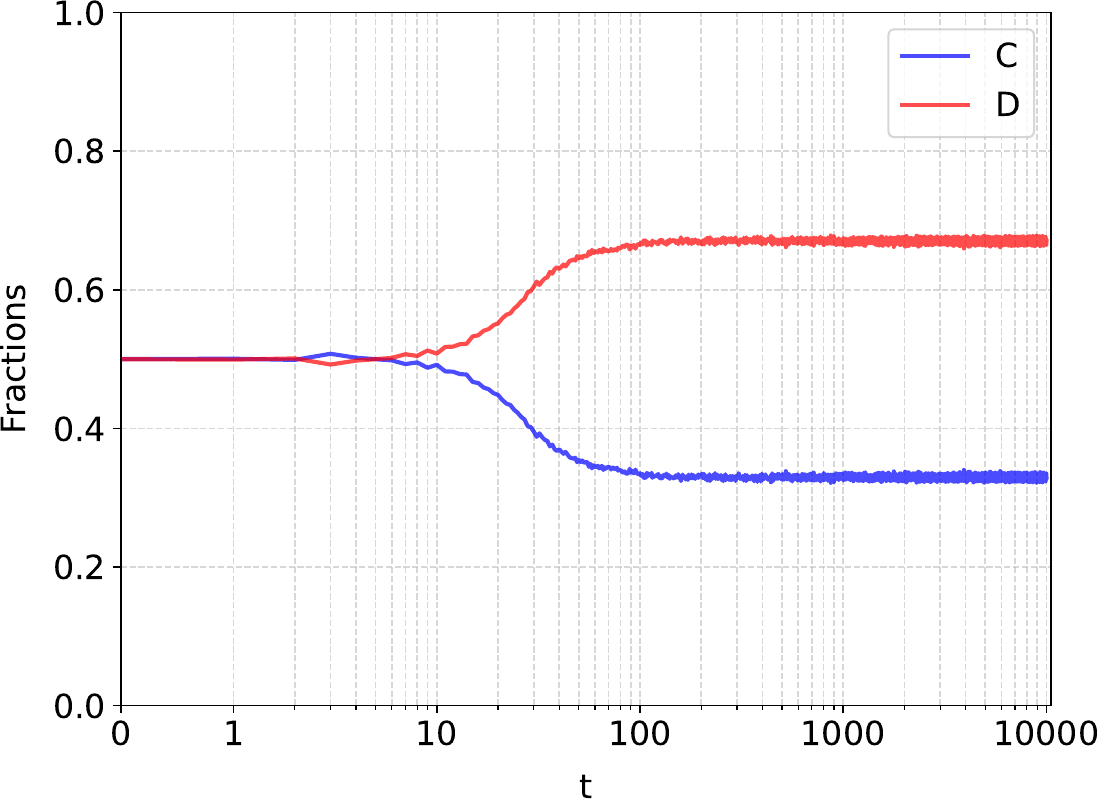}\\
			\end{minipage}
			\begin{minipage}{0.14\linewidth}
				\centering
				\fbox{\includegraphics[width=\linewidth]{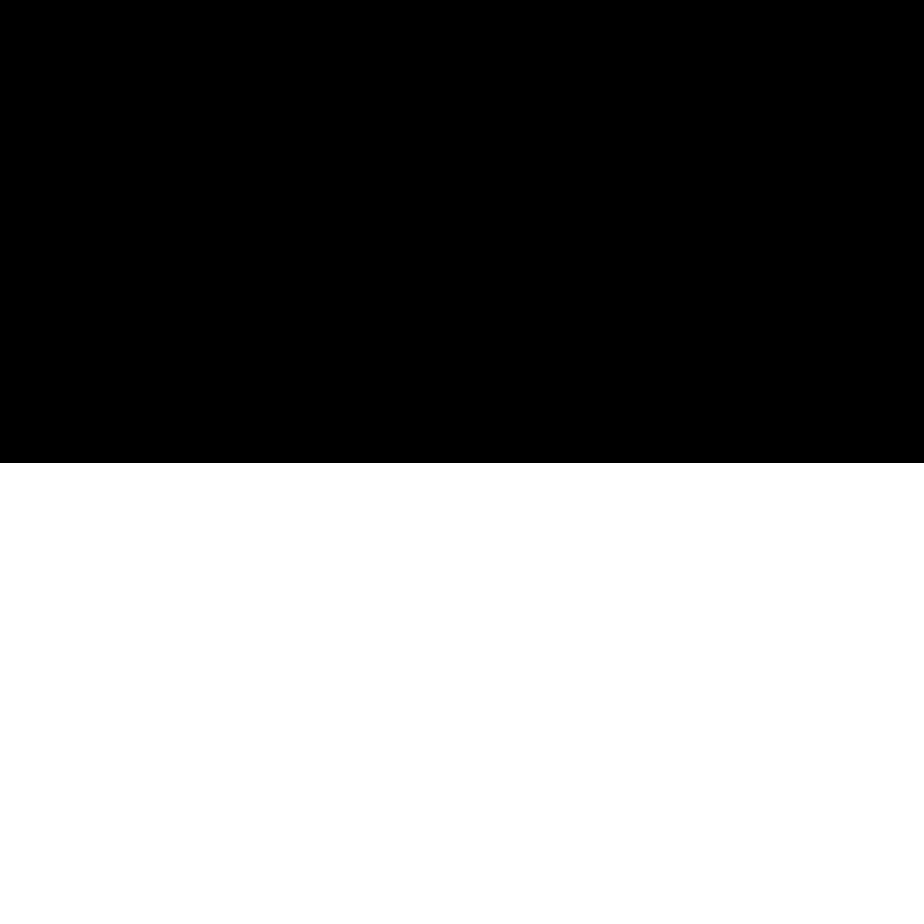}}\\
				\vspace{-2mm}
				{\footnotesize t=0}
			\end{minipage}
			\begin{minipage}{0.14\linewidth}
				\centering
				\fbox{\includegraphics[width=\linewidth]{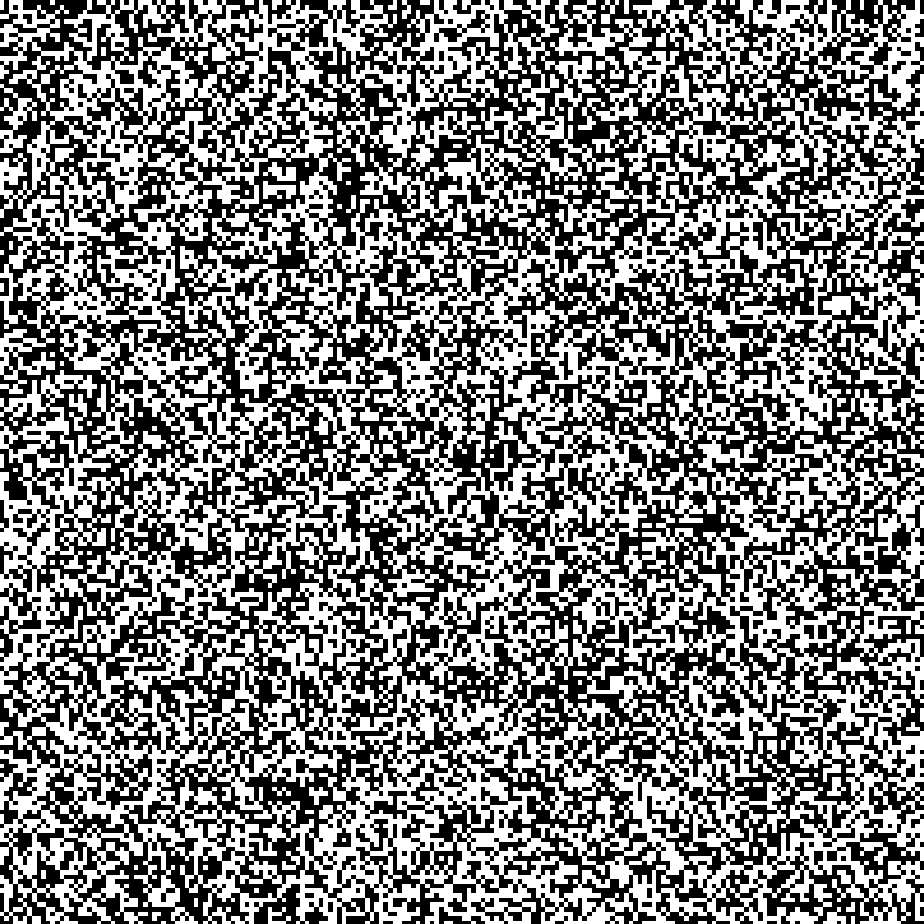}}\\
				\vspace{-2mm}
				{\footnotesize t=10}
			\end{minipage}
			\begin{minipage}{0.14\linewidth}
				\centering
				\fbox{\includegraphics[width=\linewidth]{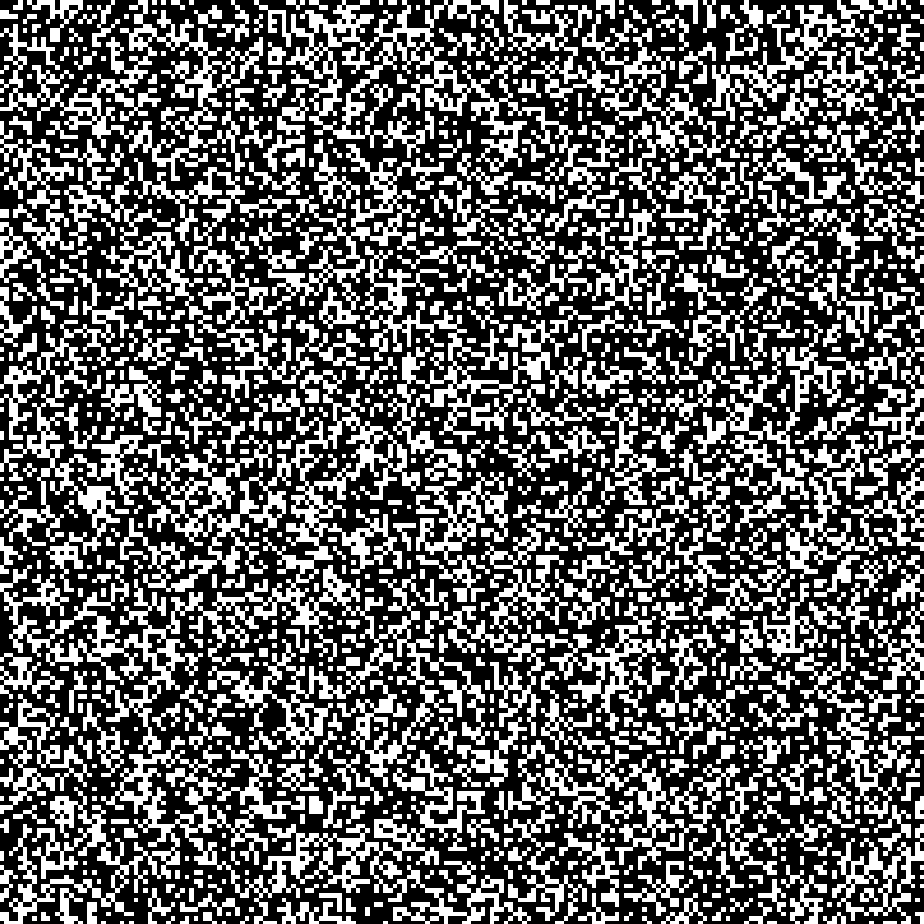}}\\
				\vspace{-2mm}
				{\footnotesize t=100}
			\end{minipage}
			\begin{minipage}{0.14\linewidth}
				\centering
				\fbox{\includegraphics[width=\linewidth]{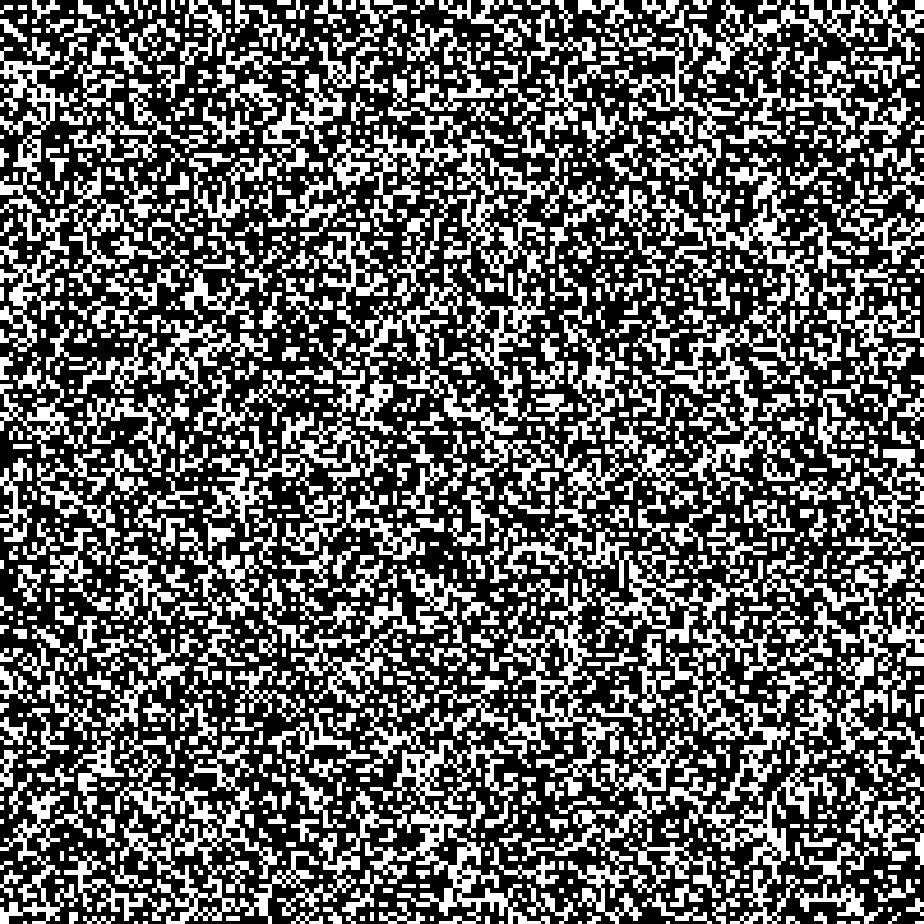}}\\
				\vspace{-2mm}
				{\footnotesize t=1000}
			\end{minipage}
			\begin{minipage}{0.14\linewidth}
				\centering
				\fbox{\includegraphics[width=\linewidth]{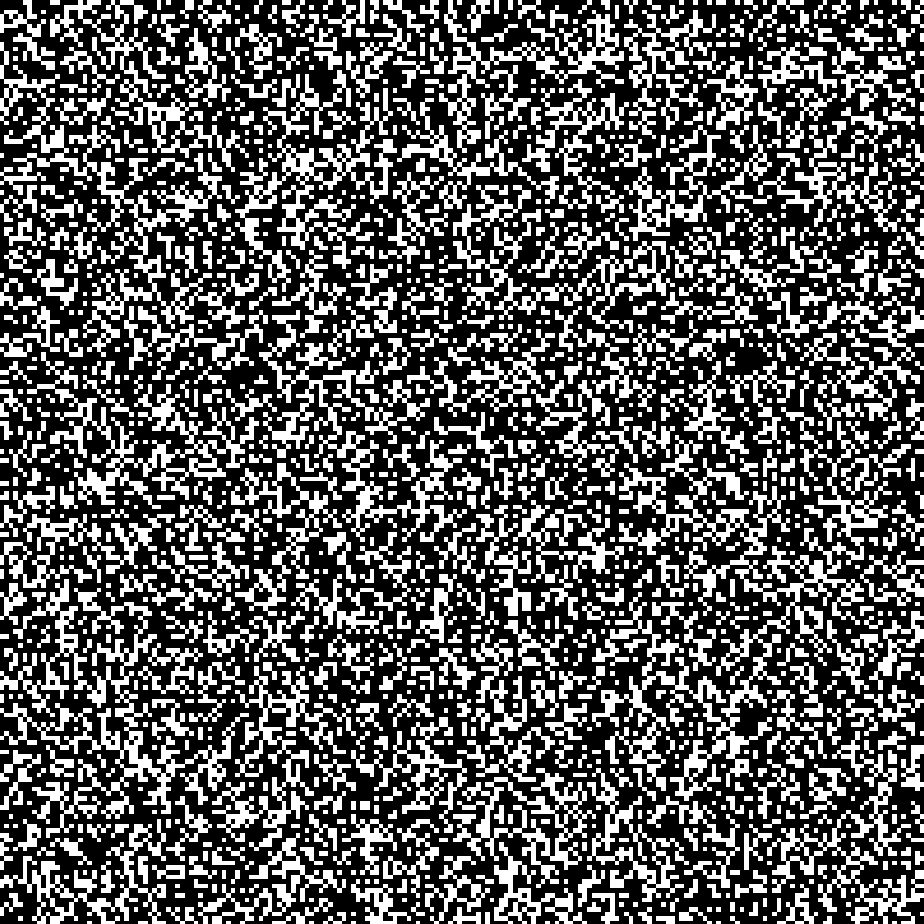}}\\
				\vspace{-2mm}
				{\footnotesize t=10000}
			\end{minipage}
			\vspace{-3mm}
			\caption*{\footnotesize (c) Q-learning}
		\end{minipage}	
		\\[2mm]
		\begin{minipage}{\linewidth}
			\begin{minipage}{0.24\linewidth}
				\centering
				\includegraphics[width=\linewidth]{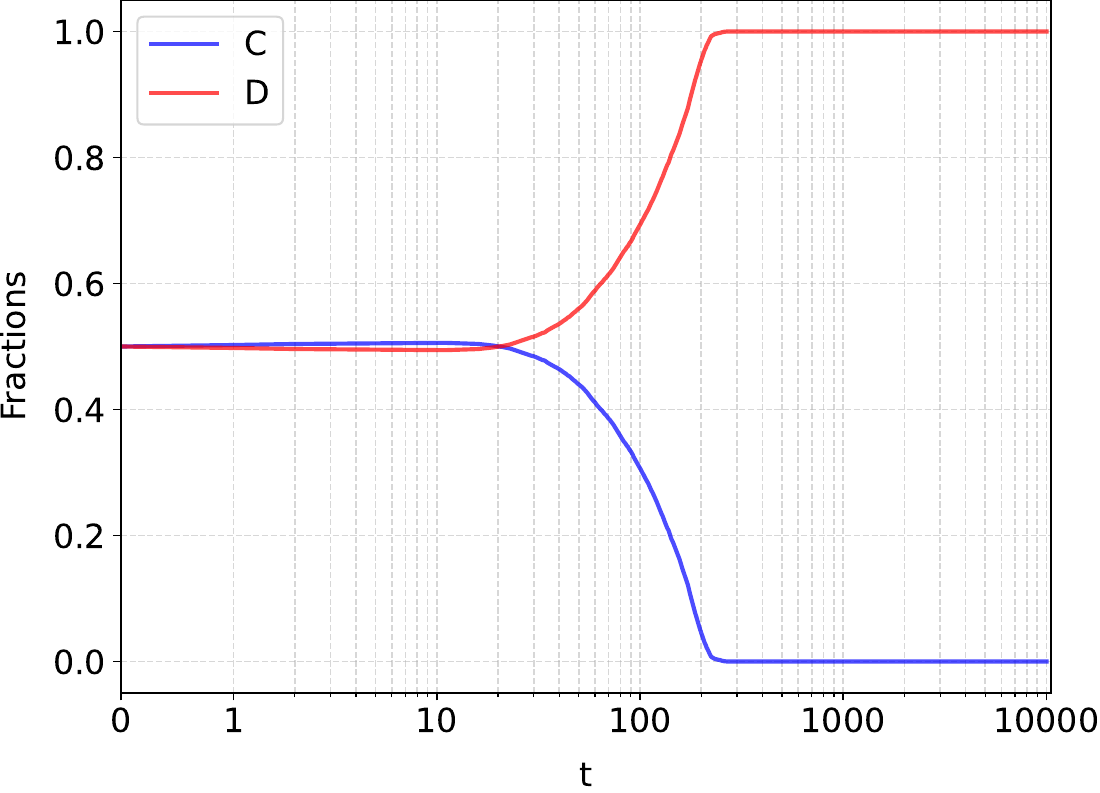}\\
			\end{minipage}
			\begin{minipage}{0.14\linewidth}
				\centering
				\fbox{\includegraphics[width=\linewidth]{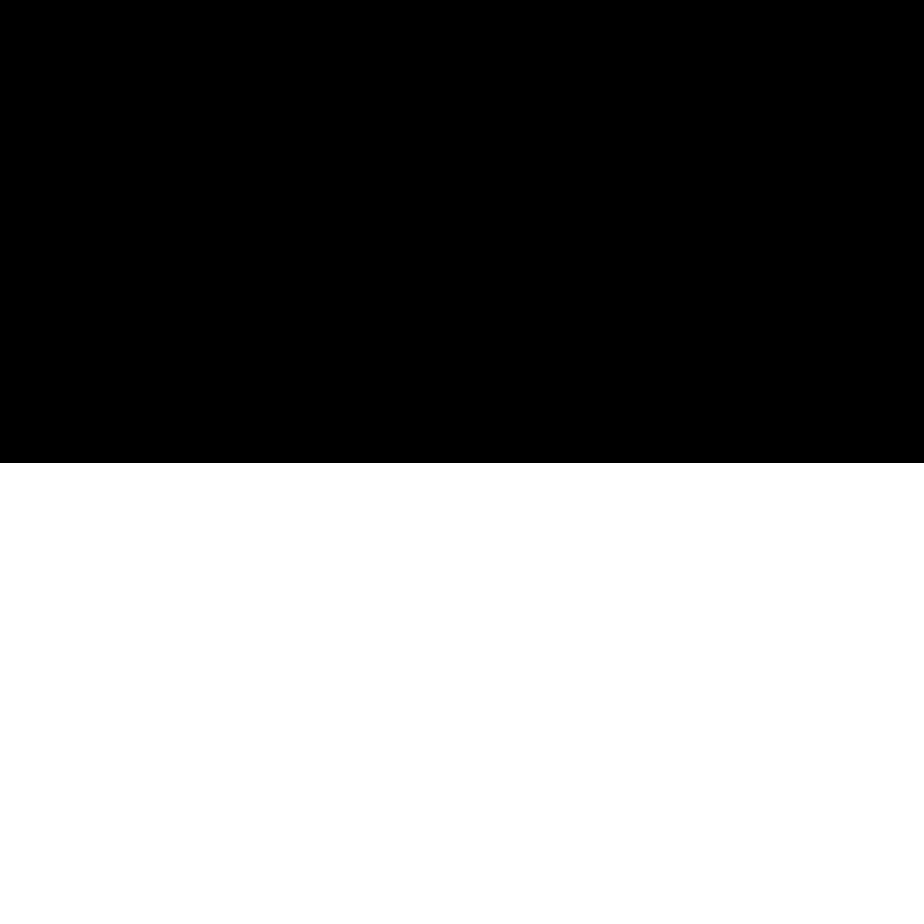}}\\
				\vspace{-2mm}
				{\footnotesize t=0}
			\end{minipage}
			\begin{minipage}{0.14\linewidth}
				\centering
				\fbox{\includegraphics[width=\linewidth]{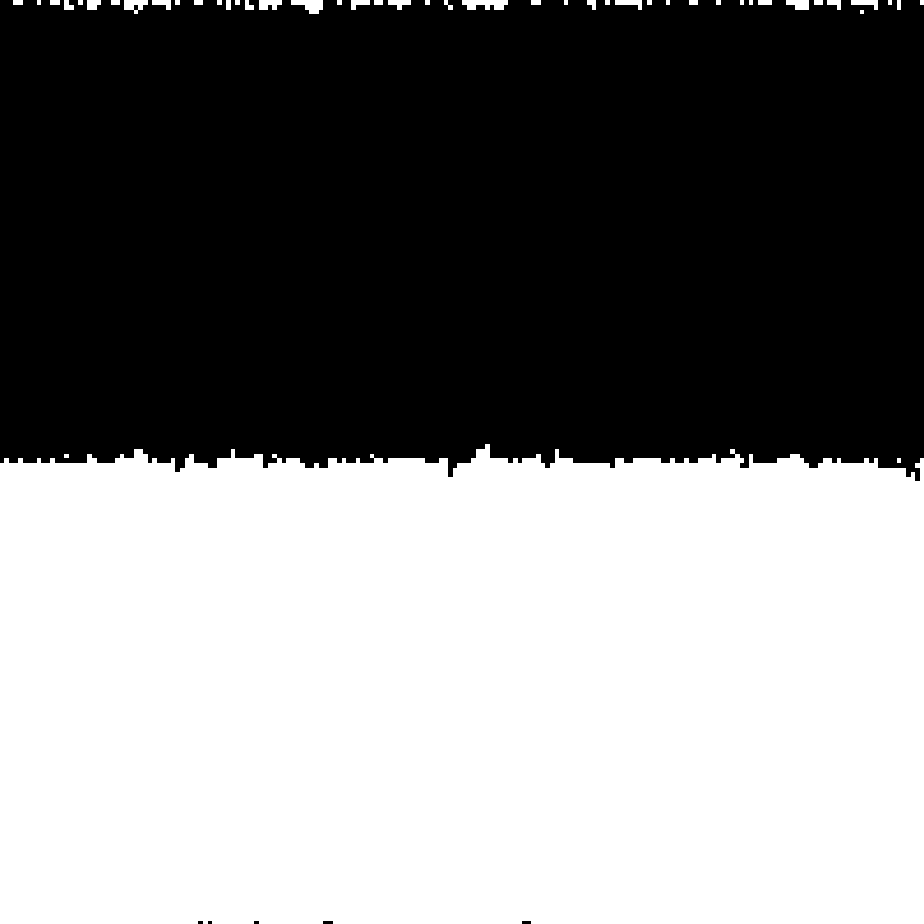}}\\
				\vspace{-2mm}
				{\footnotesize t=10}
			\end{minipage}
			\begin{minipage}{0.14\linewidth}
				\centering
				\fbox{\includegraphics[width=\linewidth]{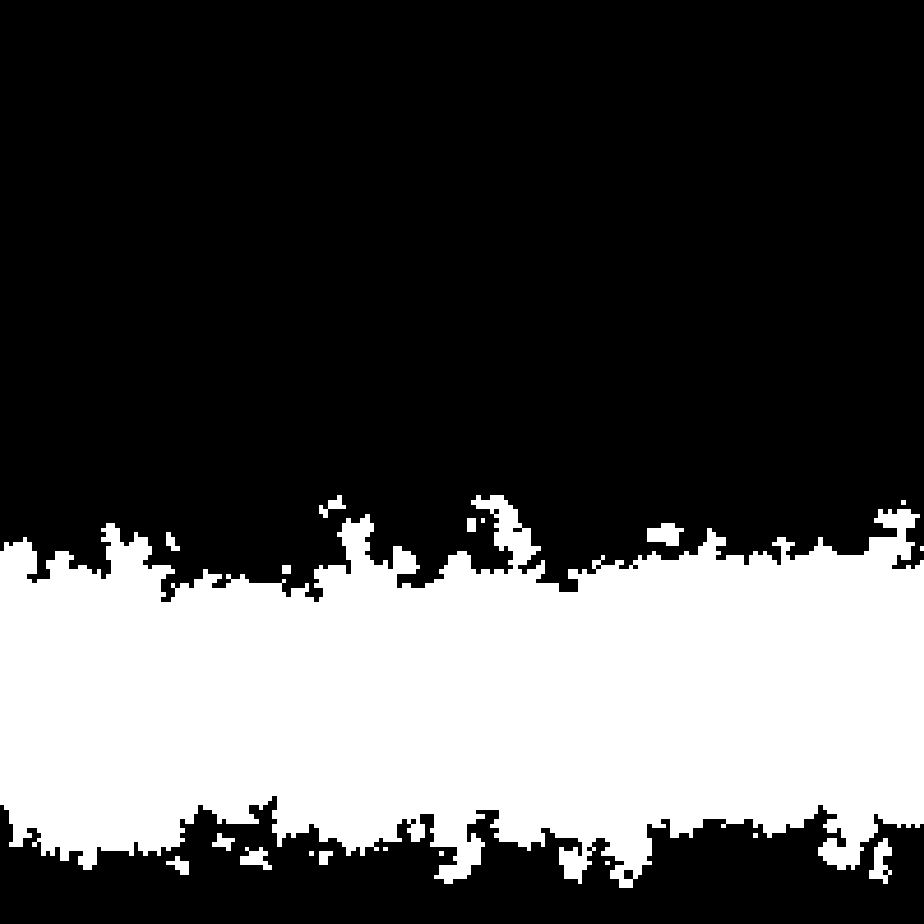}}\\
				\vspace{-2mm}
				{\footnotesize t=100}
			\end{minipage}
			\begin{minipage}{0.14\linewidth}
				\centering
				\fbox{\includegraphics[width=\linewidth]{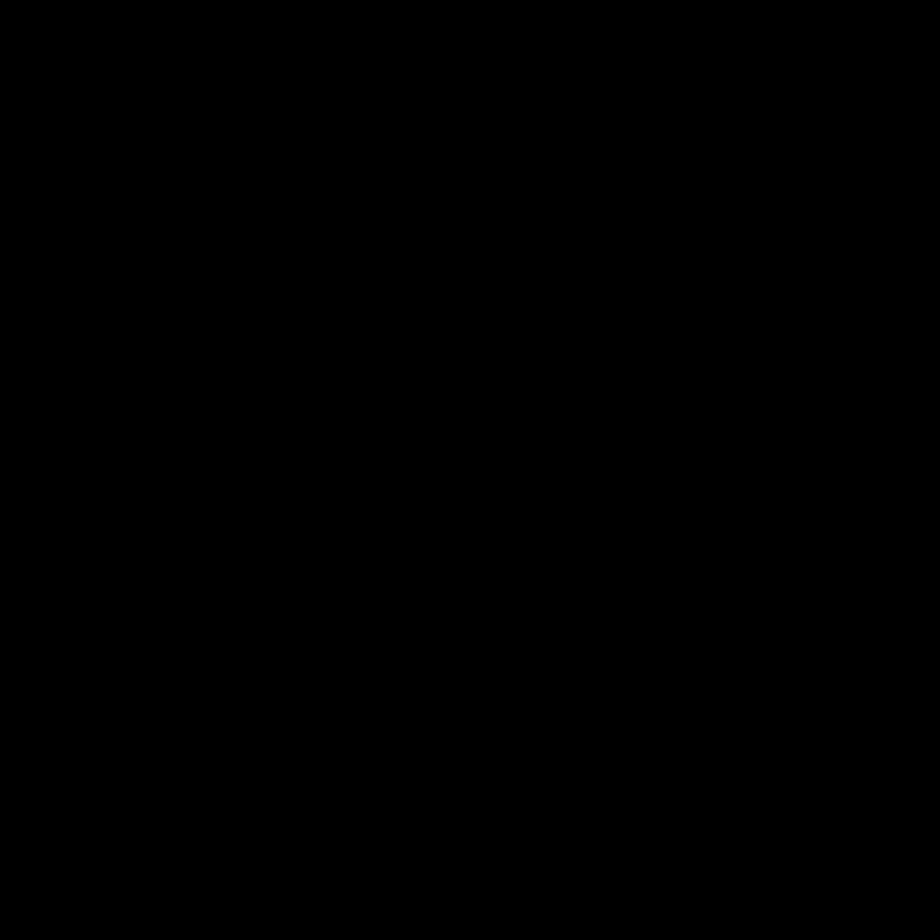}}\\
				\vspace{-2mm}
				{\footnotesize t=1000}
			\end{minipage}
			\begin{minipage}{0.14\linewidth}
				\centering
				\fbox{\includegraphics[width=\linewidth]{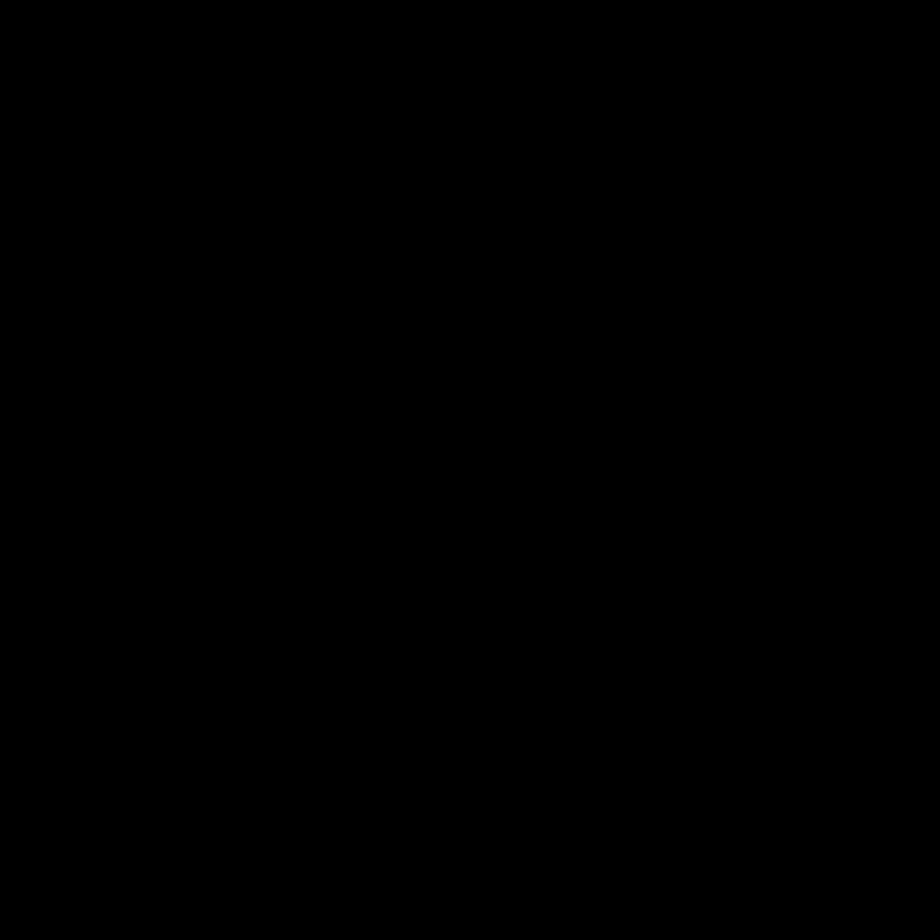}}\\
				\vspace{-2mm}
				{\footnotesize t=10000}
			\end{minipage}
			\vspace{-3mm}
			\caption*{\footnotesize (d) Fermi update rule}
		\end{minipage}	
		\caption{Simulation for TUC-PPO, PPO, Q-learning, and Fermi update rule at enhancement factor $r=3.3$. Initial conditions place defectors (black) in the upper half and cooperators (white) in the lower half. The leftmost subfigure in each row presents the temporal evolution curve. Subfigures (a) and (b) show state snapshots at iterations $t = 0, 1, 10, 100$, and $1000$. Subfigures (c) and (d) display snapshots at $t = 0, 10, 100, 1000$, and $10000$. TUC-PPO achieves rapid global cooperation, outperforming baselines that either fail to sustain cooperation (PPO/Fermi), or achieve only suboptimal cooperation (Q-learning).}
		\label{fig:PPO-TUC_uDbC_compare}
	\end{figure*}

	Figure \ref{fig:PPO-TUC_uDbC_compare} presents a comparative analysis of four algorithms: PPO-TUC, PPO, Q-learning, and the Fermi update rule. These were evaluated under an enhancement factor $r=3.3$. The simulation initializes with defectors concentrated in the upper half of the domain and cooperators in the lower half. The left subfigure displays the evolution of cooperation and defection rates over time. In this visual, the iteration count is plotted along the horizontal axis, with the fraction of cooperators and defectors represented by blue and red curves, respectively. The remaining subfigures capture spatial snapshots at key intervals, where white pixels denote cooperators and black pixels indicate defectors. By examining both temporal trends and spatial configurations, the results demonstrate significant variations in algorithmic performance within critical parameter ranges. TUC-PPO and PPO have only undergone 1,000 iterations, while Q-learning and Fermi update rules have undergone 10,000 iterations.
	
	Our TUC-PPO achieves full cooperation among all agents within just 20 iterations, demonstrating significantly faster convergence than other methods. Initially, the randomly initialized policy network leads to mixed strategies among agents.  The standard PPO without the TUC mechanism rapidly converges to complete defection within 100 iterations, with defectors eventually dominating the entire population later in training. Q-learning requires nearly 300 iterations to stabilize, yet the cooperation fraction remains below $40\%$, reflecting its inefficiency in policy optimization. The Fermi update rule, under the current enhancement factor, fails to sustain cooperation, with defectors completely overtaking the population before 300 iterations.   These results highlight TUC-PPO's superior convergence speed and stability in promoting cooperation. While the PPO and Fermi update rule succumb to defectors, and Q-learning struggles with inefficient learning, our method ensures rapid and robust cooperation emergence. The comparison underscores the critical role of temporal update coordination in multi-agent RL for evolutionary games.
	
	\subsection{Algorithm performance evaluation under varying enhancement factors $r$}
	\label{exp_r}
	
\begin{figure*}[h]
		\begin{minipage}{0.48\linewidth}
			\centering
			\includegraphics[width=\linewidth]{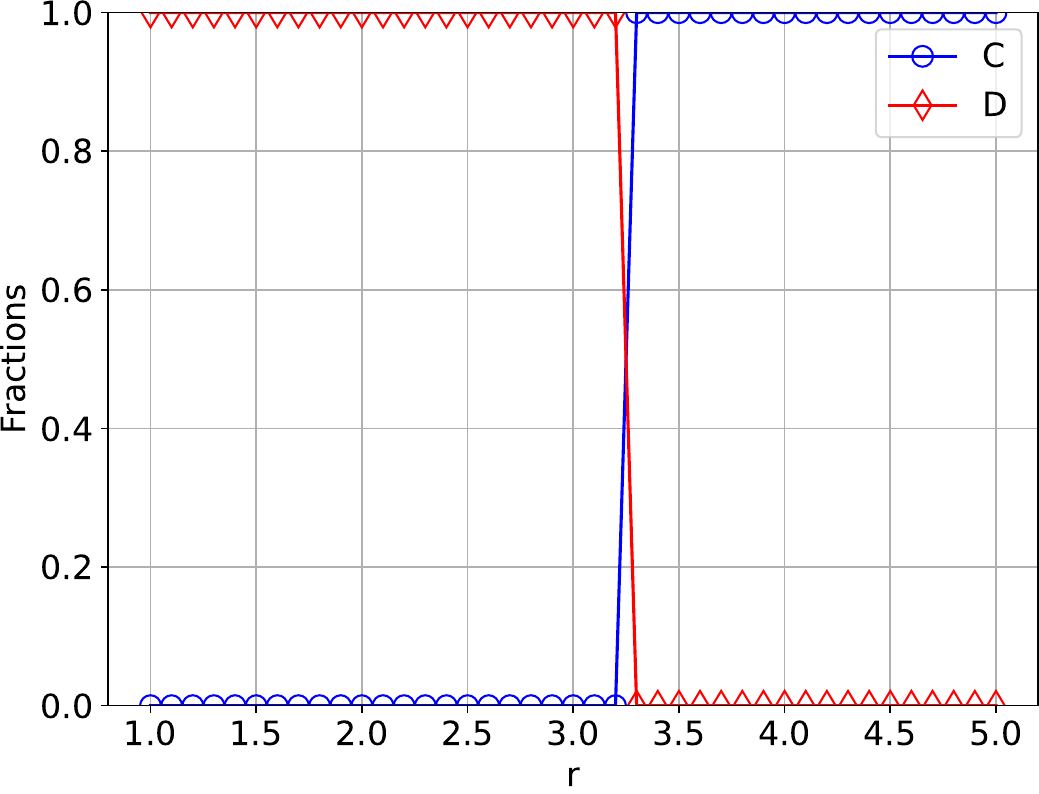}\\
			\vspace{-4mm}
			\caption*{\footnotesize (a) TUC-PPO}
		\end{minipage}
		\begin{minipage}{0.48\linewidth}
			\centering
			\includegraphics[width=\linewidth]{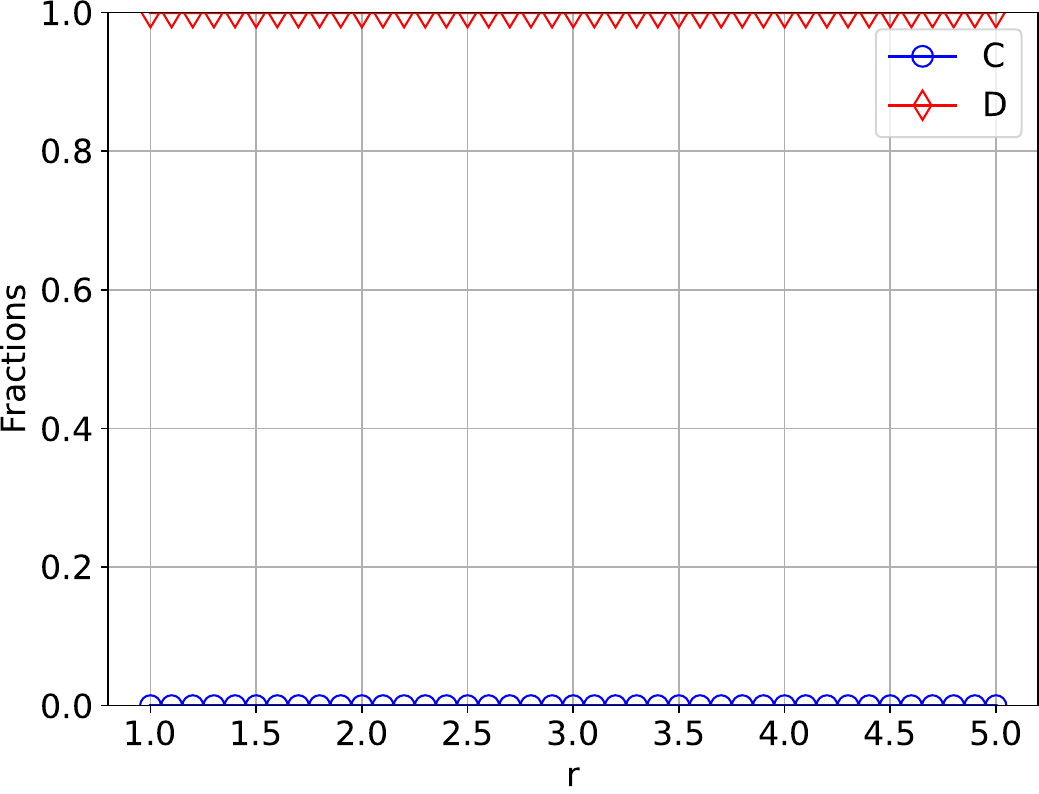}\\
			\vspace{-4mm}
			\caption*{\footnotesize (b) PPO}
		\end{minipage}
		\\
		[3mm]
		\begin{minipage}{0.48\linewidth}
			\centering
			\includegraphics[width=\linewidth]{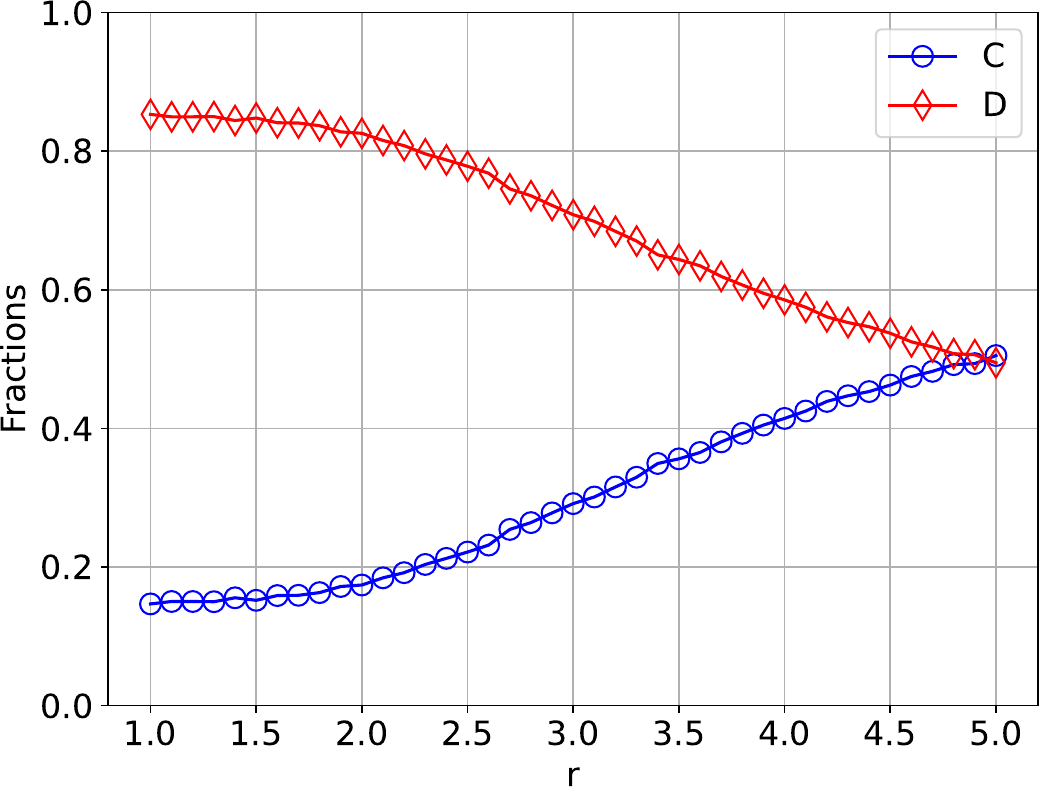}\\
			\vspace{-4mm}
			\caption*{\footnotesize (c) Q-learing}
		\end{minipage}	
		\begin{minipage}{0.48\linewidth}
			\centering
			\includegraphics[width=\linewidth]{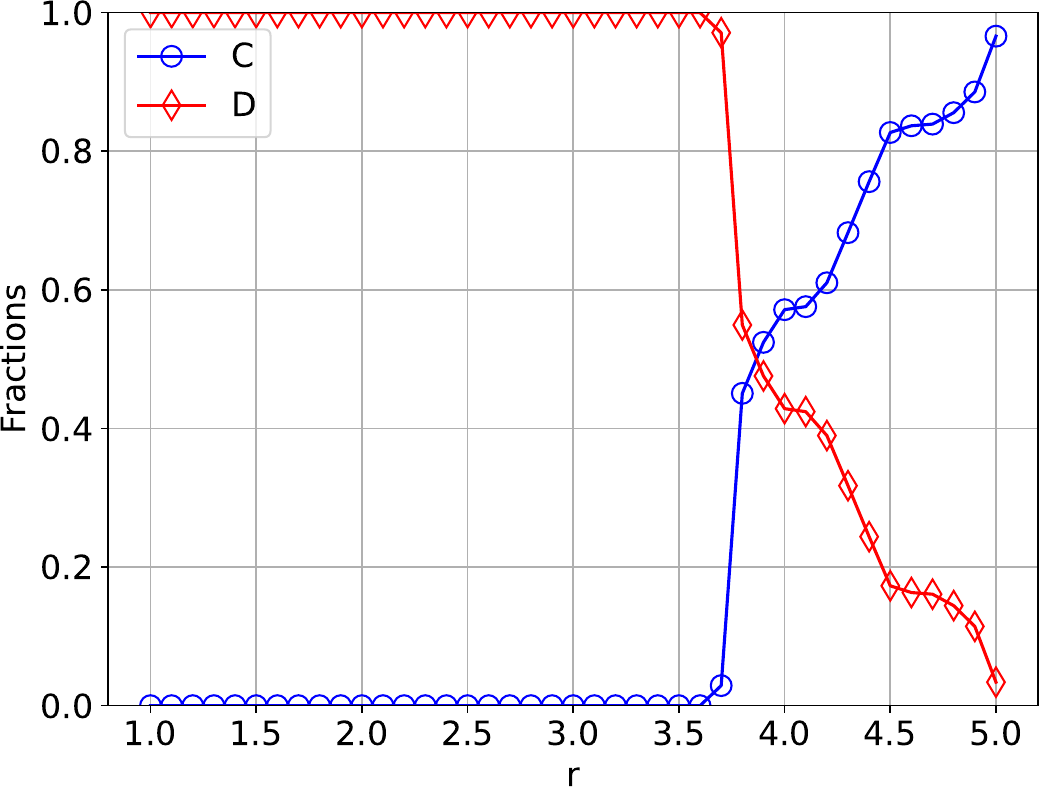}\\
			\vspace{-4mm}
			\caption*{\footnotesize (d) Fermi}
		\end{minipage}	
		\caption{Performance comparison of TUC-PPO, PPO, Q-learning, and Fermi update rule under varying enhancement factors $r$. Blue circles represent cooperators while red diamonds indicate defectors. Initial conditions positioned all defectors D in the upper half of the grid and all cooperators C in the lower half. TUC-PPO achieves full cooperation at lower enhancement factors, outperforming other methods that fail to sustain cooperation, reach only suboptimal levels, or require higher incentive thresholds.}
		\label{fig:TUC-PPO_r_uDbC_compare}
	\end{figure*}

	The experiment systematically evaluates four computational approaches including TUC-PPO, PPO, Q-learning, and Fermi dynamics. The experimental setup maintains consistent initial spatial distributions where defectors occupy the upper region and cooperators the lower region of the domain. Figure \ref{fig:TUC-PPO_r_uDbC_compare} displays the relationship between fractions measured on the vertical axis and enhancement factor $r$ shown on the horizontal axis.
	
	Our TUC-PPO framework demonstrates superior performance in promoting cooperative behavior under demanding conditions. When the enhancement factor $r\ge3.3$, all agents become cooperators by the end of 1000 iterations. This achievement originates from TUC-PPO's team utility constraint, which effectively strengthens agents' inclination toward cooperation. The algorithm achieves complete cooperation at substantially lower enhancement factors compared to traditional approaches. Standard PPO without TUC exhibits fundamental limitations in the discovery of cooperative strategies. For enhancement factors below 5.0, all agents inevitably converge to become defectors. This pathological convergence demonstrates PPO's tendency toward extreme strategy polarization. Such pathological convergence stems from the policy update mechanism amplifying initial random biases. This process drives the entire population toward complete cooperation or complete defection, eliminating intermediate stable states. Q-learning shows different behavioral characteristics, maintaining some cooperators even at relatively low $r$-values.  While the fraction of cooperators increases with higher enhancement factors, the approach fails to achieve majority cooperation, reaching only $55\%$ even at $r=5.0$. This performance ceiling reflects the inherent limitations of tabular methods in modeling continuous policy spaces and spatial agent interactions. The Fermi update rule demonstrates intermediate performance, producing cooperative when $r$ exceeds 3.7. However, this imitate update mechanism cannot guarantee full cooperation even at $r=5.0$, instead showing gradual improvement in cooperation levels. This contrasts sharply with TUC-PPO's ability to achieve complete cooperation at substantially lower enhancement factors.
	
	The comparative results yield fundamental insights about TUC-PPO's advantages. The team utility constraint successfully enforces cooperative behavior in scenarios where conventional methods fail. This constraint mechanism achieves full cooperation with remarkable efficiency, outperforming both value-based methods and local interaction rules. Most importantly, the results demonstrate how explicitly modeling team utility can resolve core limitations in multi-agent learning systems. This approach offers an effective solution for evolutionary game scenarios. These findings position TUC-PPO as an important innovation in cooperative RL through its principled team-based optimization approach.
	
	\subsection{Statistical analysis of TUC-PPO}
	\label{exp:compare_stat}
	
	To rigorously evaluate the operational reliability of TUC-PPO, we implemented three complementary statistical visualization approaches. These methods analyze distribution spread through error bars, uncover probability density patterns via violin plots, and detail precision metrics using comparative tables of $95\%$ confidence intervals. Across the enhancement factor domain where $r \in [1.0,6.0]$, we executed 50 independent experimental trials for each parameter configuration.
	
\begin{figure*}[htbp!]
		\begin{minipage}{0.48\linewidth}
			\centering
			\includegraphics[width=\linewidth]{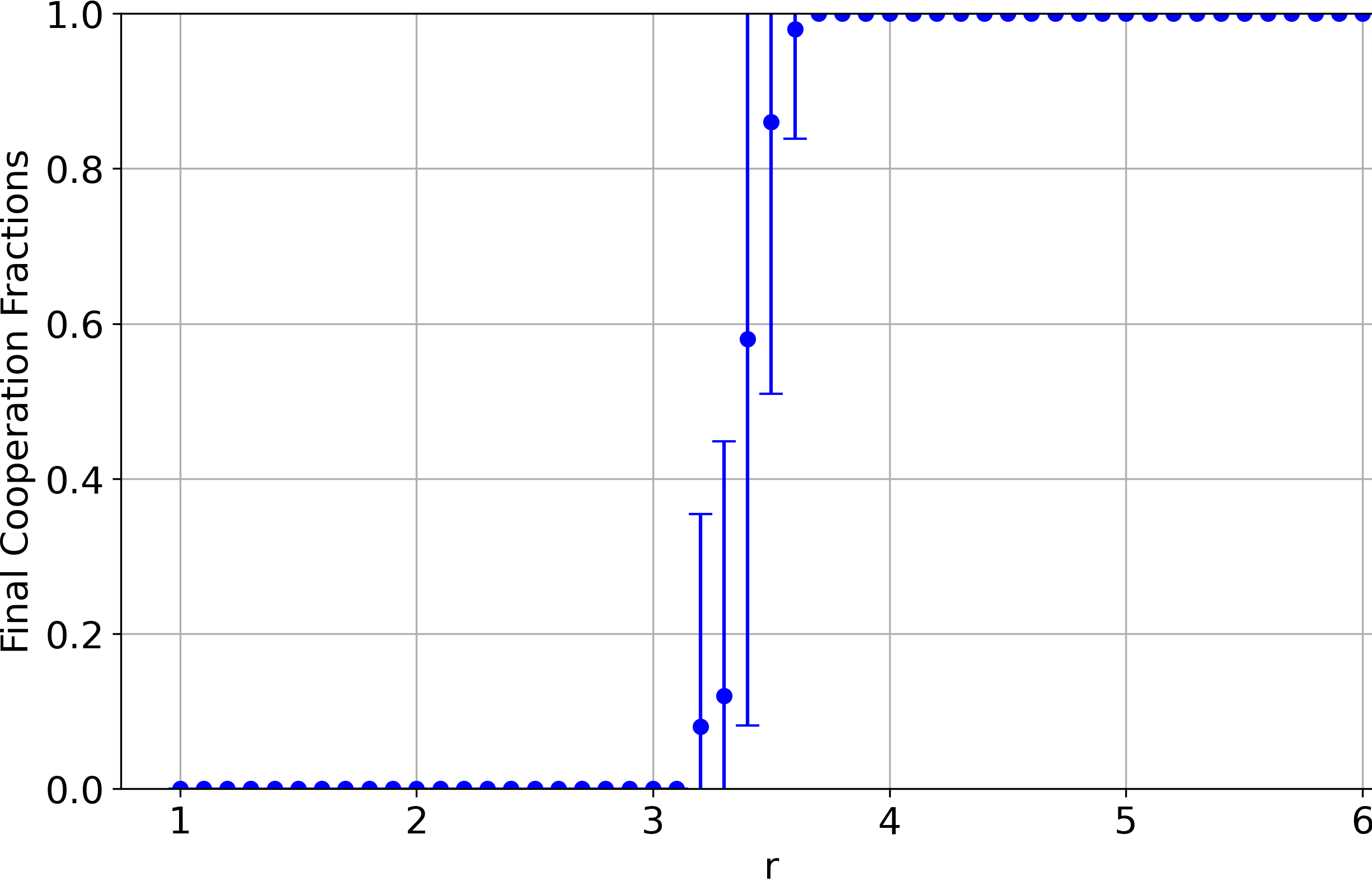}\\
			\vspace{-4mm}
			\caption*{\footnotesize (a) TUC-PPO}
		\end{minipage}
		\begin{minipage}{0.48\linewidth}
			\centering
			\includegraphics[width=\linewidth]{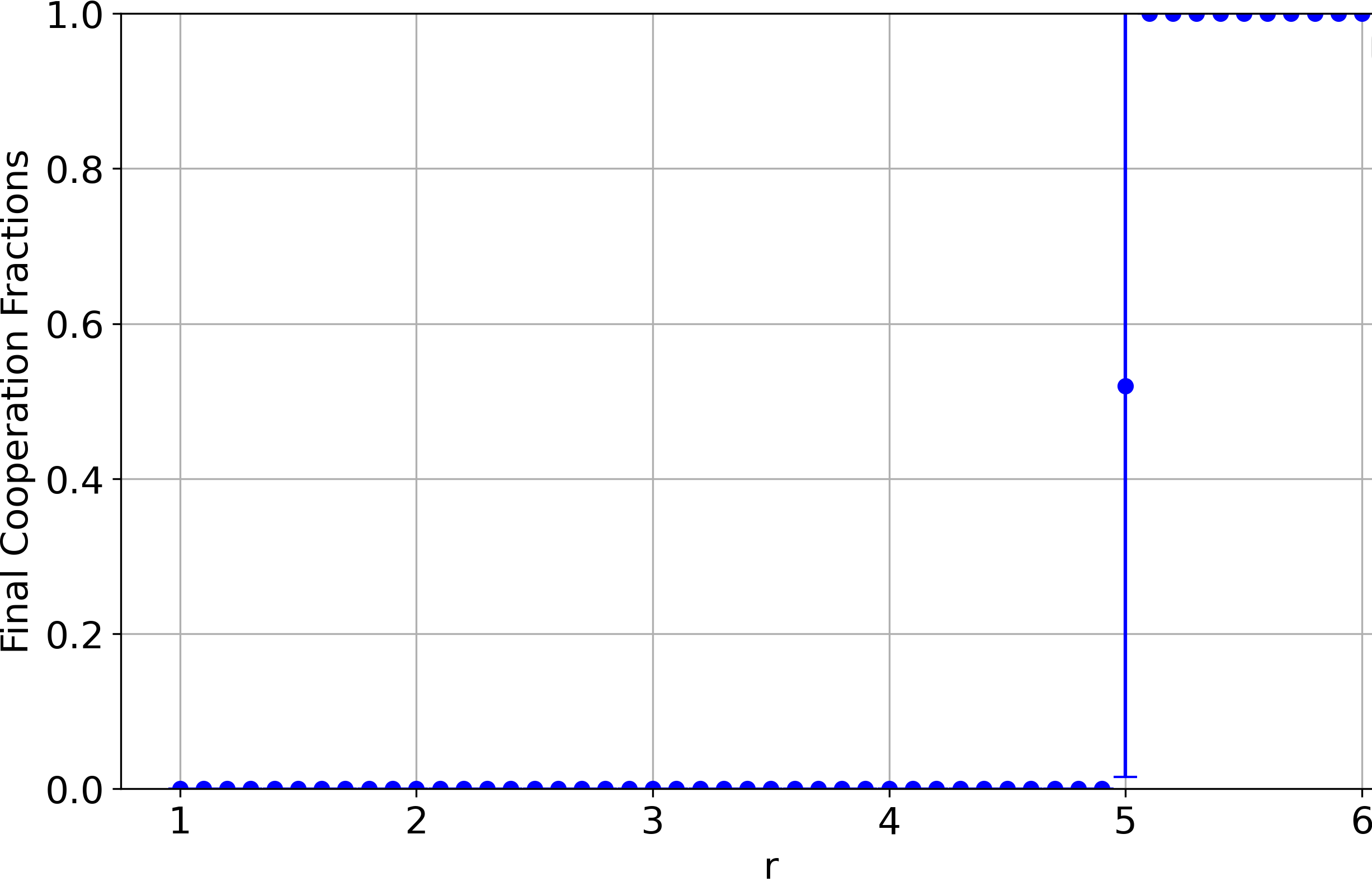}\\
			\vspace{-4mm}
			\caption*{\footnotesize (b) PPO}
		\end{minipage}
		\caption{Comparison of mean cooperation fractions and standard deviation between TUC-PPO and PPO. TUC-PPO achieves a state where all agents are collaborators at lower enhancement factors $r$ than standard PPO through the integration of the TUC.}
		\label{fig:TUC-PPO_r_stat_err}
	\end{figure*}

	Experimental executions consistently generated dichotomous outcomes for TUC-PPO and PPO, where individual runs manifested either $0\%$ or $100\%$ cooperation probabilities. This behavior produced substantial standard deviations across all enhancement factor values. Fig.~\ref{fig:TUC-PPO_r_stat_err} fundamentally distinguishes TUC-PPO from baseline PPO through error bar analysis quantifying differences in cooperation rate distributions and variability patterns. This visualization confirms the core mechanism's efficacy in diminishing cooperation barriers. Specifically, TUC-PPO achieves reliable cooperation significantly earlier than the baseline algorithm while requiring lower enhancement factors.
	﻿
	
\begin{figure*}[htbp!]
		\centering
		\begin{minipage}{0.8\linewidth}
			\centering
			\includegraphics[width=\linewidth]{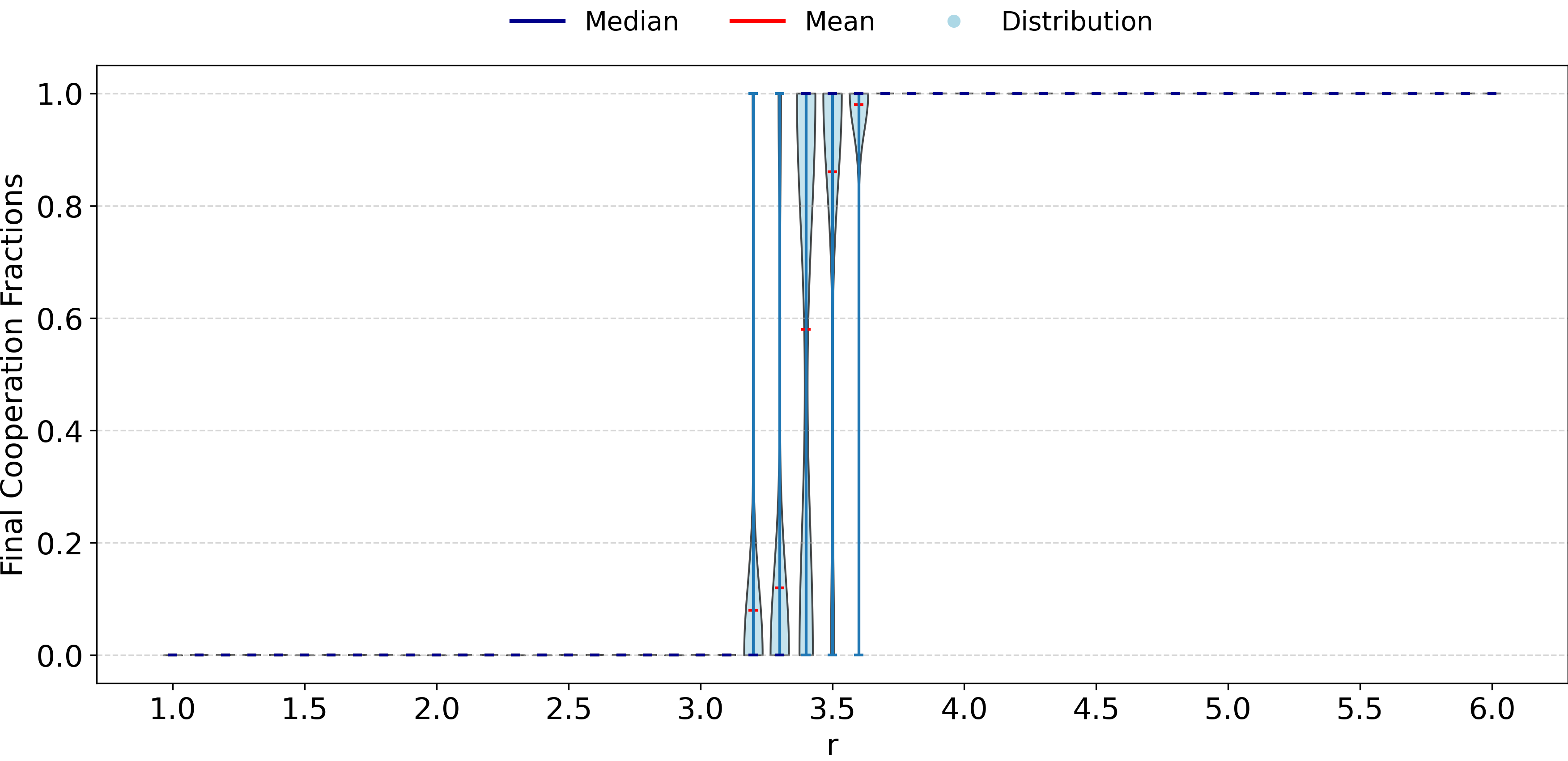}\\
			\vspace{-4mm}
			\caption*{\footnotesize (a) TUC-PPO}
		\end{minipage}
		\\[1mm]
		\begin{minipage}{0.8\linewidth}
			\centering
			\includegraphics[width=\linewidth]{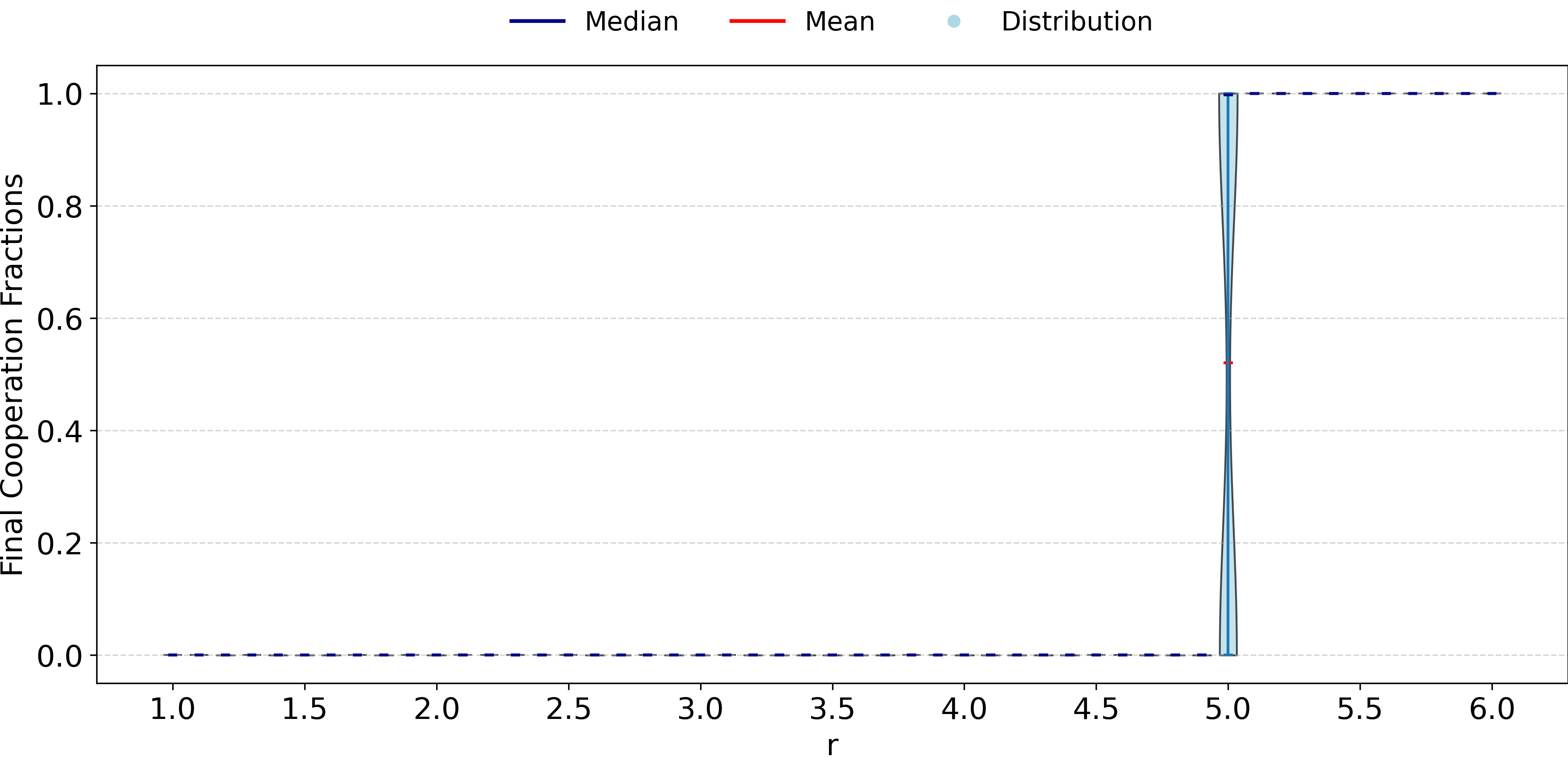}\\
			\vspace{-4mm}
			\caption*{\footnotesize (b) PPO}
		\end{minipage}	
		\caption{Probability density of cooperation fractions for TUC-PPO and standard PPO across enhancement factors $r \in [1.0, 6.0]$. Violin plots show that compared to PPO, TUC-PPO enables all agents to ultimately select the cooperative strategy at smaller enhancement factors.}
		\label{fig:TUC-PPO_r_stat_vio}
	\end{figure*}

	Violin plot analysis (Fig.~\ref{fig:TUC-PPO_r_stat_vio}) reveals TUC-PPO's superior cooperation dynamics compared to baseline PPO. At enhancement factors below 3.2, TUC-PPO maintains stable $0\%$ cooperation. Between $r=3.2$ and $3.6$, its cooperation distribution progressively approaches optimal values. Beyond r=3.6, TUC-PPO achieves perfect $100\%$ cooperation stability. This performance contrasts with baseline PPO, which sustains $0\%$ cooperation below $r=5$ and reaches stable $100\%$ cooperation above this threshold. At the critical $r=5$ point, baseline PPO achieves approximately half the cooperation rate of TUC-PPO. The distributions demonstrate TUC-PPO's decisive advantages through establishing stable cooperation at lower enhancement factors while maintaining superior performance at critical thresholds and transitioning more efficiently to optimal cooperation.
	
As shown in Table~\ref{tab:CI_comparison}, the confidence interval analysis confirms two critical behavioral mechanisms affecting convergence. First, NaN intervals emerge when algorithms achieve perfect deterministic outcomes ($100\%$ cooperation or $0\%$ cooperation) across all trials. This results in zero standard deviation, which invalidates confidence interval computation. Second, the $[0.00, 0.00]$ intervals represent scenarios where exploration strategies prevent absolute convergence. During 1000 training iterations, PPO's exploration mechanisms cause strategy deviations that maintain minimal but non-zero cooperation or defection probabilities. This proves that 1000 iterations cannot achieve complete convergence for PPO, but it is already very close. These stochastic policy decisions prevent either perfect cooperation or complete defection in final states. The results demonstrate TUC-PPO's superior adaptability in social dilemmas. Moreover, PPO requires substantially higher enhancement factors ($r>5.0$) to achieve meaningful cooperation and suffers from convergence instability at critical thresholds. In contrast, TUC-PPO establishes reliable cooperation patterns at lower enhancement factors ($r\ge3.6$) with smoother probabilistic transitions. This performance advantage stems from TUC-PPO's architectural capability to maintain cooperative equilibria despite exploration noise. Comparatively, PPO's strategy exhibits greater vulnerability to exploration-induced deviations that delay convergence and reduce cooperation reliability.
	
	\begin{table}[h]
		\centering
		\footnotesize
		\caption{$95\%$ confidence intervals comparison for cooperation fractions}
		\label{tab:CI_comparison}
		\resizebox{\textwidth}{!}{ 
			\begin{tabular}{@{}c*{7}{S[table-format=1.2]@{\,--\,}S[table-format=1.2]}@{}}
				\toprule
				r & \multicolumn{2}{c}{3.1} & \multicolumn{2}{c}{3.2} & \multicolumn{2}{c}{3.3} & \multicolumn{2}{c}{3.4} & \multicolumn{2}{c}{3.5} & \multicolumn{2}{c}{3.6} & \multicolumn{2}{c}{3.7} \\
				\cmidrule(lr){2-3} \cmidrule(lr){4-5} \cmidrule(lr){6-7} \cmidrule(lr){8-9} \cmidrule(lr){10-11} \cmidrule(lr){12-13} \cmidrule(lr){14-15}
				TUC-PPO & nan & nan & 0.00 & 0.16 & 0.03 & 0.21 & 0.44 & 0.72 & 0.76 & 0.96 & 0.94 & 1.02 &  nan & nan \\
				PPO & nan & nan & 0.00 & 0.00 & 0.00 & 0.00  & 0.00 & 0.00  & 0.00 & 0.00 & 0.00 & 0.00  & 0.00 & 0.00 \\
				\midrule
				
				r & \multicolumn{2}{c}{3.8} & \multicolumn{2}{c}{3.9} & \multicolumn{2}{c}{4.0} & \multicolumn{2}{c}{4.1} & \multicolumn{2}{c}{4.2} & \multicolumn{2}{c}{4.3} & \multicolumn{2}{c}{4.4} \\
				\cmidrule(lr){2-3} \cmidrule(lr){4-5} \cmidrule(lr){6-7} \cmidrule(lr){8-9} \cmidrule(lr){10-11} \cmidrule(lr){12-13} \cmidrule(lr){14-15}
				TUC-PPO &  nan & nan &  nan & nan &  nan & nan &  nan & nan & nan & nan & nan & nan & nan & nan \\
				PPO &nan & nan  & nan & nan  & 0.00 & 0.00 & 0.00 & 0.00 & 0.00 & 0.00  & 0.00 & 0.00  & 0.00 & 0.00 \\
				\midrule
				
				r & \multicolumn{2}{c}{4.5} & \multicolumn{2}{c}{4.6} & \multicolumn{2}{c}{4.7} & \multicolumn{2}{c}{4.8} & \multicolumn{2}{c}{4.9} & \multicolumn{2}{c}{5.0} & \multicolumn{2}{c}{5.1} \\
				\cmidrule(lr){2-3} \cmidrule(lr){4-5} \cmidrule(lr){6-7} \cmidrule(lr){8-9} \cmidrule(lr){10-11} \cmidrule(lr){12-13} \cmidrule(lr){14-15}
				TUC-PPO & nan & nan & nan & nan & nan & nan & nan & nan & nan & nan& nan & nan & nan & nan \\
				PPO & nan & nan & nan & nan  & 0.00 & 0.00  & 0.00 & 0.00 & 0.00 & 0.00 & 0.38 & 0.66 & nan & nan \\
				\bottomrule
			\end{tabular}
		}
	\end{table}
	
	\subsection{TUC-PPO with half-and-half initialization}
	\label{exp_hh}
	
\begin{figure*}[htbp!]
		\begin{minipage}{0.45\linewidth}
			\begin{minipage}{\linewidth}
				\centering
				\includegraphics[width=\linewidth]{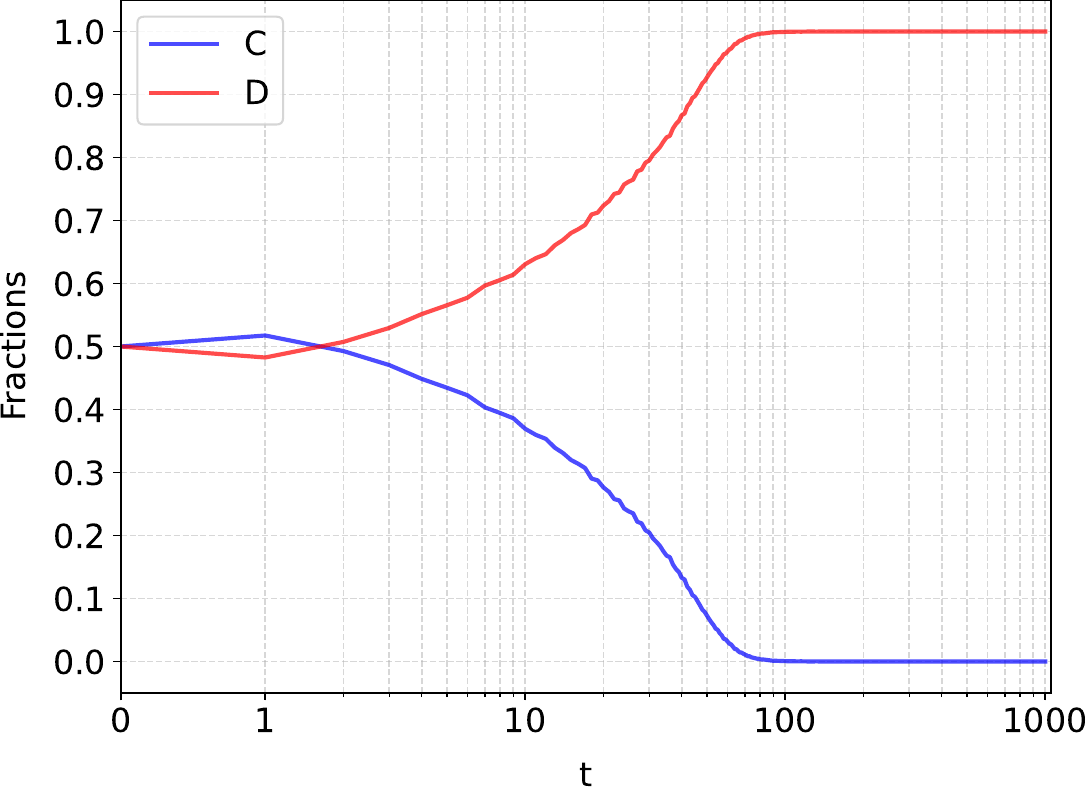}\\
			\end{minipage}
			\vspace{2mm}
			\\
			\begin{minipage}{0.188\linewidth}
				\centering
				\fbox{\includegraphics[width=\linewidth]{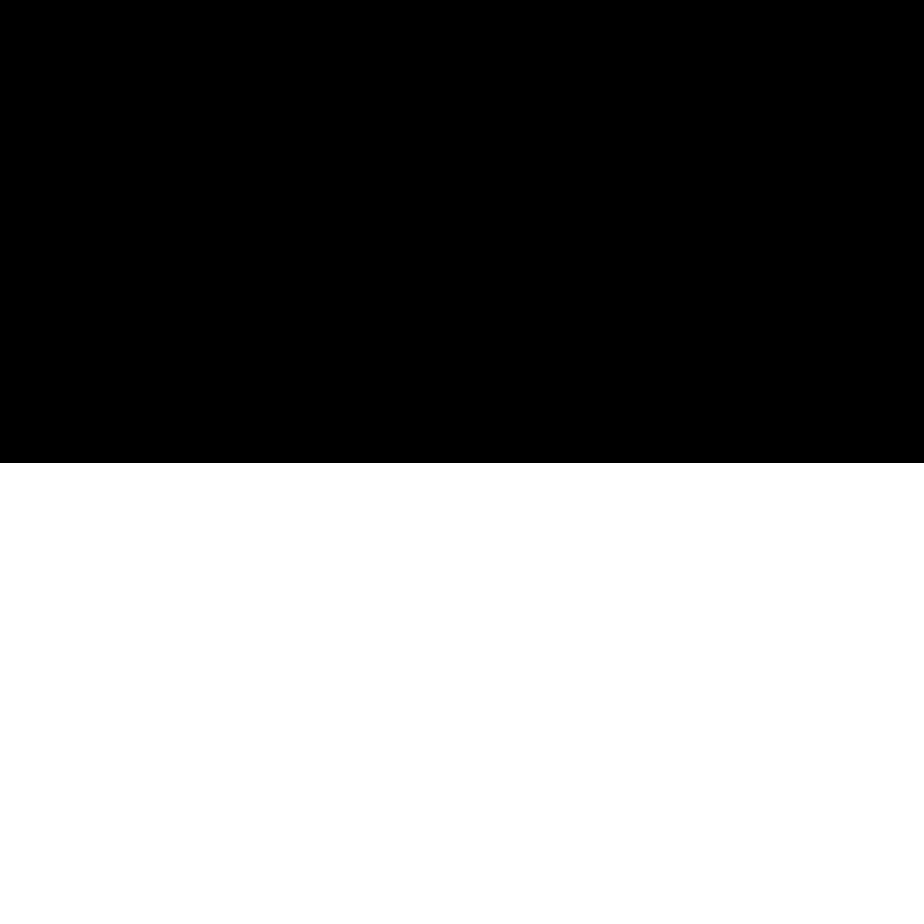}}\\
				\vspace{-2mm}
				{\footnotesize t=0}
			\end{minipage}
			\begin{minipage}{0.188\linewidth}
				\centering
				\fbox{\includegraphics[width=\linewidth]{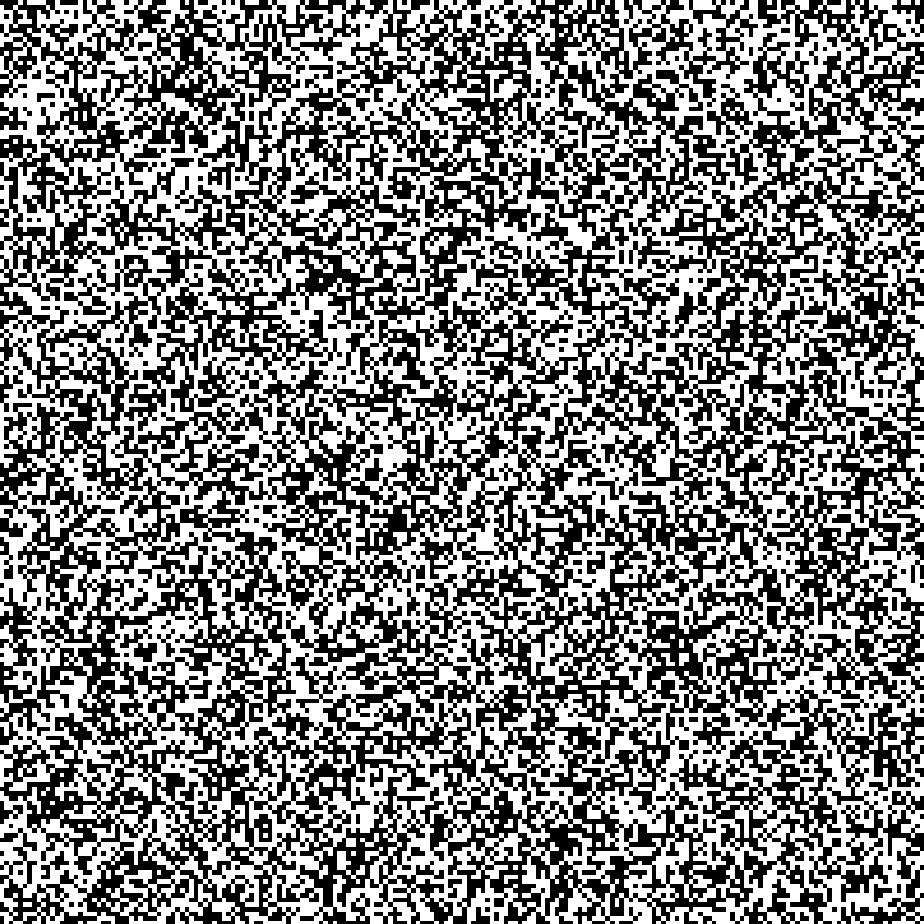}}\\
				\vspace{-2mm}
				{\footnotesize t=1}
			\end{minipage}
			\begin{minipage}{0.188\linewidth}
				\centering
				\fbox{\includegraphics[width=\linewidth]{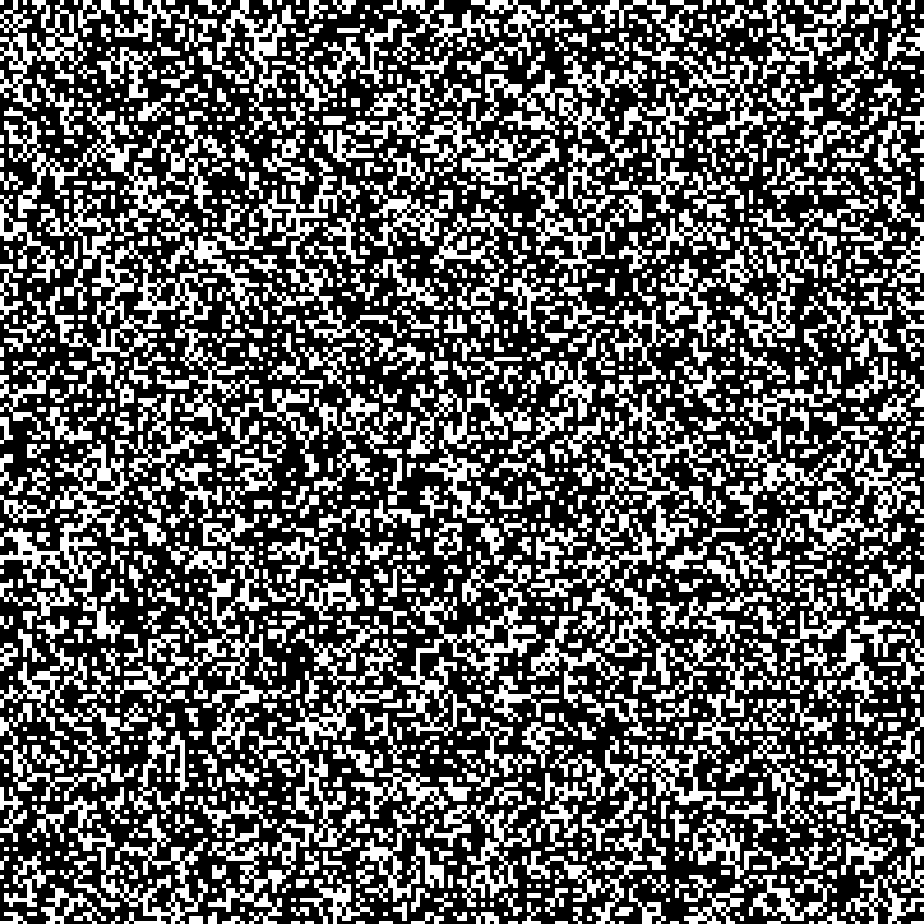}}\\
				\vspace{-2mm}
				{\footnotesize t=10}
			\end{minipage}
			\begin{minipage}{0.188\linewidth}
				\centering
				\fbox{\includegraphics[width=\linewidth]{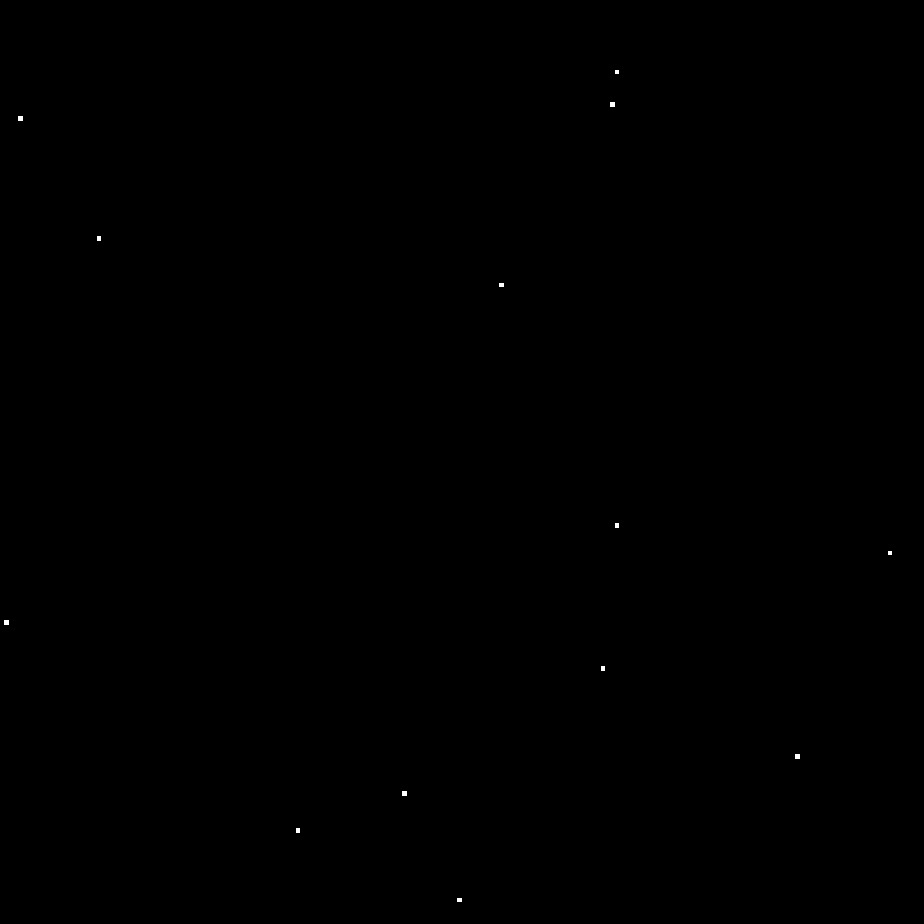}}\\
				\vspace{-2mm}
				{\footnotesize t=100}
			\end{minipage}
			\begin{minipage}{0.188\linewidth}
				\centering
				\fbox{\includegraphics[width=\linewidth]{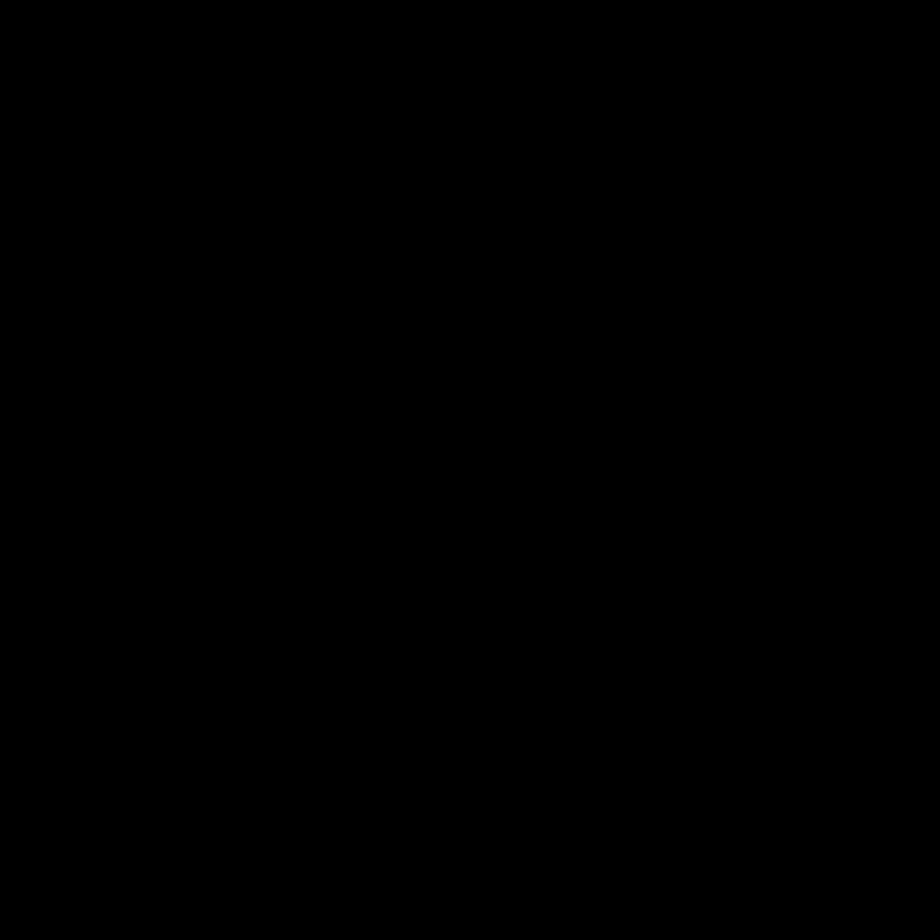}}\\
				\vspace{-2mm}
				{\footnotesize t=1000}
			\end{minipage}
			\vspace{-2mm}
			\caption*{\footnotesize (a) r=3.0}
		\end{minipage}
		\hfill
		\begin{minipage}{0.45\linewidth}
			\begin{minipage}{\linewidth}
				\centering
				\includegraphics[width=\linewidth]{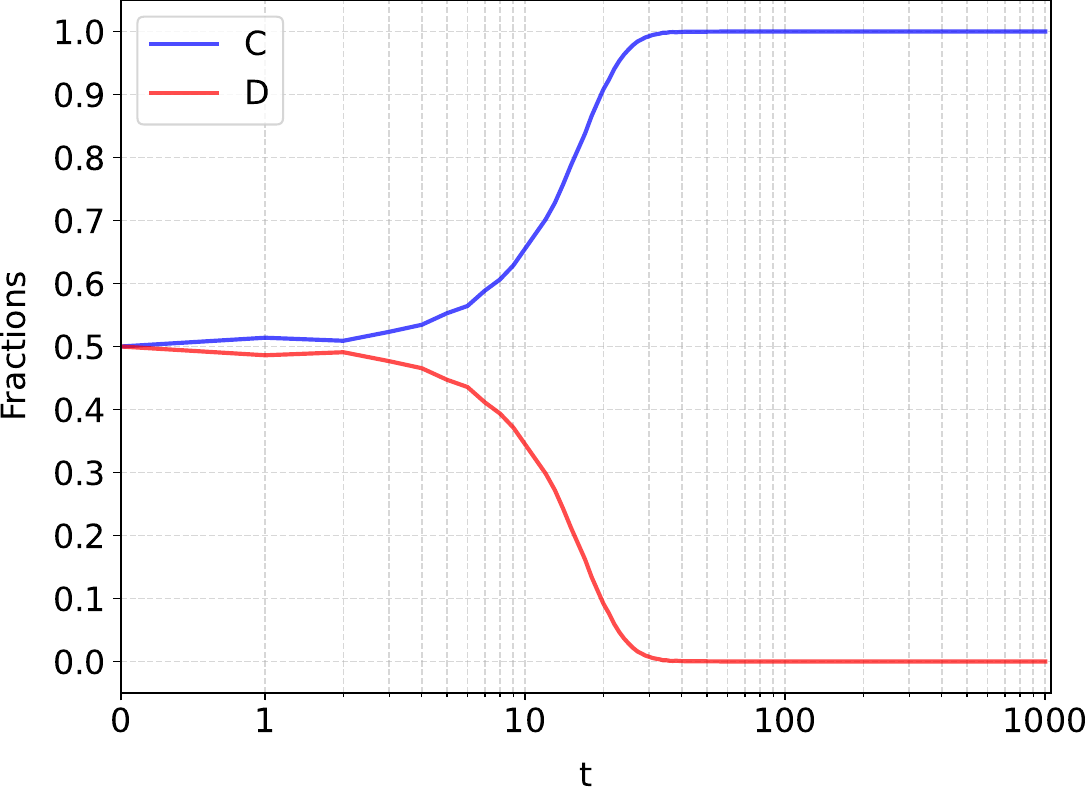}\\
			\end{minipage}
			\vspace{2mm}
			\\
			\begin{minipage}{0.188\linewidth}
				\centering
				\fbox{\includegraphics[width=\linewidth]{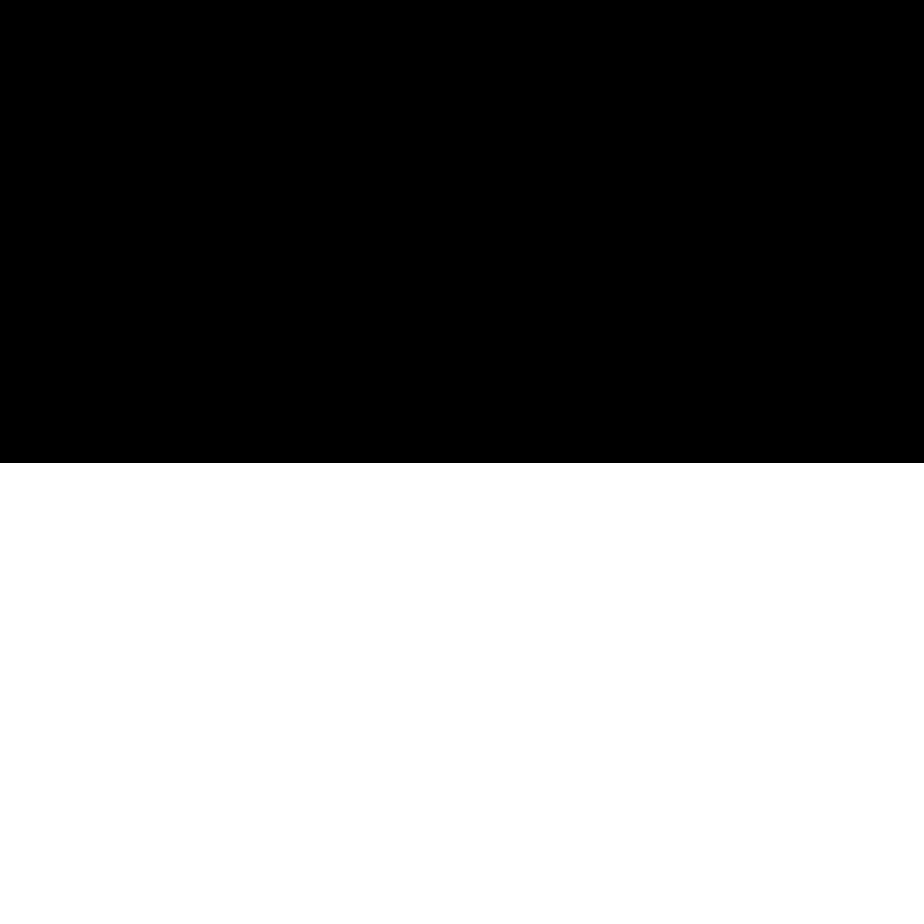}}\\
				\vspace{-2mm}
				{\footnotesize t=0}
			\end{minipage}
			\begin{minipage}{0.188\linewidth}
				\centering
				\fbox{\includegraphics[width=\linewidth]{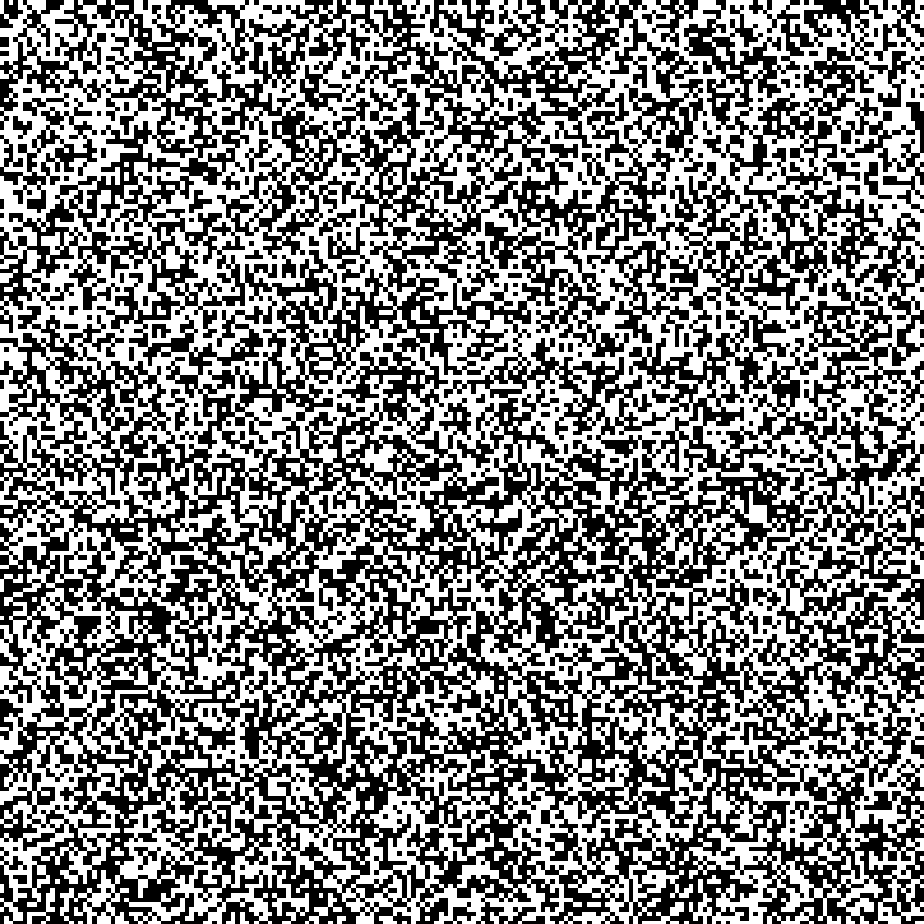}}\\
				\vspace{-2mm}
				{\footnotesize t=1}
			\end{minipage}
			\begin{minipage}{0.188\linewidth}
				\centering
				\fbox{\includegraphics[width=\linewidth]{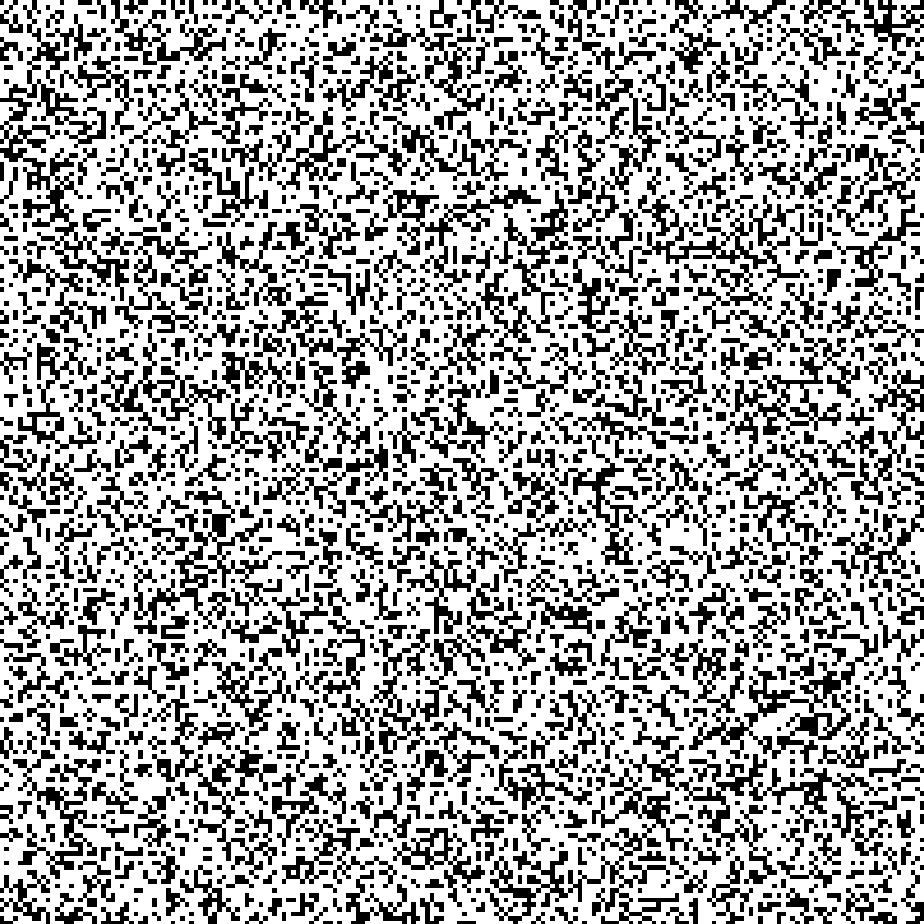}}\\
				\vspace{-2mm}
				{\footnotesize t=10}
			\end{minipage}
			\begin{minipage}{0.188\linewidth}
				\centering
				\fbox{\includegraphics[width=\linewidth]{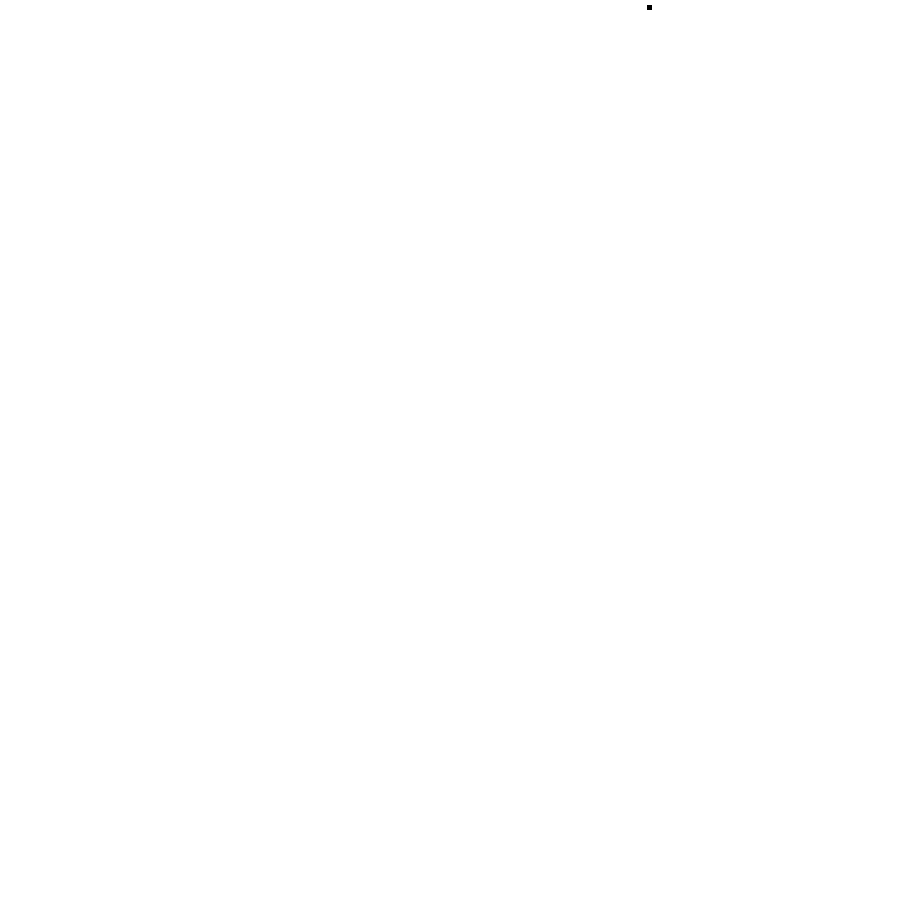}}\\
				\vspace{-2mm}
				{\footnotesize t=100}
			\end{minipage}
			\begin{minipage}{0.188\linewidth}
				\centering
				\fbox{\includegraphics[width=\linewidth]{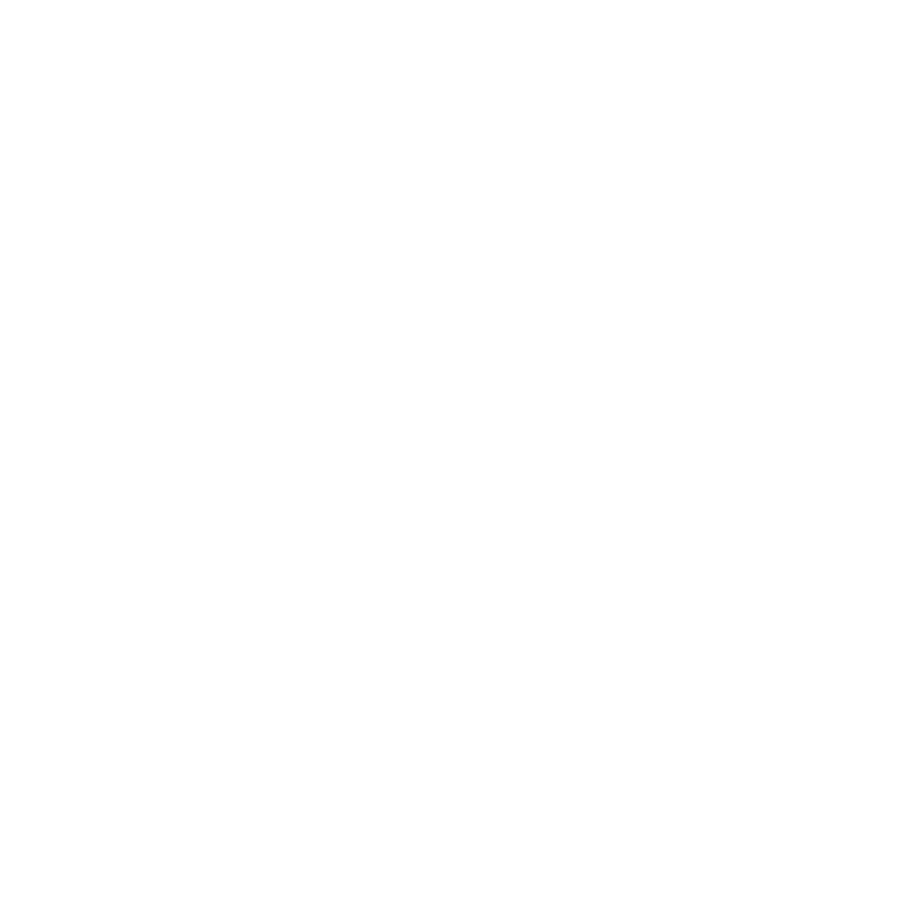}}\\
				\vspace{-2mm}
				{\footnotesize t=1000}
			\end{minipage}
			\vspace{-2mm}
			\caption*{\footnotesize (b) r=3.5}
		\end{minipage}
		\\
		[2mm]
		\begin{minipage}{\linewidth}
				\begin{minipage}{0.188\linewidth}
					\centering
					\includegraphics[width=\linewidth]{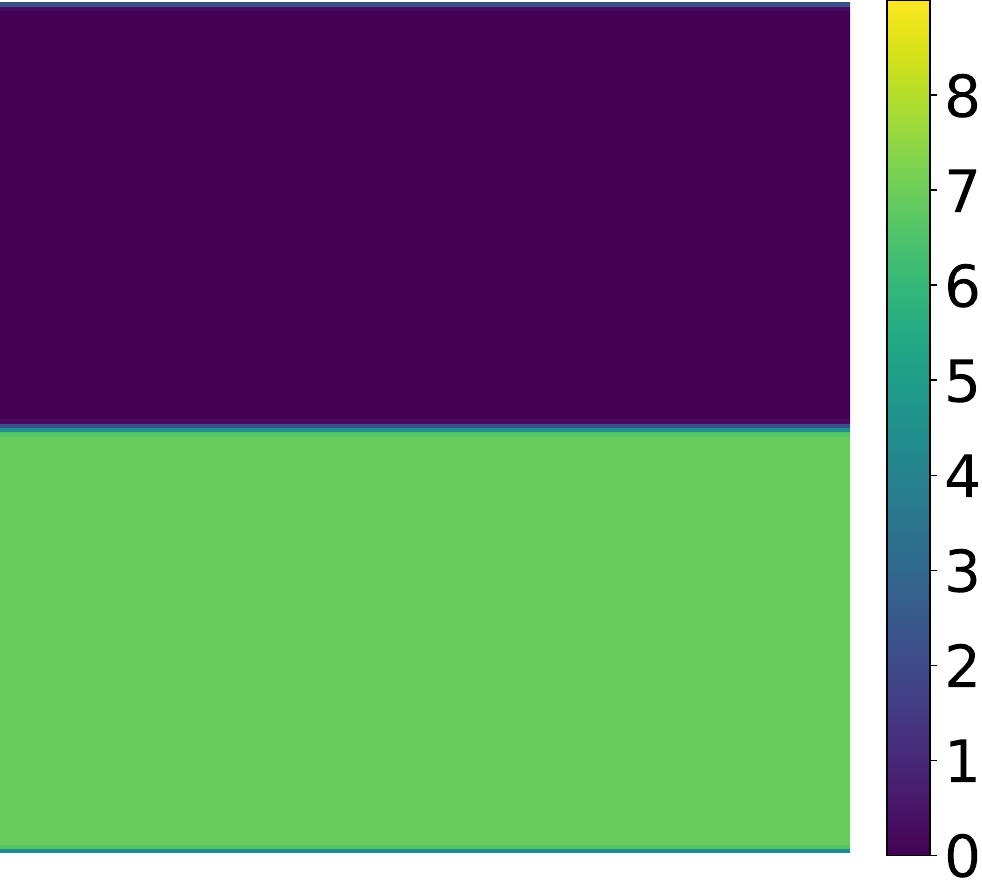}\\
					\vspace{-2mm}
					{\footnotesize t=0}
				\end{minipage}
				\hfill
				\begin{minipage}{0.188\linewidth}
					\centering
					\includegraphics[width=\linewidth]{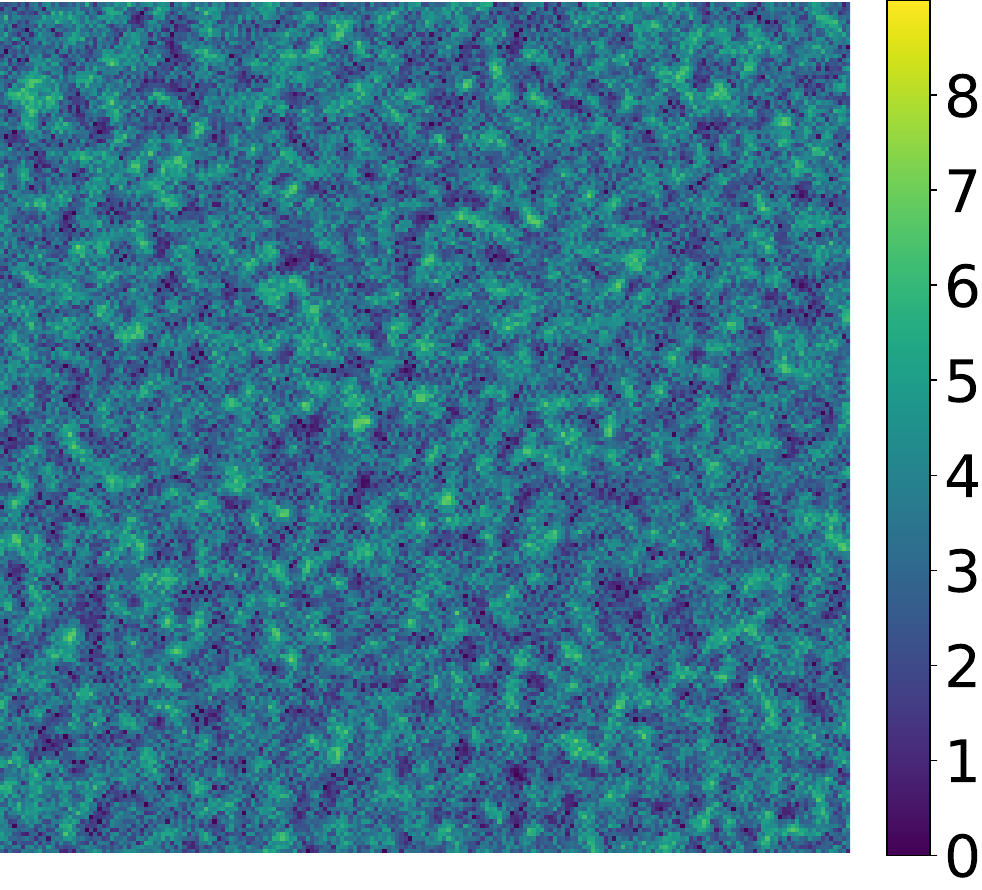}\\
					\vspace{-2mm}
					{\footnotesize t=1}
				\end{minipage}
				\hfill
				\begin{minipage}{0.188\linewidth}
					\centering
					\includegraphics[width=\linewidth]{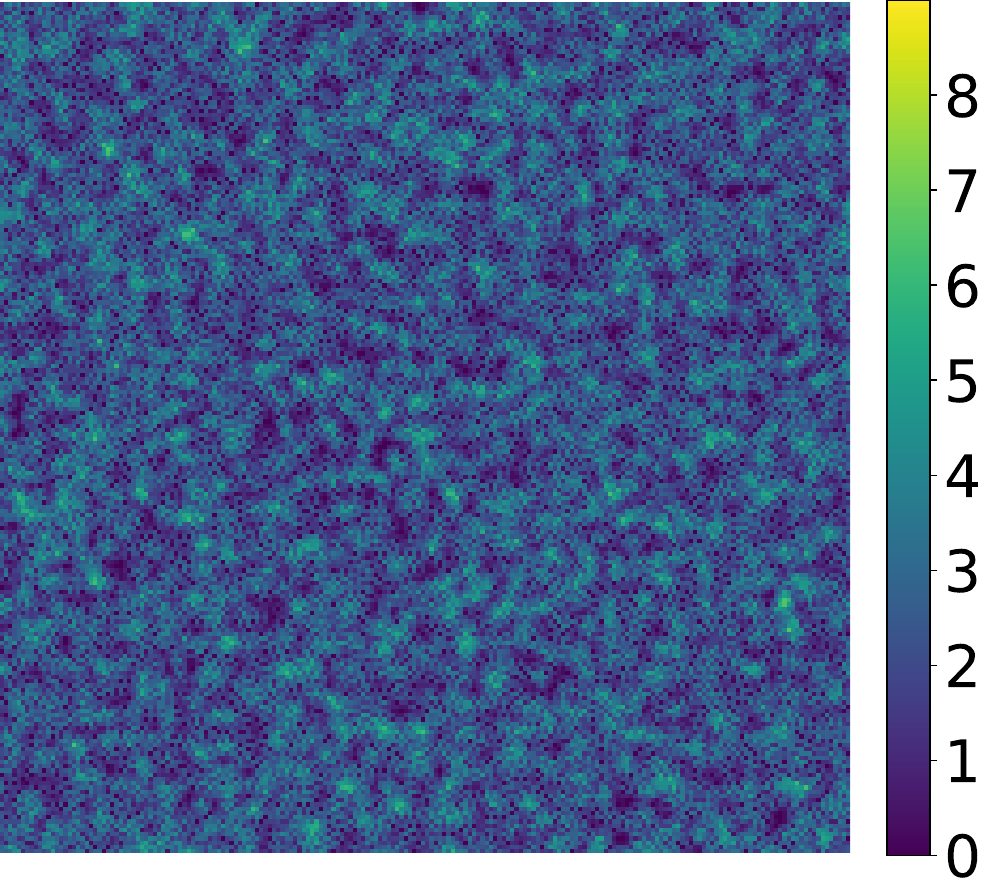}\\
					\vspace{-2mm}
					{\footnotesize t=10}
				\end{minipage}
				\hfill
				\begin{minipage}{0.188\linewidth}
					\centering
					\includegraphics[width=\linewidth]{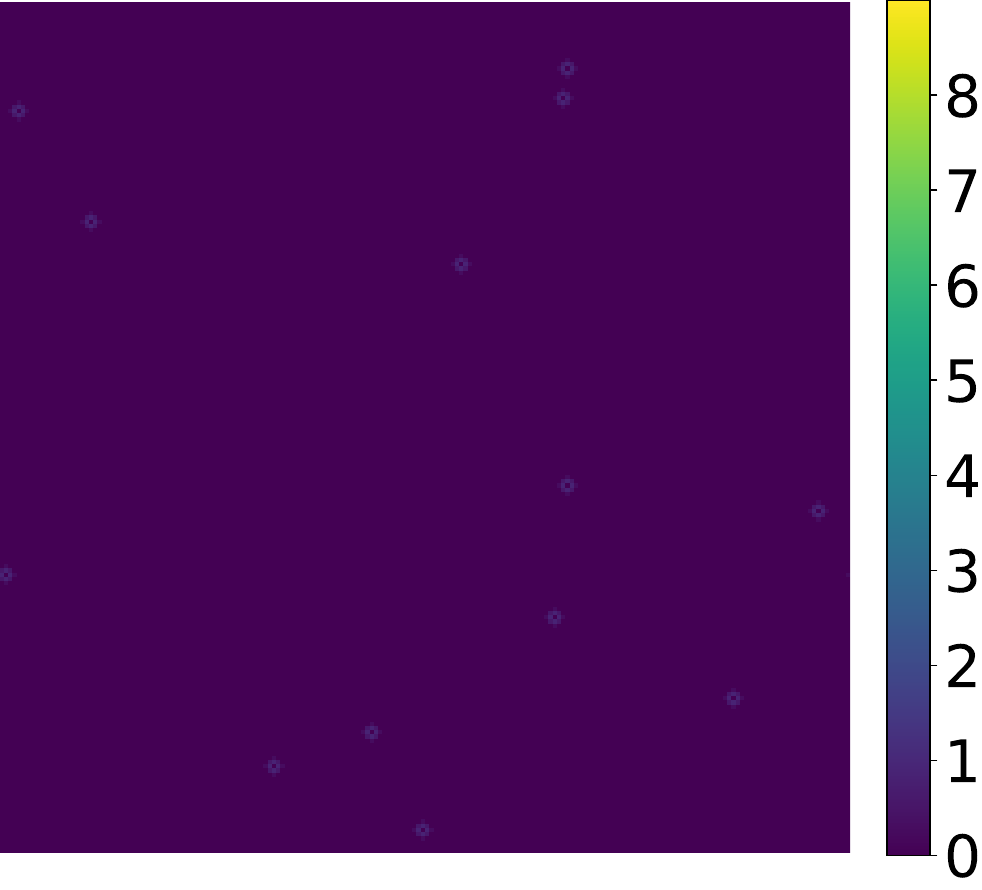}\\
					\vspace{-2mm}
					{\footnotesize t=100}
				\end{minipage}
				\hfill
				\begin{minipage}{0.188\linewidth}
					\centering
					\includegraphics[width=\linewidth]{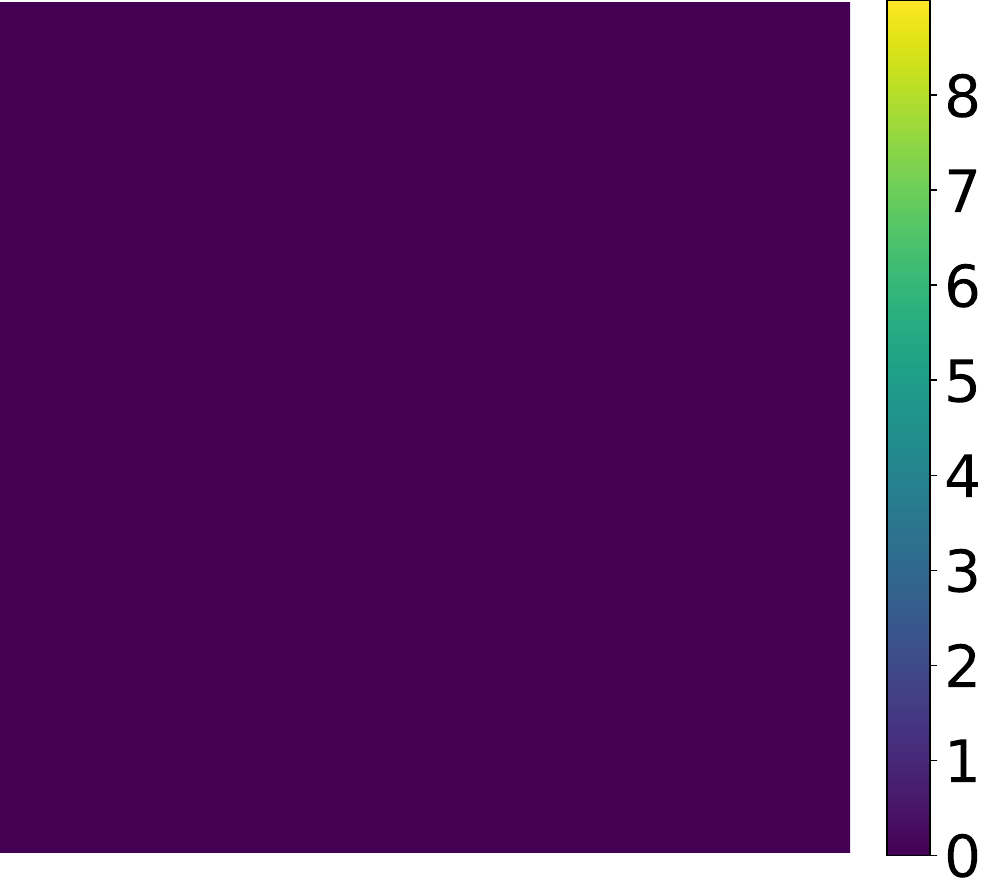}\\
					\vspace{-2mm}
					{\footnotesize t=1000}
				\end{minipage}
				\vspace{-2mm}
				\caption*{\footnotesize (c) r=3.0 (Payoff heatmaps)}
			\end{minipage}
				\\
			[2mm]
			\begin{minipage}{\linewidth}
				\begin{minipage}{0.188\linewidth}
					\centering
					\includegraphics[width=\linewidth]{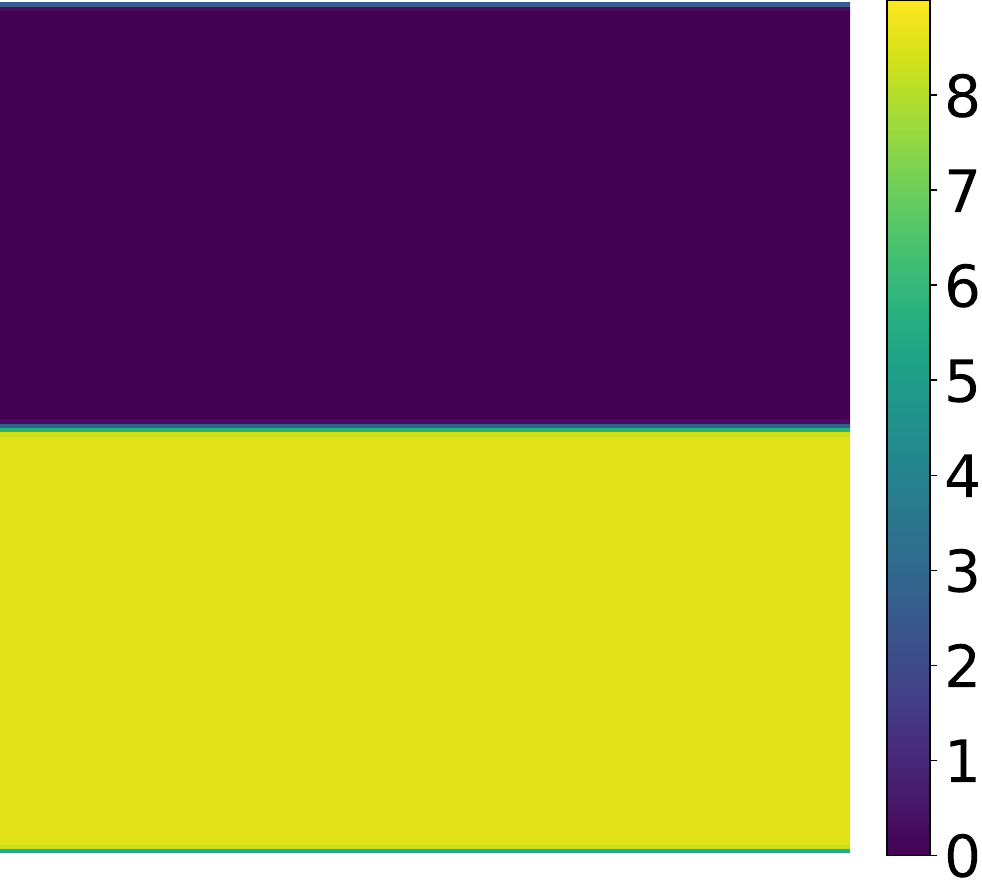}\\
					\vspace{-2mm}
					{\footnotesize t=0}
				\end{minipage}
				\hfill
				\begin{minipage}{0.188\linewidth}
					\centering
					\includegraphics[width=\linewidth]{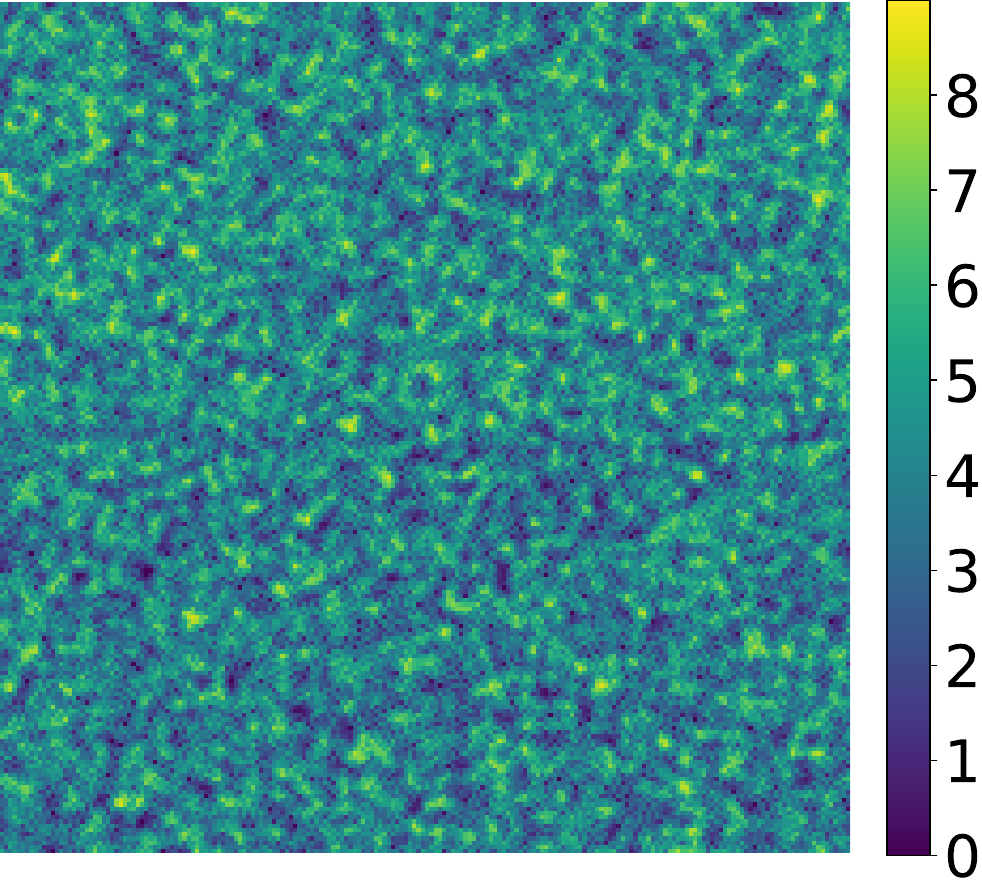}\\
					\vspace{-2mm}
					{\footnotesize t=1}
				\end{minipage}
				\hfill
				\begin{minipage}{0.188\linewidth}
					\centering
					\includegraphics[width=\linewidth]{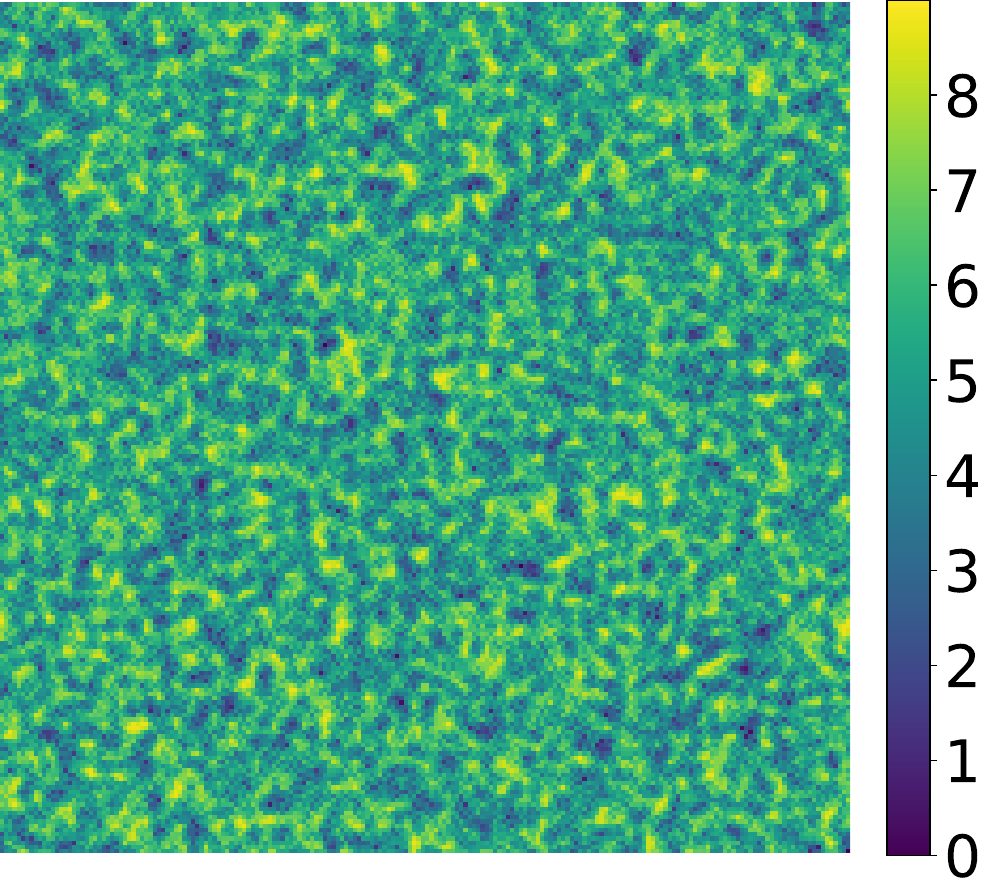}\\
					\vspace{-2mm}
					{\footnotesize t=10}
				\end{minipage}
				\hfill
				\begin{minipage}{0.188\linewidth}
					\centering
					\includegraphics[width=\linewidth]{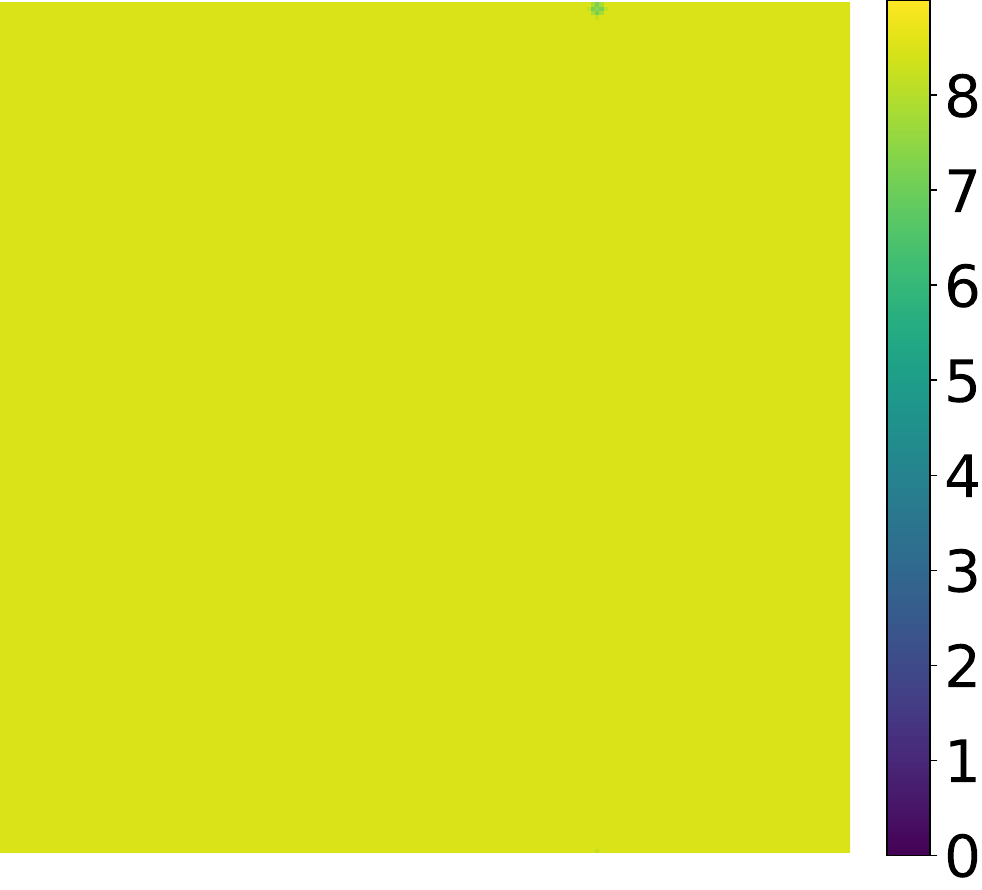}\\
					\vspace{-2mm}
					{\footnotesize t=100}
				\end{minipage}
				\hfill
				\begin{minipage}{0.188\linewidth}
					\centering
					\includegraphics[width=\linewidth]{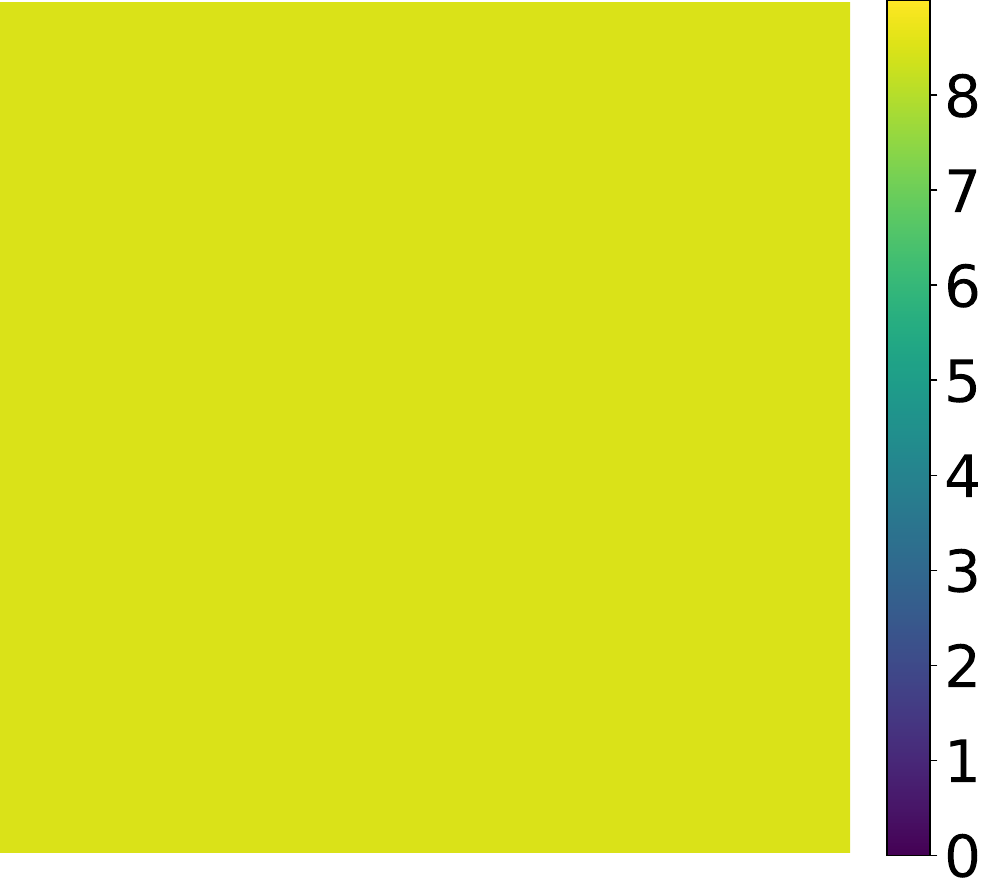}\\
					\vspace{-2mm}
					{\footnotesize t=1000}
				\end{minipage}
				\vspace{-2mm}
				\caption*{\footnotesize (d) r=3.5 (Payoff heatmaps)}
			\end{minipage}
		\caption{Evolution of cooperation and payoffs in SPGG using TUC-PPO. Initial conditions place defectors in the upper half of the grid and cooperators in the lower half. (\textbf{a, b}) Temporal evolution of cooperation (blue curves) and defection (red curves), with corresponding strategy snapshots below (black blocks: defectors, white blocks: cooperators). (\textbf{c, d}) Payoff heatmaps at key iterations, showing individual agent earnings. (\textbf{a, c}) Results for enhancement factor $r = 3.0$; (\textbf{b, d}) Results for $r = 3.5$. Convergence properties demonstrate that TUC-PPO enables all agents to rapidly stabilize in pure cooperation or defection states within minimal iterations. This process features near-absent oscillations during strategy evolution.}
		\label{fig:TUC-PPO_uDbC}
	\end{figure*}

	Fig.~\ref{fig:TUC-PPO_uDbC} shows the evolution curve for TUC-PPO under spatially partitioned initial conditions. The study implements a controlled initialization protocol where defectors and cooperators are systematically allocated to distinct grid regions before training. Specifically, defector agents populate the upper half of the spatial domain while cooperators occupy the lower half, establishing well-defined initial strategy boundaries. The investigation examines system dynamics across two enhancement factor values: $r = 3.0$ and $r = 4.0$. These parameter selections enable comparative analysis of cooperation emergence under different environmental conditions. The experimental output combines quantitative temporal metrics with qualitative spatial representations to capture both macroscopic and microscopic evolutionary patterns. The experimental output incorporates three complementary visualization modalities to capture different aspects of system dynamics. First, temporal progression curves track the evolution of strategy. Cooperator and defector populations are represented through distinct colored trajectories plotted against the iteration count. Second, spatial strategy distributions display agent-type configurations at key iterations, using high-contrast markers to indicate individual decisions. Third, payoff heatmaps visualize the immediate reward landscape, employing a color gradient to represent each agent's current earnings. This tripartite visualization framework enables simultaneous examination of temporal trends, spatial patterns, and economic incentives throughout the evolutionary process.

	The experimental results reveal a critical dependence of cooperative behavior on the enhancement factor $r$. Figs.~\ref{fig:TUC-PPO_uDbC} (a) and (c) demonstrate the $r=3.0$ scenario, where the fractions of cooperators decay almost monotonically, with all agents adopting defector strategies within 100 iterations. Spatial snapshots document the progressive territorial expansion of defectors, while payoff heatmaps reveal a corresponding decline in individual rewards as cooperation diminishes. Conversely, Figs.~\ref{fig:TUC-PPO_uDbC} (b) and (d) illustrate the $r=3.5$ scenario where the enhancement factor crosses the critical threshold for cooperation sustainability. Here we observe rapid convergence to universal cooperation, with complete conversion occurring before 40 iterations. The visualization panels demonstrate cooperators systematically replacing defectors, accompanied by steadily improving payoff distributions across the population. This dichotomy highlights the crucial role of the enhancement factor in determining system equilibrium. The results confirm that TUC-PPO successfully maintains cooperative equilibrium when $r=3.5$, while systems inevitably collapse to defection when $r=3.0$. The payoff heatmaps provide direct evidence that cooperative states yield superior economic outcomes compared to defective equilibria.
	
	\subsection{TUC-PPO with bernoulli random initialization}
	\label{exp_b}
	
\begin{figure*}[htbp!]
		\begin{minipage}{0.45\linewidth}
			\begin{minipage}{\linewidth}
				\centering
				\includegraphics[width=\linewidth]{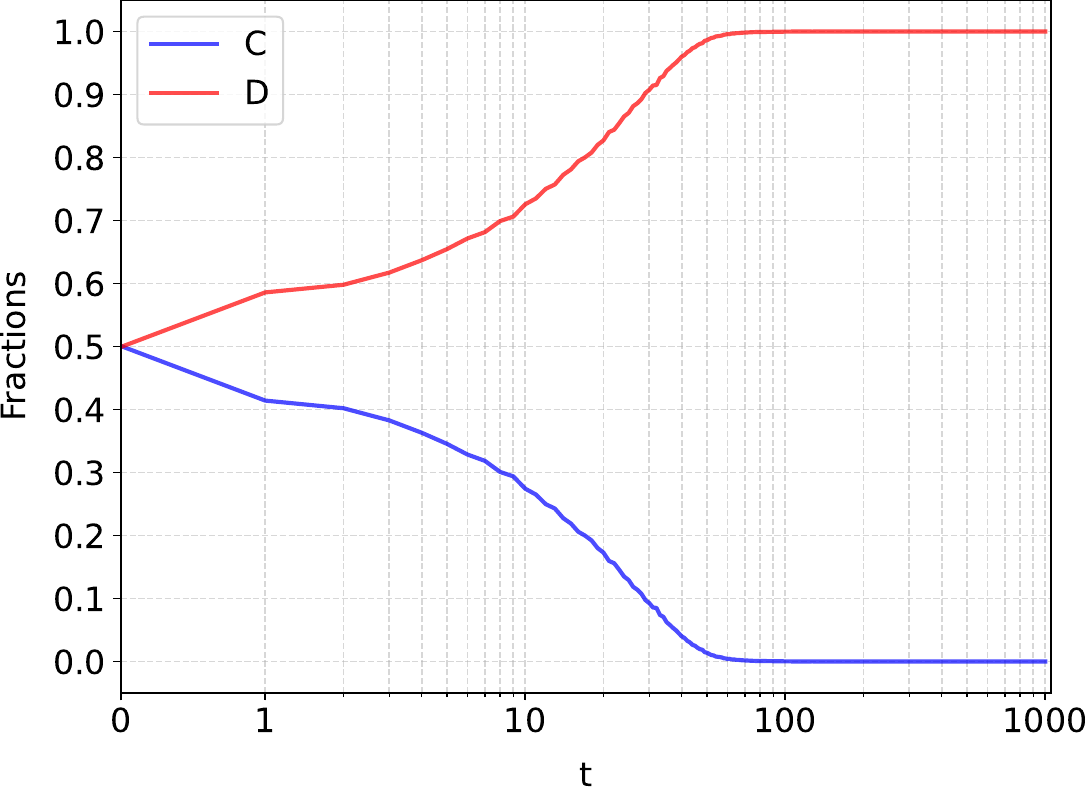}\\
			\end{minipage}
			\vspace{2mm}
			\\
			\begin{minipage}{0.188\linewidth}
				\centering
				\fbox{\includegraphics[width=\linewidth]{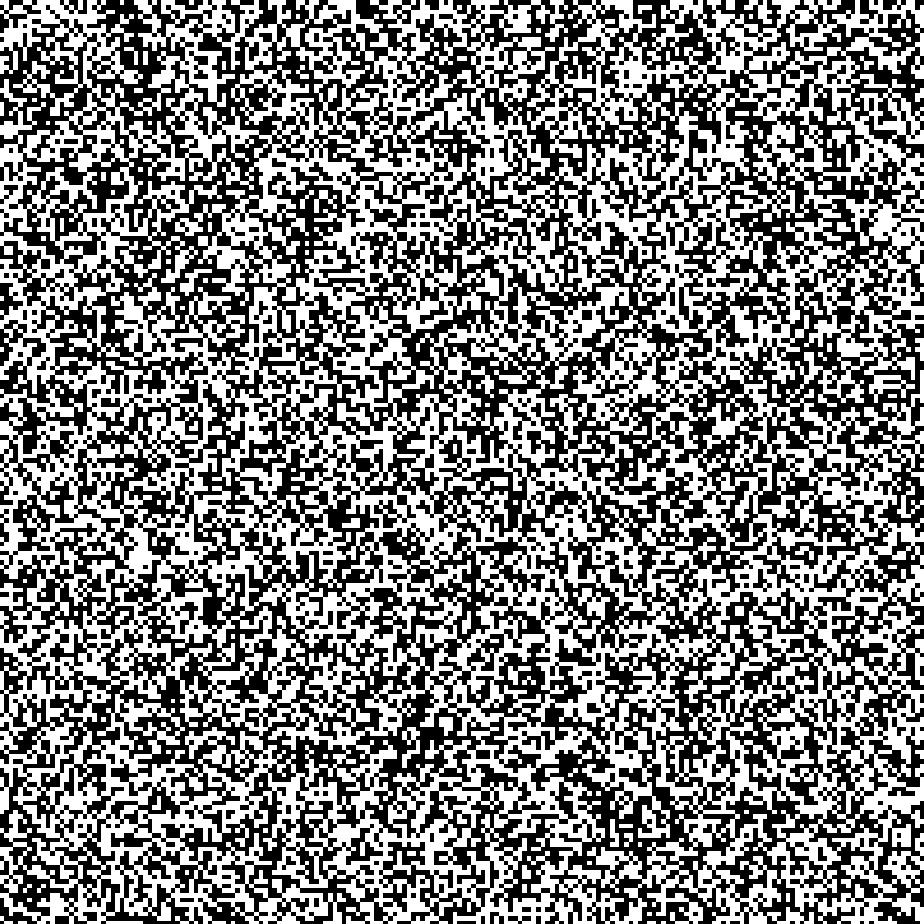}}\\
				\vspace{-2mm}
				{\footnotesize t=0}
			\end{minipage}
			\begin{minipage}{0.188\linewidth}
				\centering
				\fbox{\includegraphics[width=\linewidth]{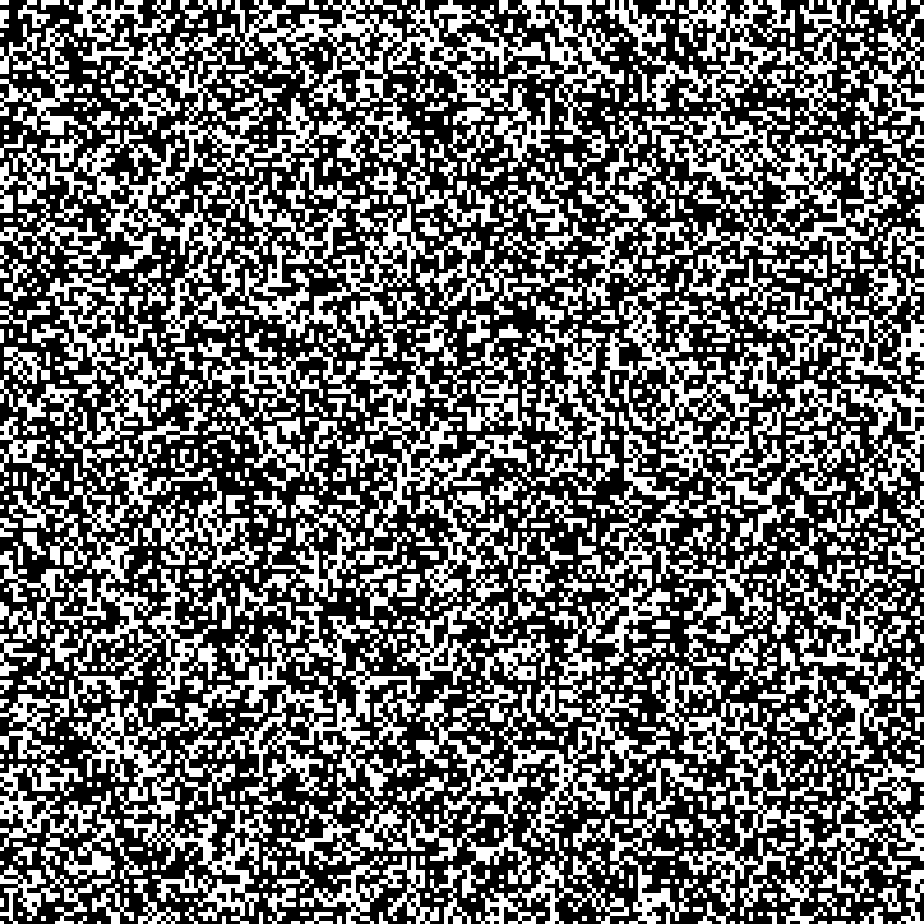}}\\
				\vspace{-2mm}
				{\footnotesize t=1}
			\end{minipage}
			\begin{minipage}{0.188\linewidth}
				\centering
				\fbox{\includegraphics[width=\linewidth]{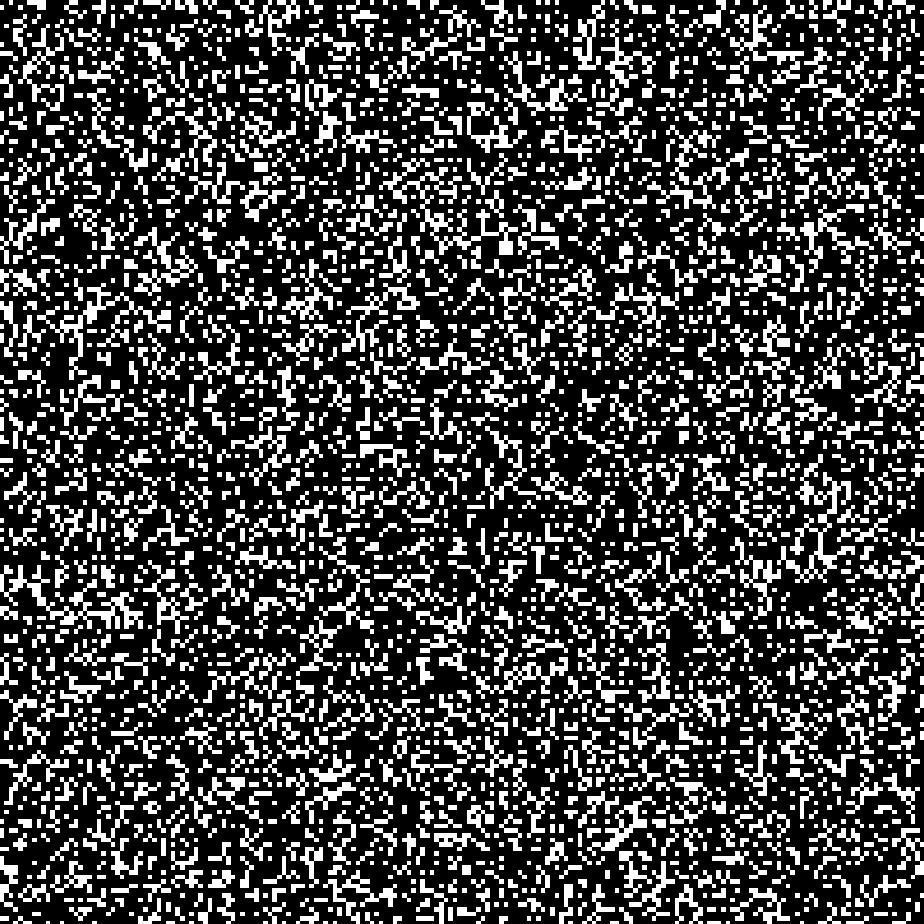}}\\
				\vspace{-2mm}
				{\footnotesize t=10}
			\end{minipage}
			\begin{minipage}{0.188\linewidth}
				\centering
				\fbox{\includegraphics[width=\linewidth]{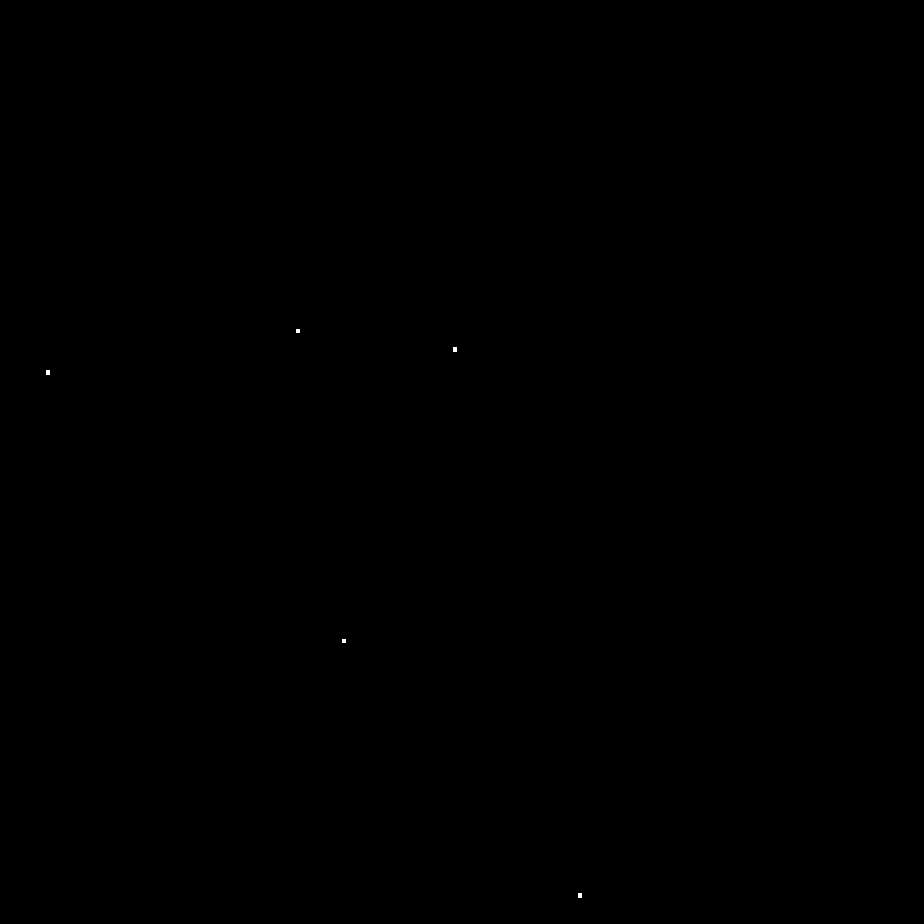}}\\
				\vspace{-2mm}
				{\footnotesize t=100}
			\end{minipage}
			\begin{minipage}{0.188\linewidth}
				\centering
				\fbox{\includegraphics[width=\linewidth]{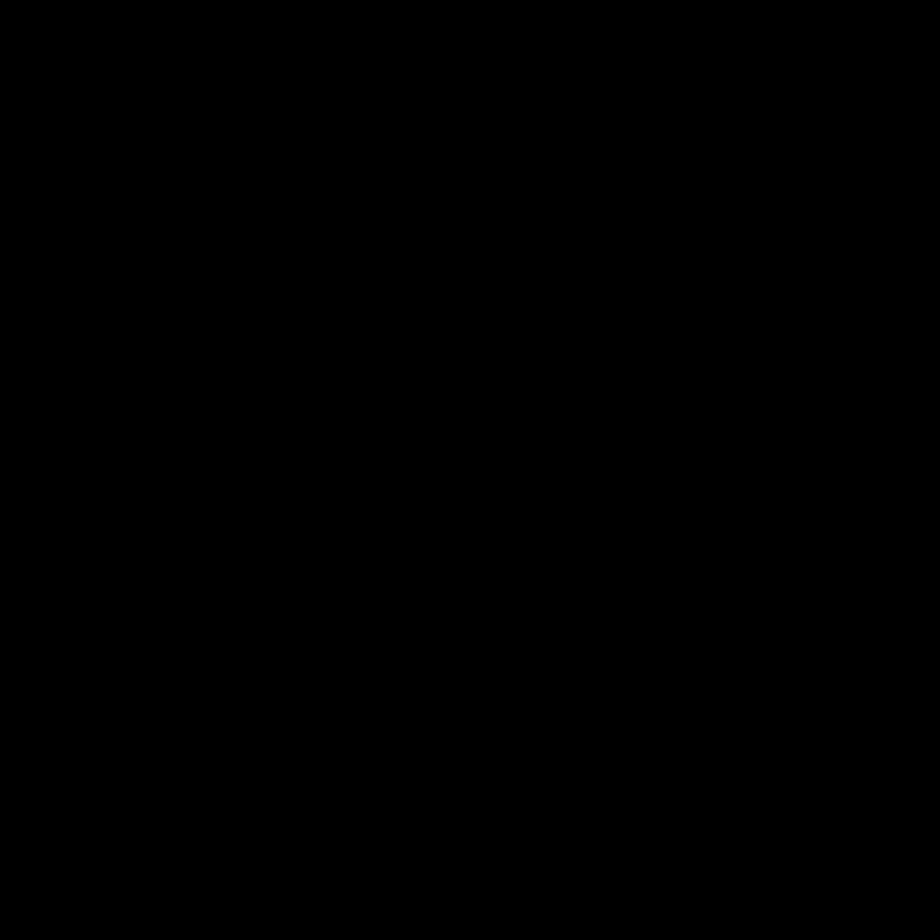}}\\
				\vspace{-2mm}
				{\footnotesize t=1000}
			\end{minipage}
			\vspace{-2mm}
			\caption*{\footnotesize (a) r=3.0}
		\end{minipage}
		\hfill
		\begin{minipage}{0.45\linewidth}
			\begin{minipage}{\linewidth}
				\centering
				\includegraphics[width=\linewidth]{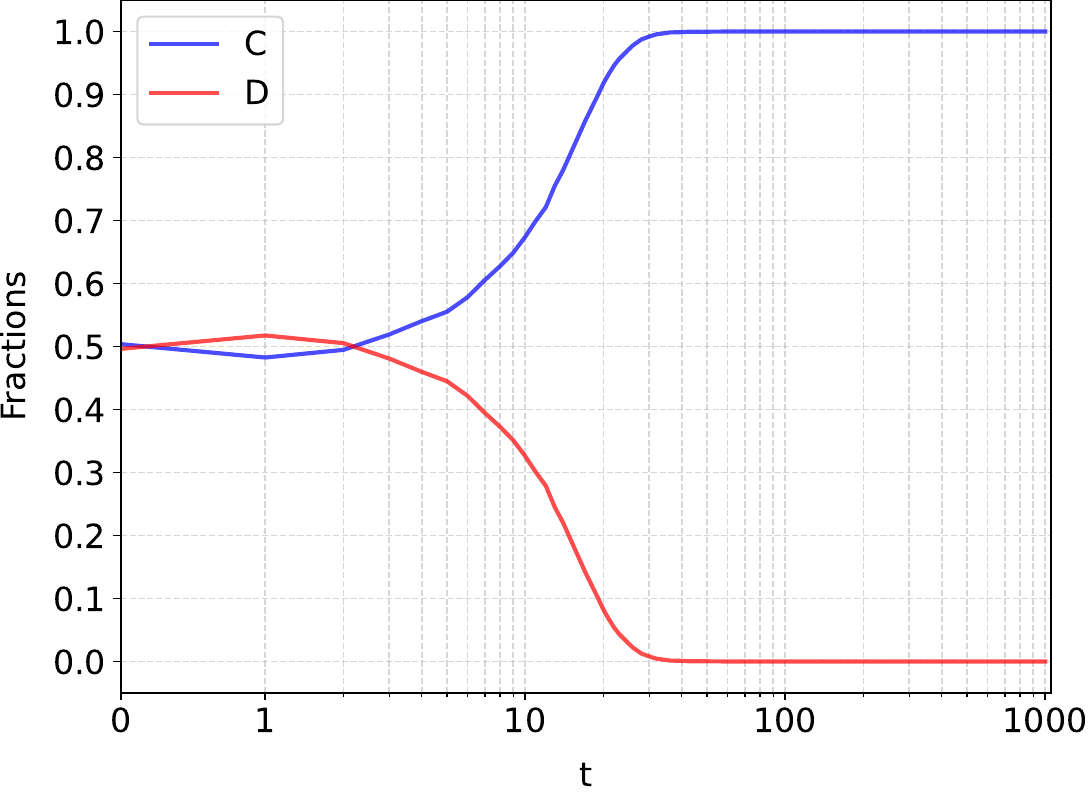}\\
			\end{minipage}
			\vspace{2mm}
			\\
			\begin{minipage}{0.188\linewidth}
				\centering
				\fbox{\includegraphics[width=\linewidth]{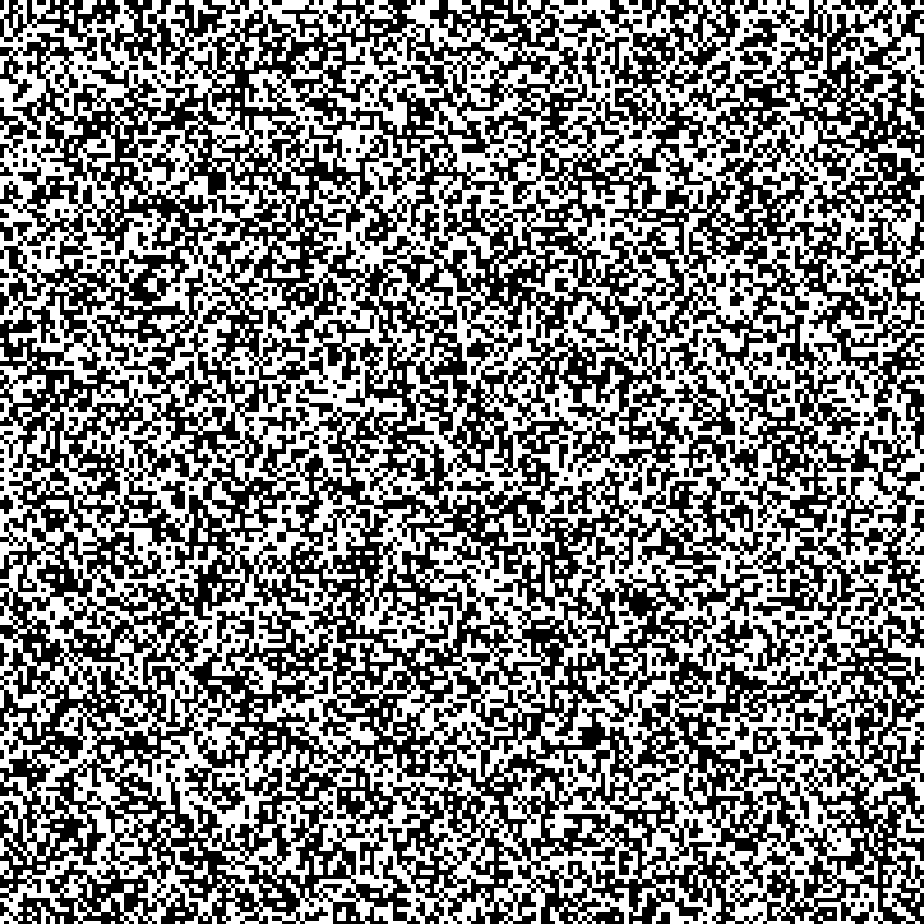}}\\
				\vspace{-2mm}
				{\footnotesize t=0}
			\end{minipage}
			\begin{minipage}{0.188\linewidth}
				\centering
				\fbox{\includegraphics[width=\linewidth]{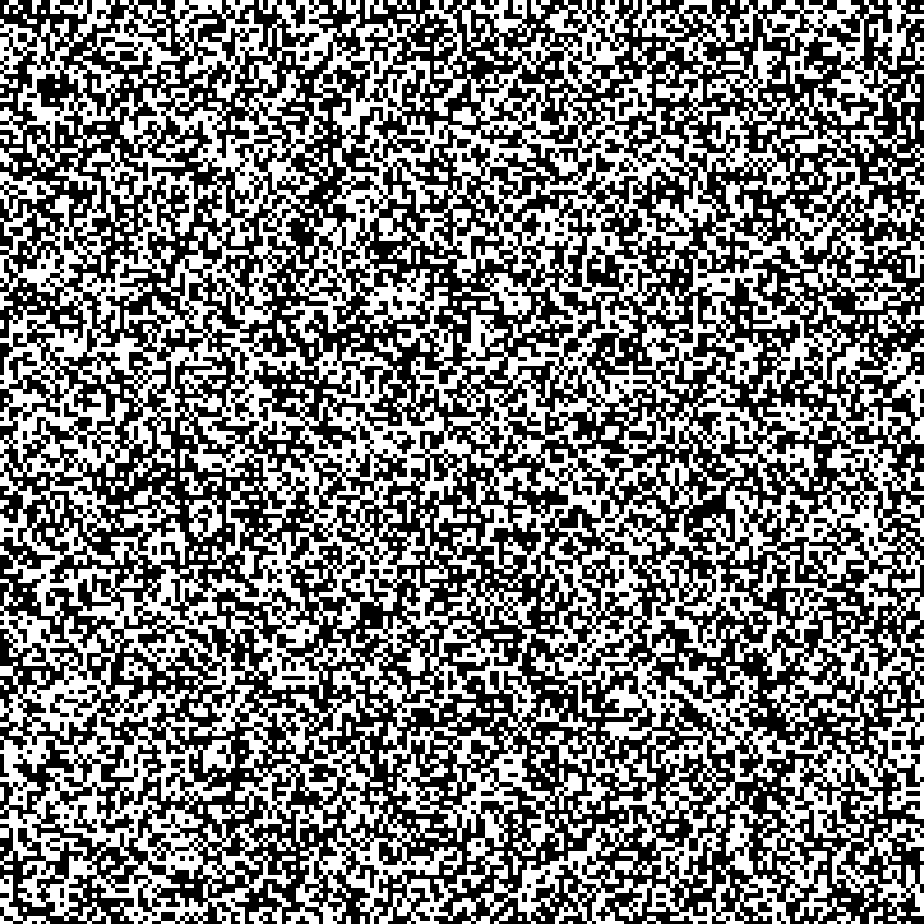}}\\
				\vspace{-2mm}
				{\footnotesize t=1}
			\end{minipage}
			\begin{minipage}{0.188\linewidth}
				\centering
				\fbox{\includegraphics[width=\linewidth]{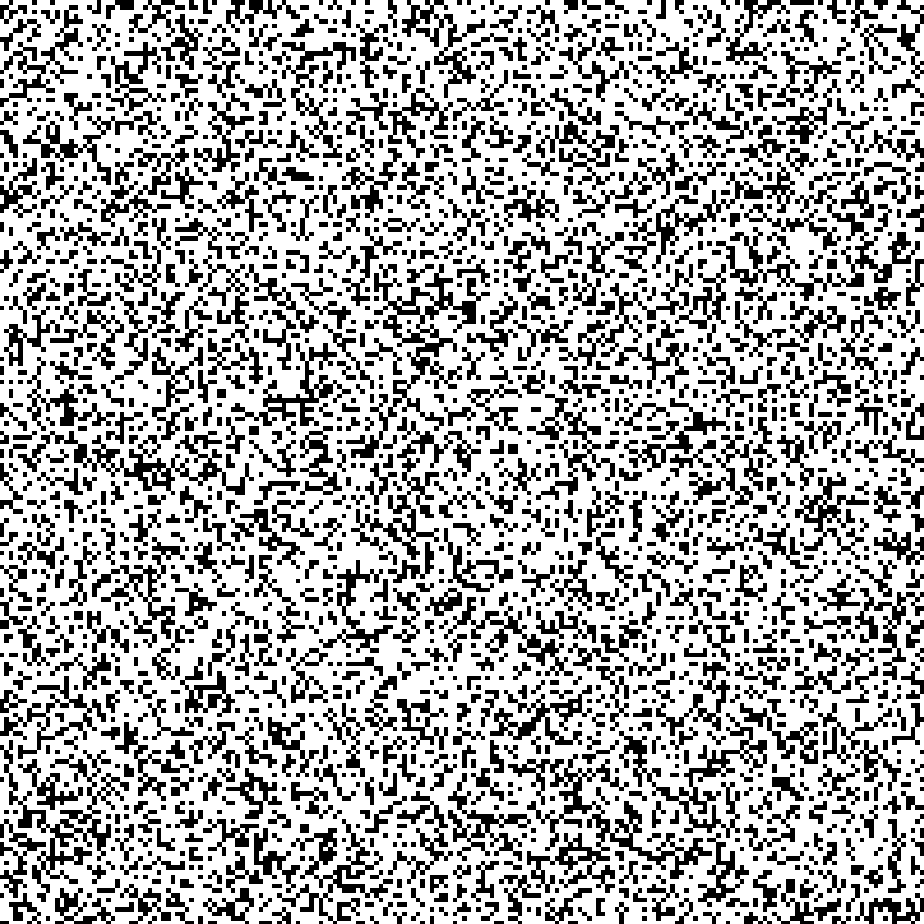}}\\
				\vspace{-2mm}
				{\footnotesize t=10}
			\end{minipage}
			\begin{minipage}{0.188\linewidth}
				\centering
				\fbox{\includegraphics[width=\linewidth]{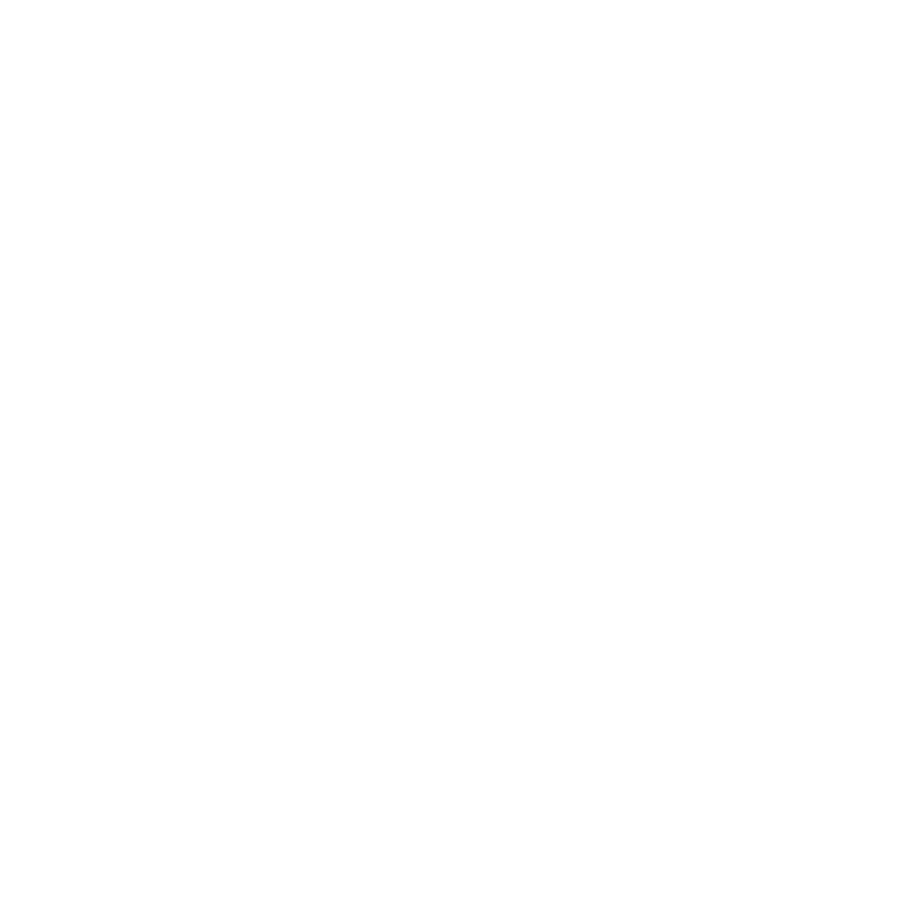}}\\
				\vspace{-2mm}
				{\footnotesize t=100}
			\end{minipage}
			\begin{minipage}{0.188\linewidth}
				\centering
				\fbox{\includegraphics[width=\linewidth]{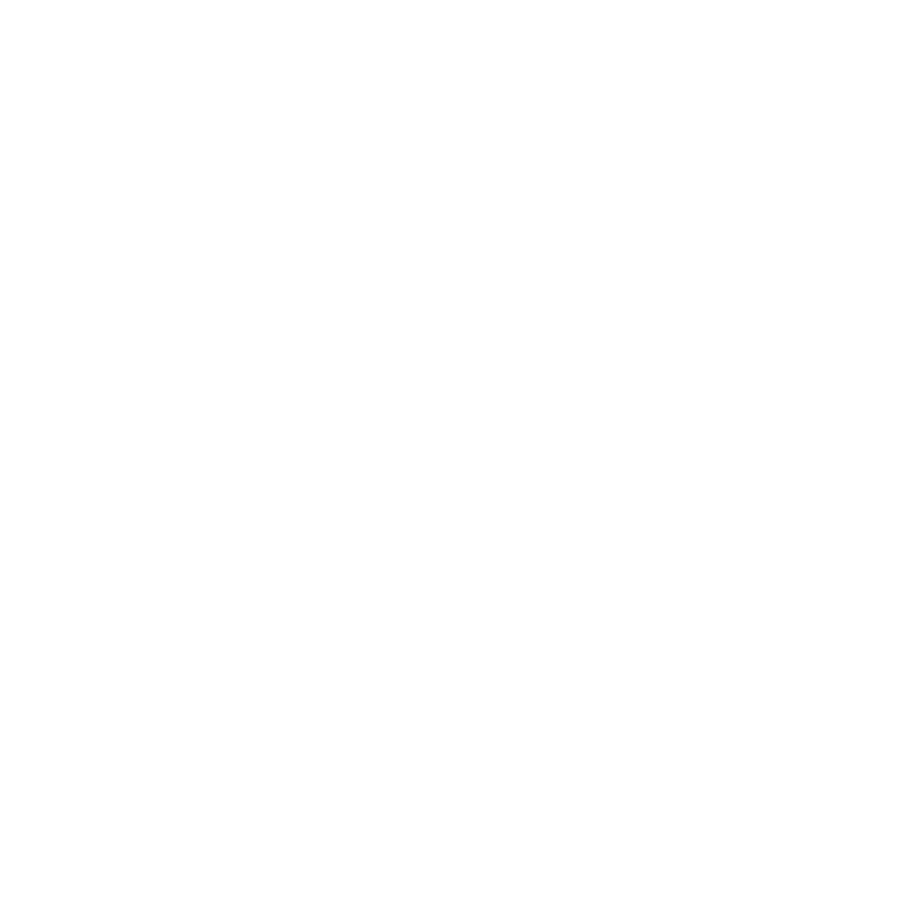}}\\
				\vspace{-2mm}
				{\footnotesize t=1000}
			\end{minipage}
			\vspace{-2mm}
			\caption*{\footnotesize (b) r=3.5}
		\end{minipage}
		\\
		[2mm]
		\begin{minipage}{\linewidth}
			\begin{minipage}{0.188\linewidth}
				\centering
				\includegraphics[width=\linewidth]{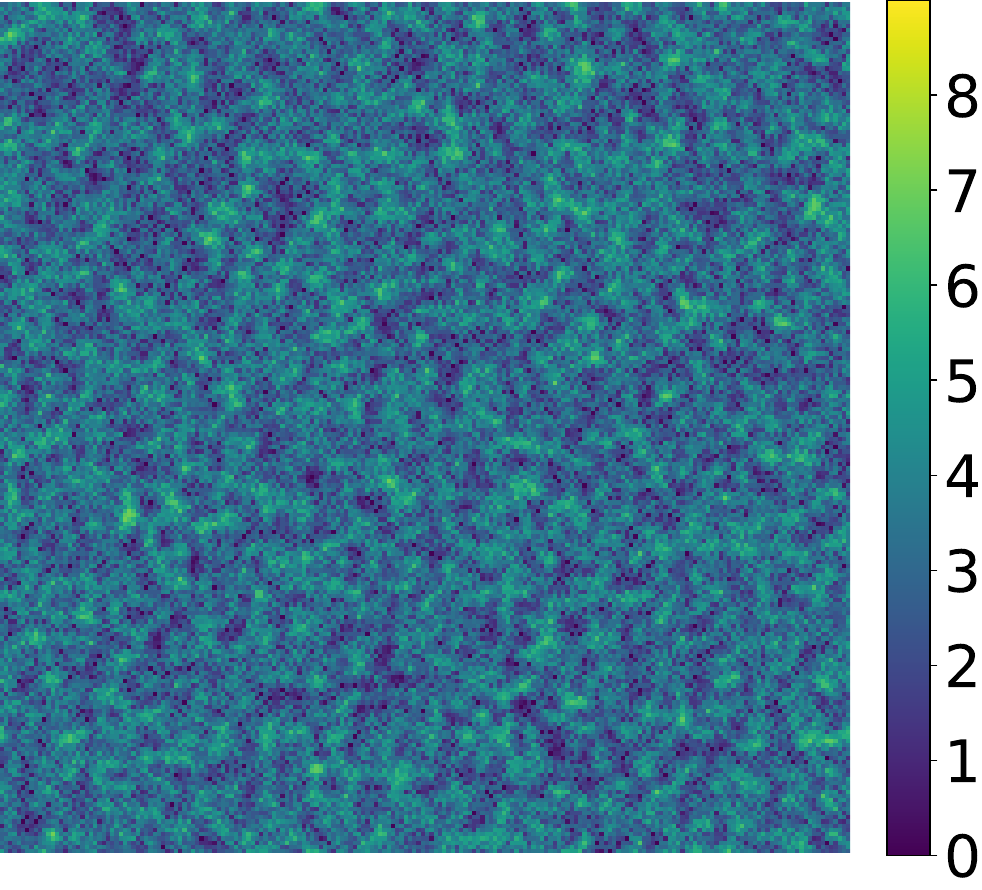}\\
				\vspace{-2mm}
				{\footnotesize t=0}
			\end{minipage}
			\hfill
			\begin{minipage}{0.188\linewidth}
				\centering
				\includegraphics[width=\linewidth]{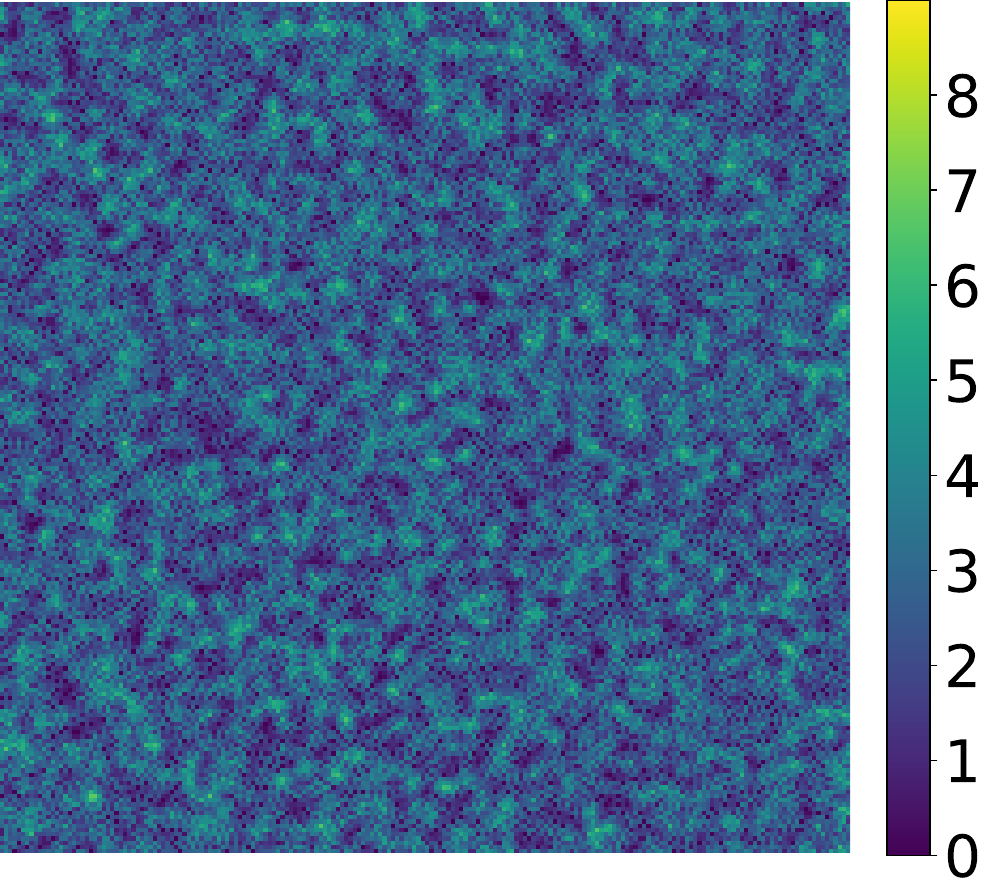}\\
				\vspace{-2mm}
				{\footnotesize t=1}
			\end{minipage}
			\hfill
			\begin{minipage}{0.188\linewidth}
				\centering
				\includegraphics[width=\linewidth]{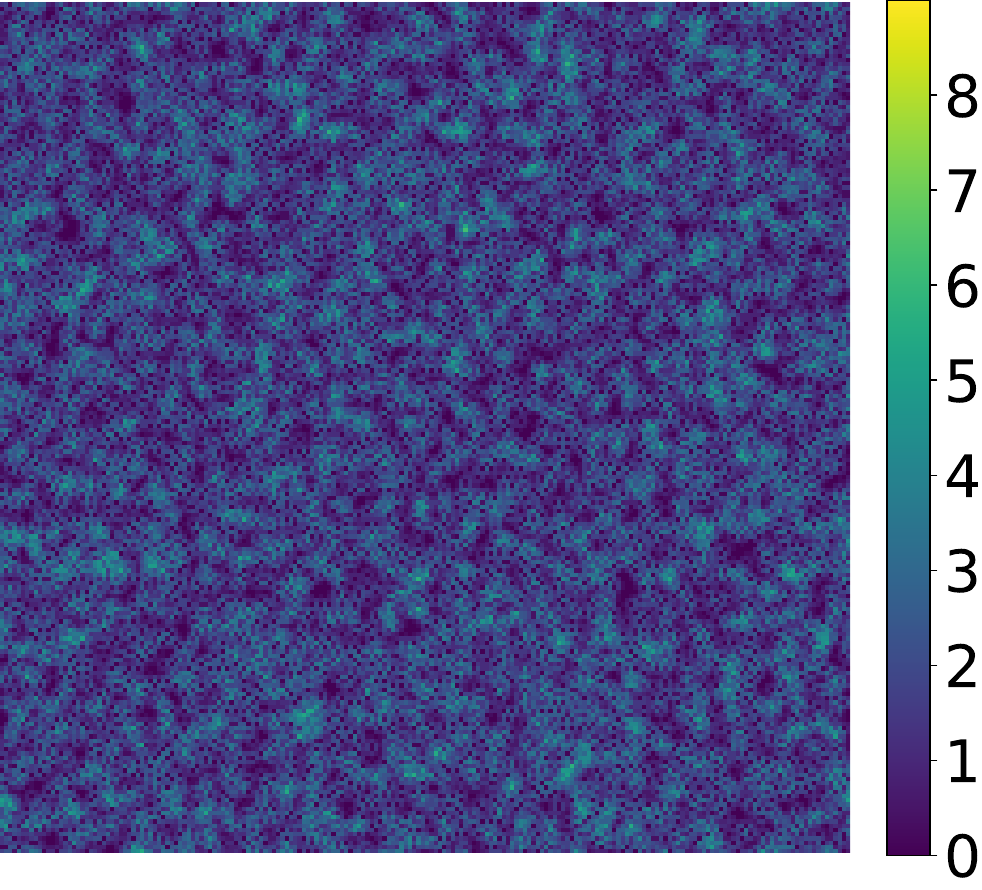}\\
				\vspace{-2mm}
				{\footnotesize t=10}
			\end{minipage}
			\hfill
			\begin{minipage}{0.188\linewidth}
				\centering
				\includegraphics[width=\linewidth]{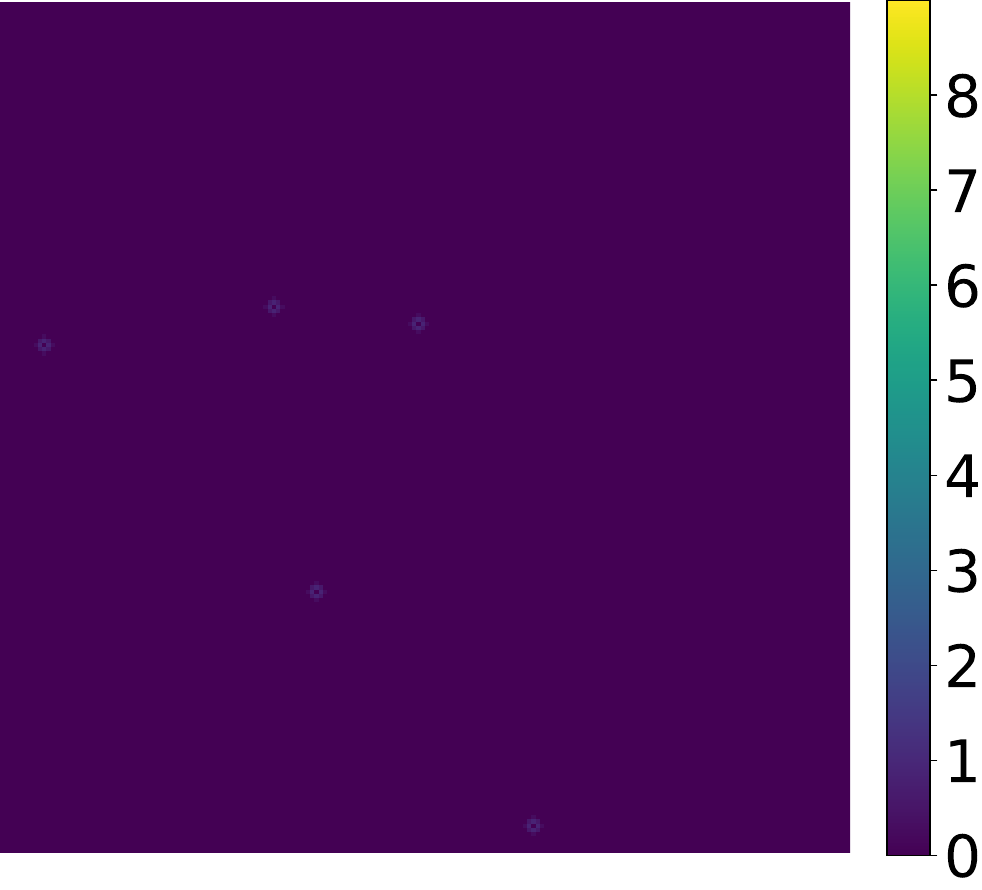}\\
				\vspace{-2mm}
				{\footnotesize t=100}
			\end{minipage}
			\hfill
			\begin{minipage}{0.188\linewidth}
				\centering
				\includegraphics[width=\linewidth]{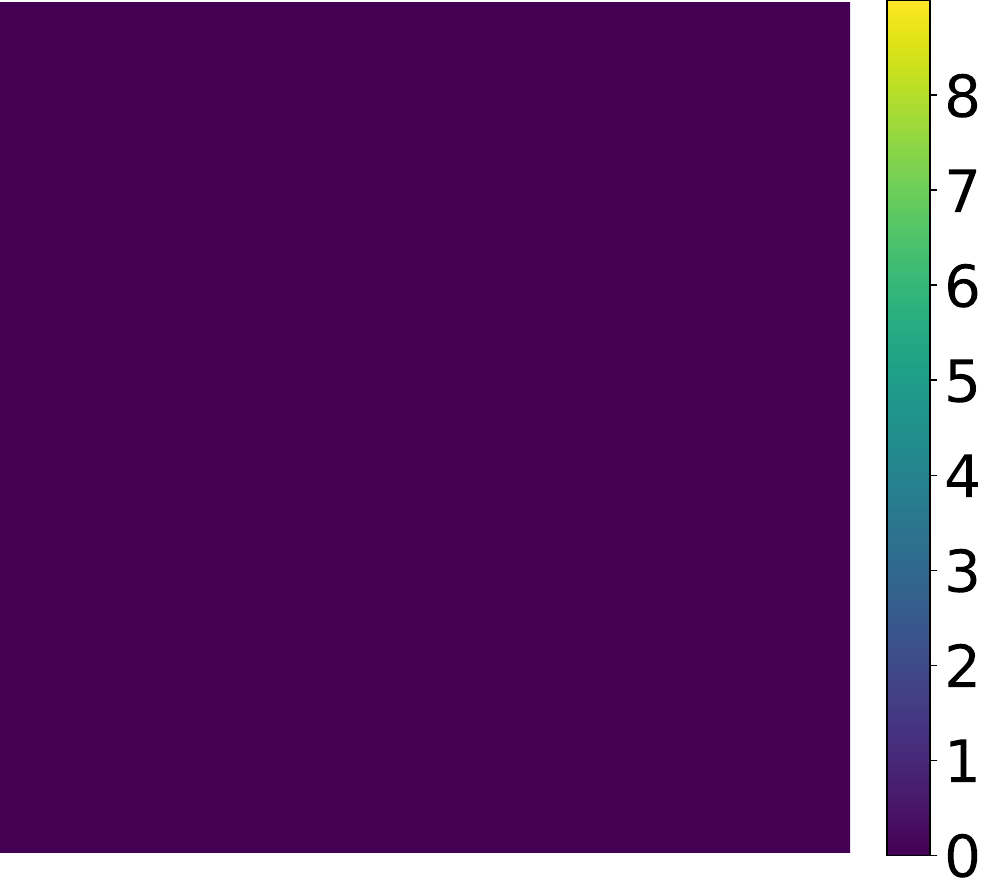}\\
				\vspace{-2mm}
				{\footnotesize t=1000}
			\end{minipage}
			\vspace{-2mm}
			\caption*{\footnotesize (c) r=3.0 (Payoff heatmaps)}
		\end{minipage}
		\\
		[2mm]
		\begin{minipage}{\linewidth}
			\begin{minipage}{0.188\linewidth}
				\centering
				\includegraphics[width=\linewidth]{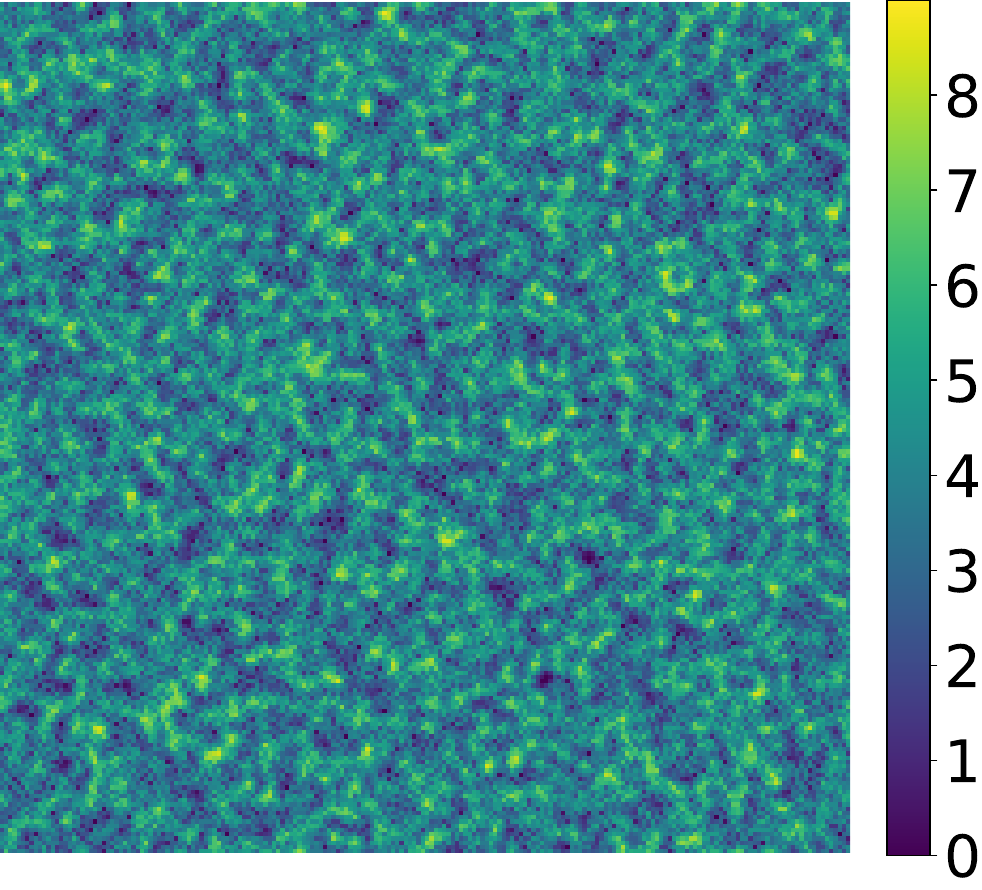}\\
				\vspace{-2mm}
				{\footnotesize t=0}
			\end{minipage}
			\hfill
			\begin{minipage}{0.188\linewidth}
				\centering
				\includegraphics[width=\linewidth]{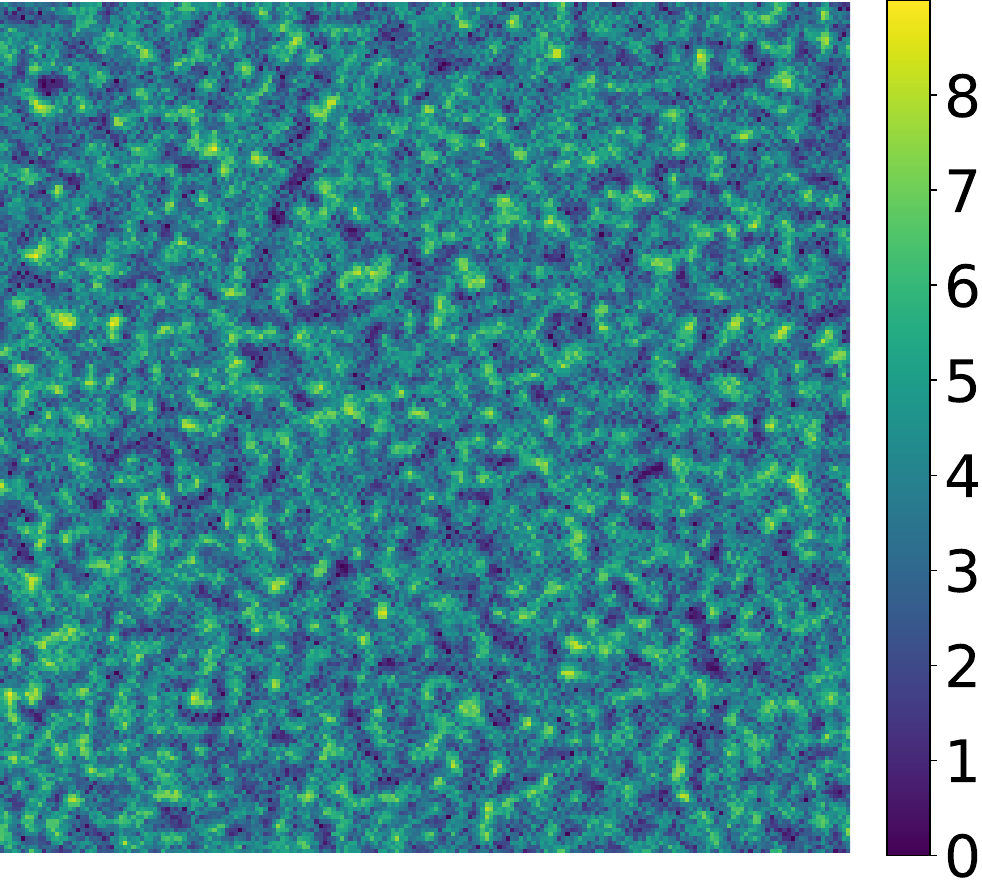}\\
				\vspace{-2mm}
				{\footnotesize t=1}
			\end{minipage}
			\hfill
			\begin{minipage}{0.188\linewidth}
				\centering
				\includegraphics[width=\linewidth]{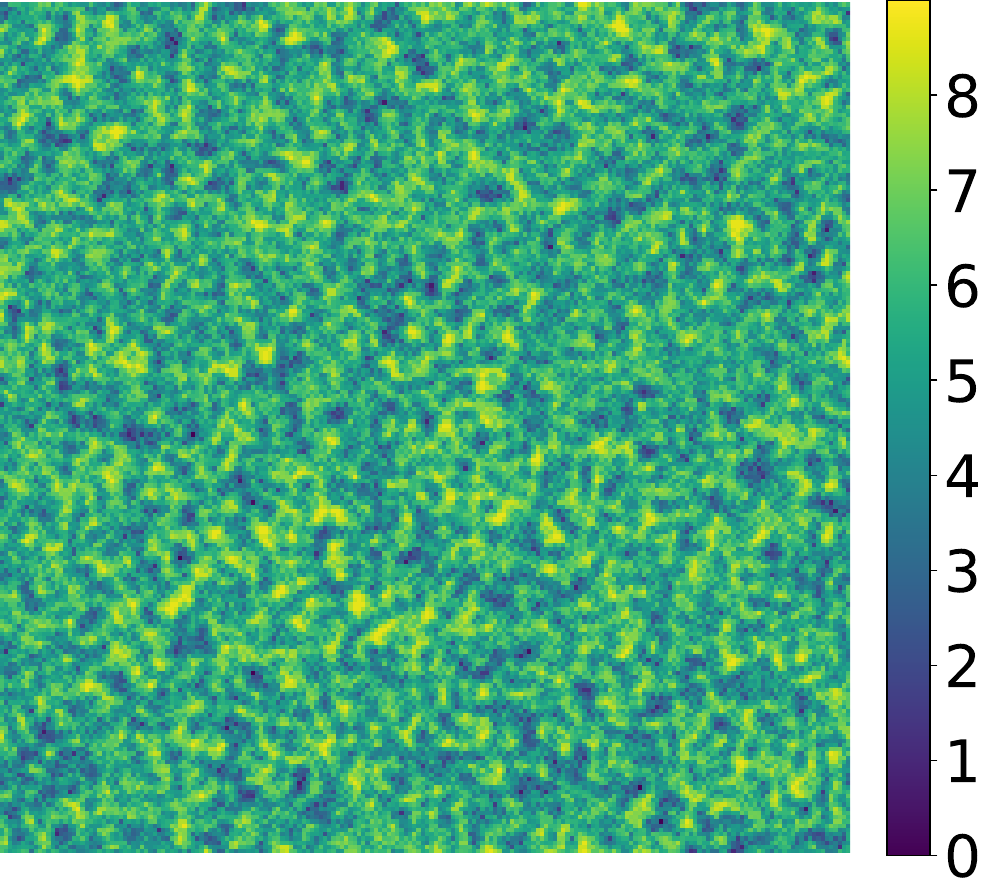}\\
				\vspace{-2mm}
				{\footnotesize t=10}
			\end{minipage}
			\hfill
			\begin{minipage}{0.188\linewidth}
				\centering
				\includegraphics[width=\linewidth]{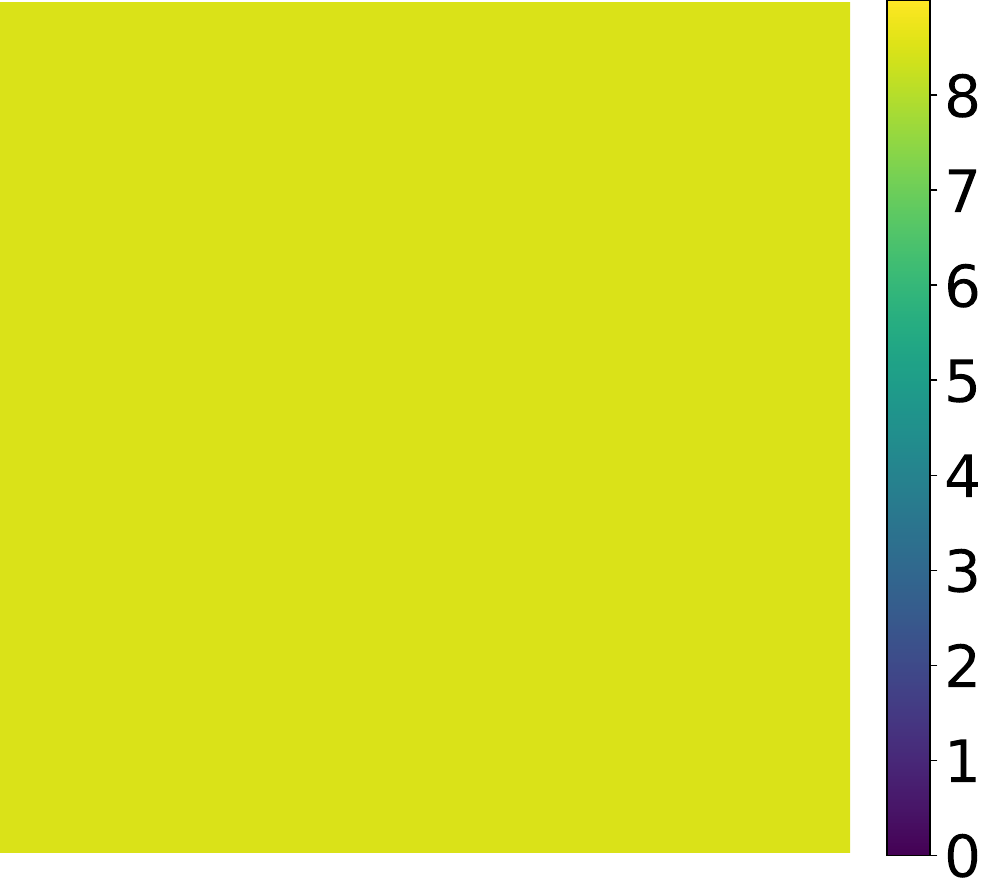}\\
				\vspace{-2mm}
				{\footnotesize t=100}
			\end{minipage}
			\hfill
			\begin{minipage}{0.188\linewidth}
				\centering
				\includegraphics[width=\linewidth]{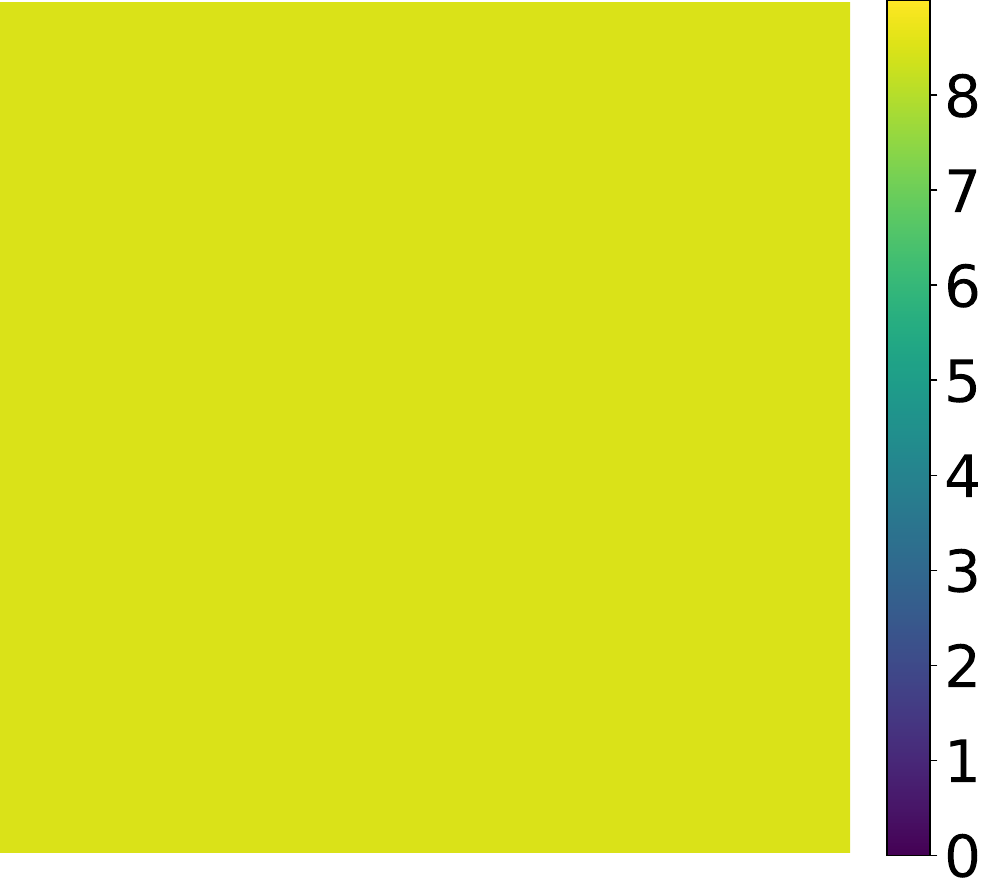}\\
				\vspace{-2mm}
				{\footnotesize t=1000}
			\end{minipage}
			\vspace{-2mm}
			\caption*{\footnotesize (d) r=3.5 (Payoff heatmaps)}
		\end{minipage}
		\caption{Evolution of cooperation and payoffs in SPGG using TUC-PPO with random initialization. Agent strategies are initially assigned according to a Bernoulli distribution, with equal probabilities for cooperation and defection. \textbf{(a, b)} Temporal evolution of cooperation is shown in blue, and defection is shown in red, accompanied by strategy distribution snapshots where defectors appear as black blocks and cooperators as white blocks. \textbf{(c, d)} Payoff heatmaps displaying individual agent rewards at selected iterations. \textbf{(a, c)} Results for enhancement factor $r = 3.0$ while \textbf{(b, d)} show $r = 3.5$ conditions. Convergence to cooperation proves markedly more challenging under random initialization than with half-and-half initialization.}
		\label{fig:TUC-PPO_bernoulli}
	\end{figure*}

	This experimental series investigates TUC-PPO dynamics under randomized initial conditions. Agent strategies are assigned following a Bernoulli distribution with equal probability $p=0.5$ for both cooperation and defection strategies. Fig.~\ref{fig:TUC-PPO_bernoulli} systematically compares the evolutionary results in different enhancement factors $r$. Each subfigure contains three integrated visualization components. The upper section displays temporal evolution profiles, plotting the fractions of cooperators (shown in blue) and defectors (shown in red) against the iteration count $t$. The middle section presents spatial strategy distributions at key evolutionary stages, using white pixels to represent cooperators and black pixels for defectors. The lower section provides payoff heatmaps that visualize individual agent rewards.
	
	Fig.~\ref{fig:TUC-PPO_bernoulli} with Bernoulli-distributed initial strategies shows a fundamental dependence of cooperation dynamics on the enhancement factor $r$. Under the $r=3.0$ condition, the cooperation fraction exhibits a steady decline over time, with all agents transitioning to defector strategies within 80 iterations. Spatial evolution patterns reveal progressively increasing black regions representing defectors, while the payoff heatmaps show corresponding deterioration in individual rewards as cooperation diminishes. In contrast, the $r=3.5$ scenario shows rapid system convergence, with complete adoption of cooperative strategies occurring before 40 iterations. The visualization panels demonstrate a swift transition to uniform cooperation, accompanied by significantly improved payoff distributions throughout the population. These results underscore the pivotal role of the enhancement factor in governing system evolution from random initial conditions. The TUC-PPO algorithm successfully drives the population to full cooperation when r=3.5, while inevitably leading to universal defection when $r=3.0$. The payoff heatmaps provide conclusive evidence that cooperative equilibria generate superior economic outcomes compared to defective states.
	
	\subsection{TUC-PPO with all-defectors initialization}
	\label{exp_ad}
	
\begin{figure*}[htbp!]
			\begin{minipage}{0.45\linewidth}
			\begin{minipage}{\linewidth}
				\centering
				\includegraphics[width=\linewidth]{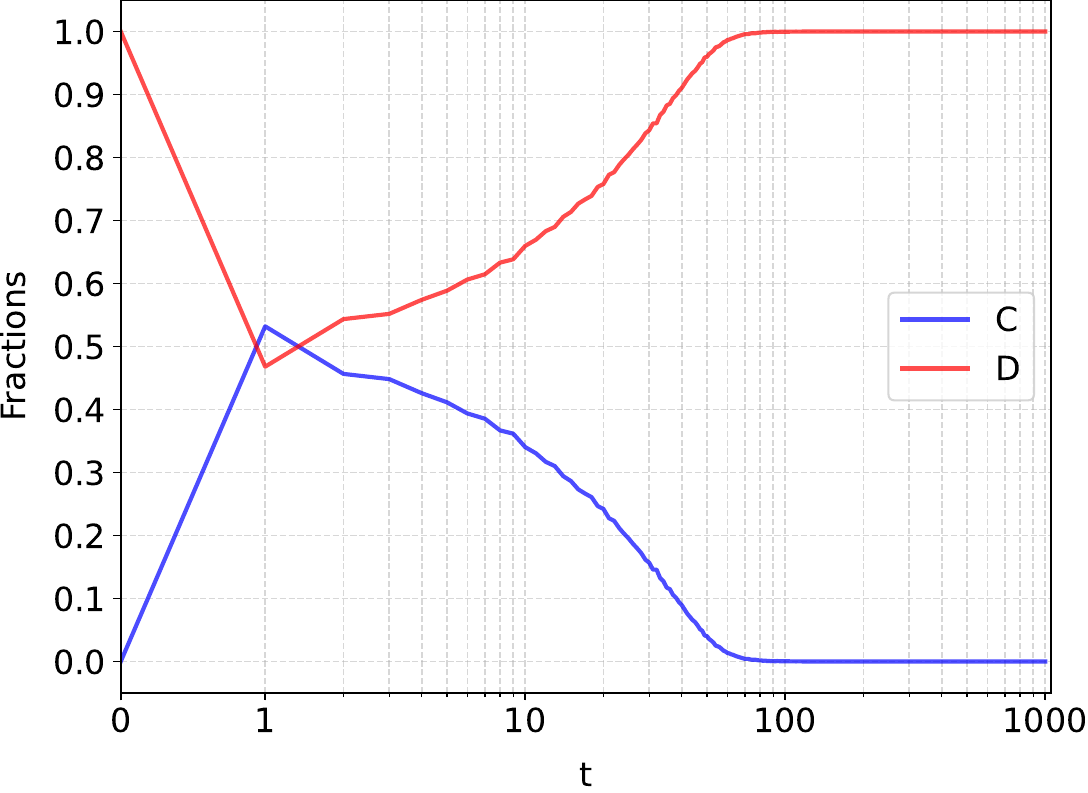}\\
			\end{minipage}
			\vspace{2mm}
			\\
			\begin{minipage}{0.188\linewidth}
				\centering
				\fbox{\includegraphics[width=\linewidth]{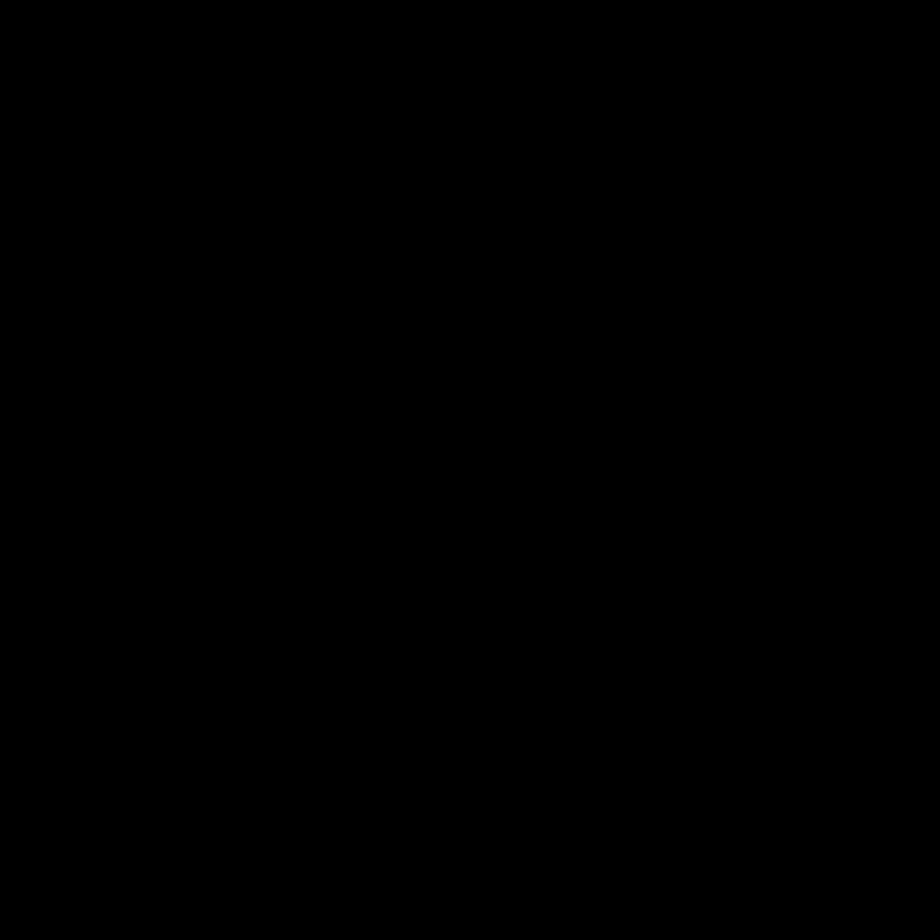}}\\
				\vspace{-2mm}
				{\footnotesize t=0}
			\end{minipage}
			\begin{minipage}{0.188\linewidth}
				\centering
				\fbox{\includegraphics[width=\linewidth]{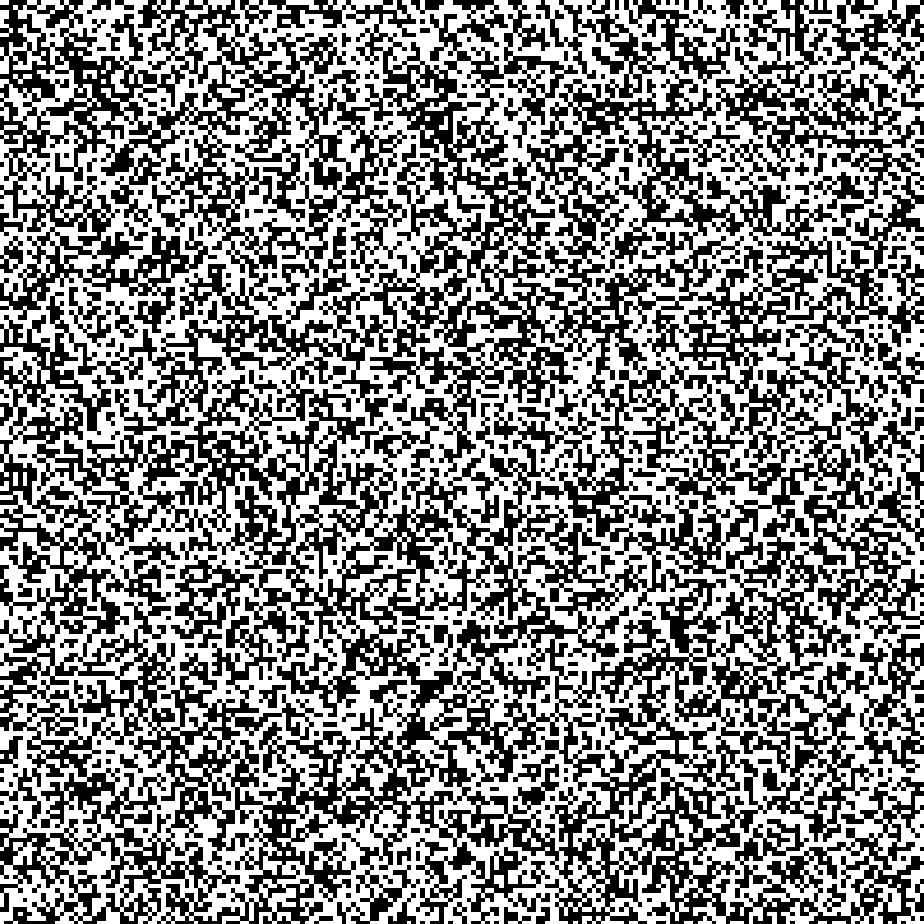}}\\
				\vspace{-2mm}
				{\footnotesize t=1}
			\end{minipage}
			\begin{minipage}{0.188\linewidth}
				\centering
				\fbox{\includegraphics[width=\linewidth]{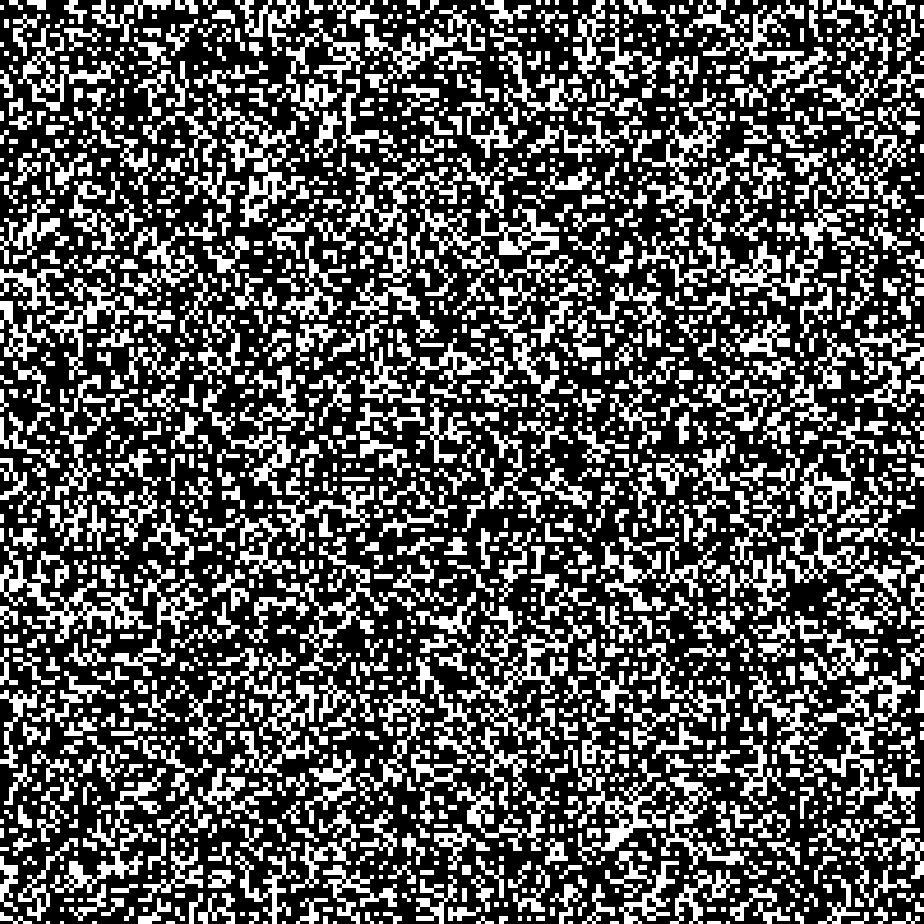}}\\
				\vspace{-2mm}
				{\footnotesize t=10}
			\end{minipage}
			\begin{minipage}{0.188\linewidth}
				\centering
				\fbox{\includegraphics[width=\linewidth]{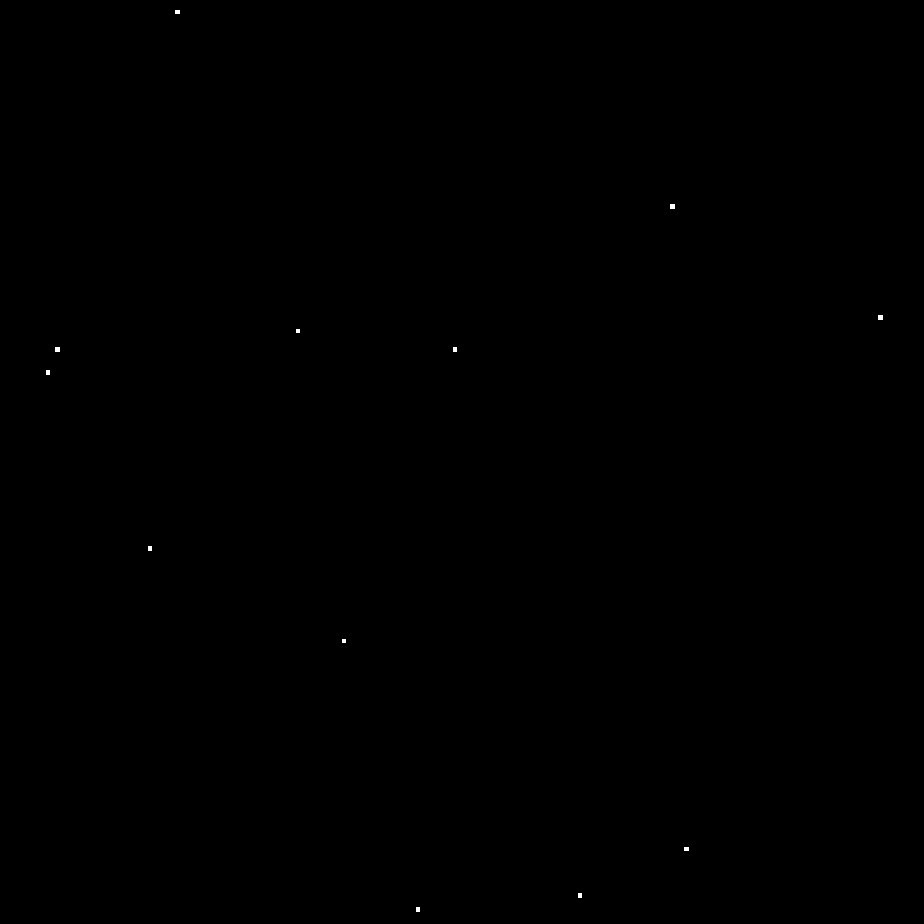}}\\
				\vspace{-2mm}
				{\footnotesize t=100}
			\end{minipage}
			\begin{minipage}{0.188\linewidth}
				\centering
				\fbox{\includegraphics[width=\linewidth]{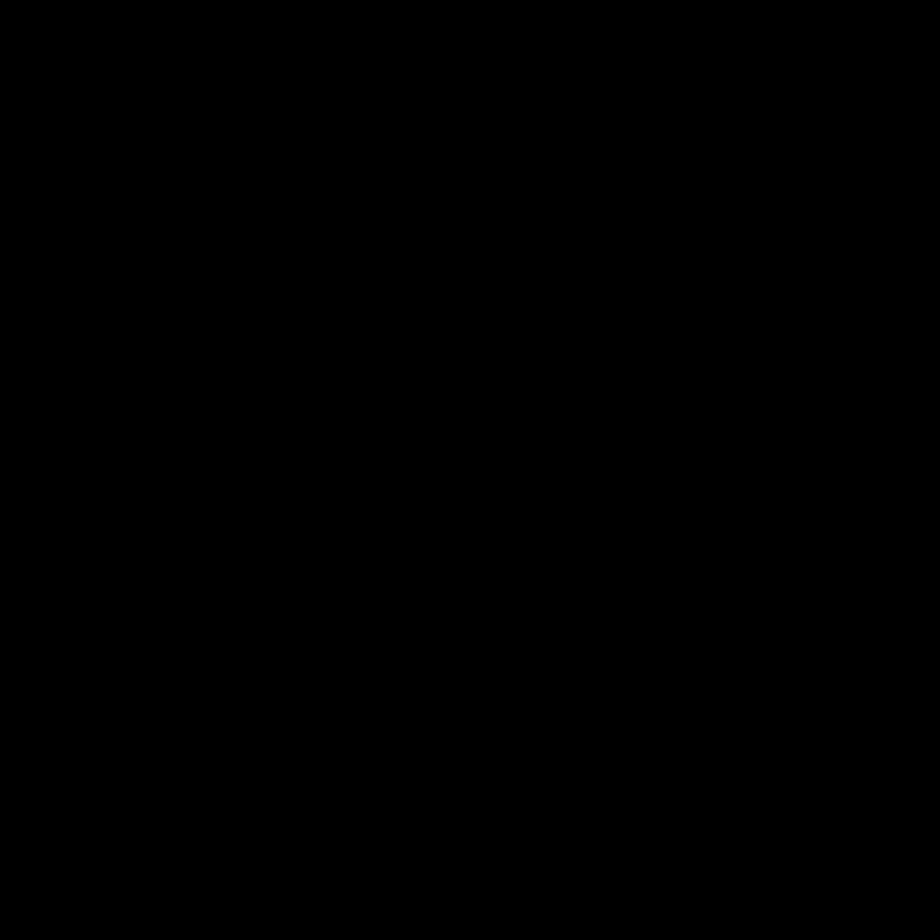}}\\
				\vspace{-2mm}
				{\footnotesize t=1000}
			\end{minipage}
			\vspace{-2mm}
			\caption*{\footnotesize (a) r=3.0}
		\end{minipage}
		\hfill
		\begin{minipage}{0.45\linewidth}
			\begin{minipage}{\linewidth}
				\centering
				\includegraphics[width=\linewidth]{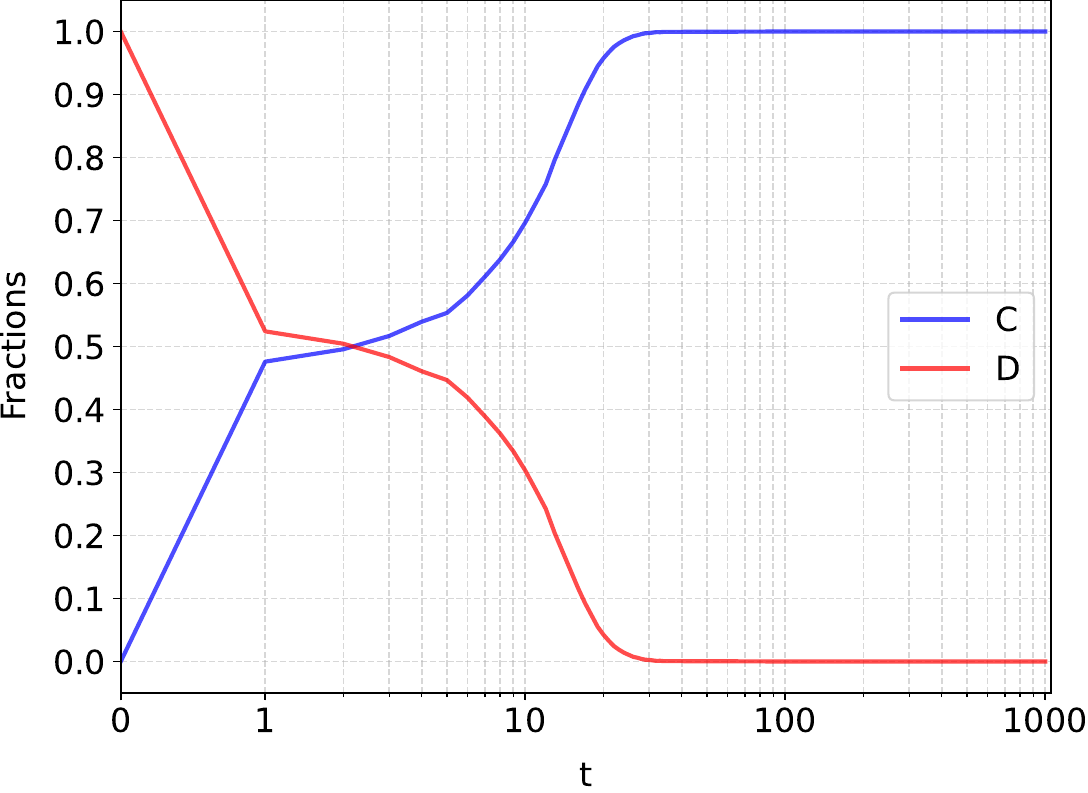}\\
			\end{minipage}
			\vspace{2mm}
			\\
			\begin{minipage}{0.188\linewidth}
				\centering
				\fbox{\includegraphics[width=\linewidth]{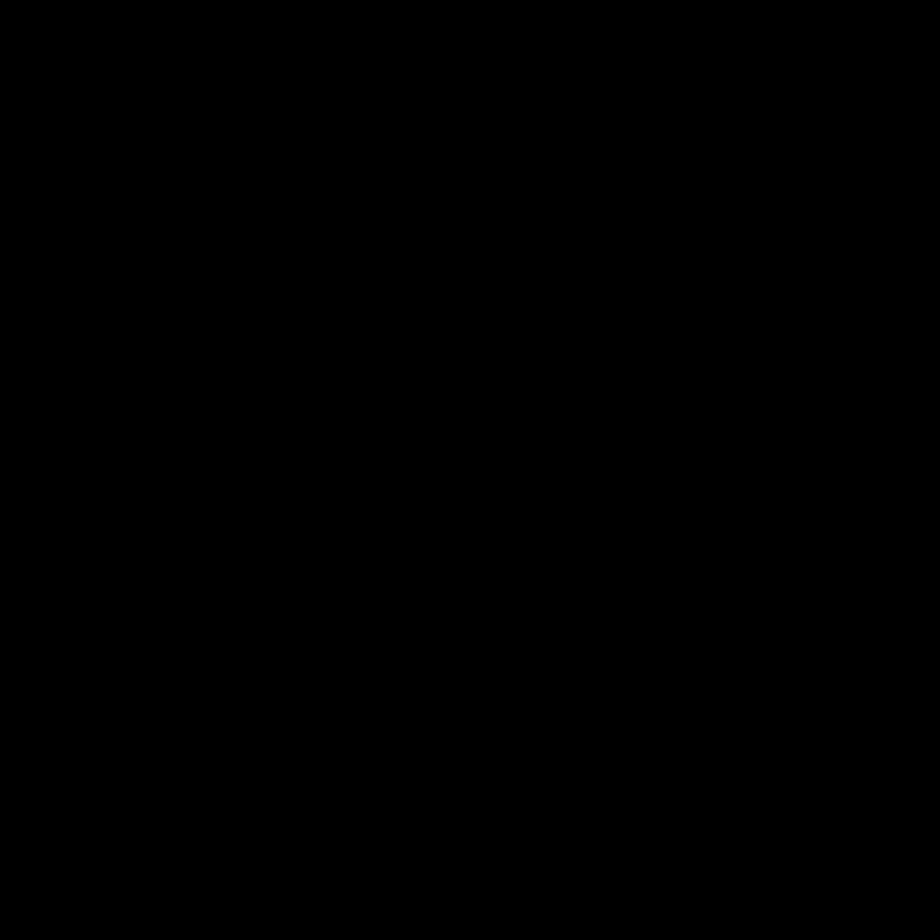}}\\
				\vspace{-2mm}
				{\footnotesize t=0}
			\end{minipage}
			\begin{minipage}{0.188\linewidth}
				\centering
				\fbox{\includegraphics[width=\linewidth]{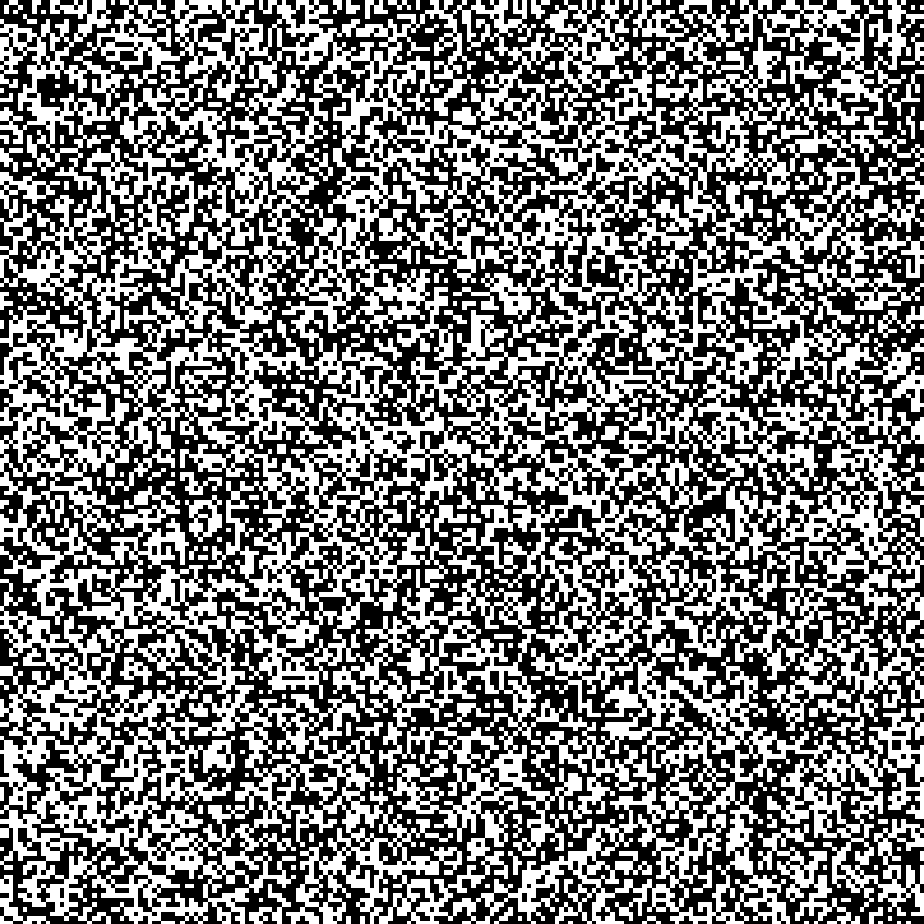}}\\
				\vspace{-2mm}
				{\footnotesize t=1}
			\end{minipage}
			\begin{minipage}{0.188\linewidth}
				\centering
				\fbox{\includegraphics[width=\linewidth]{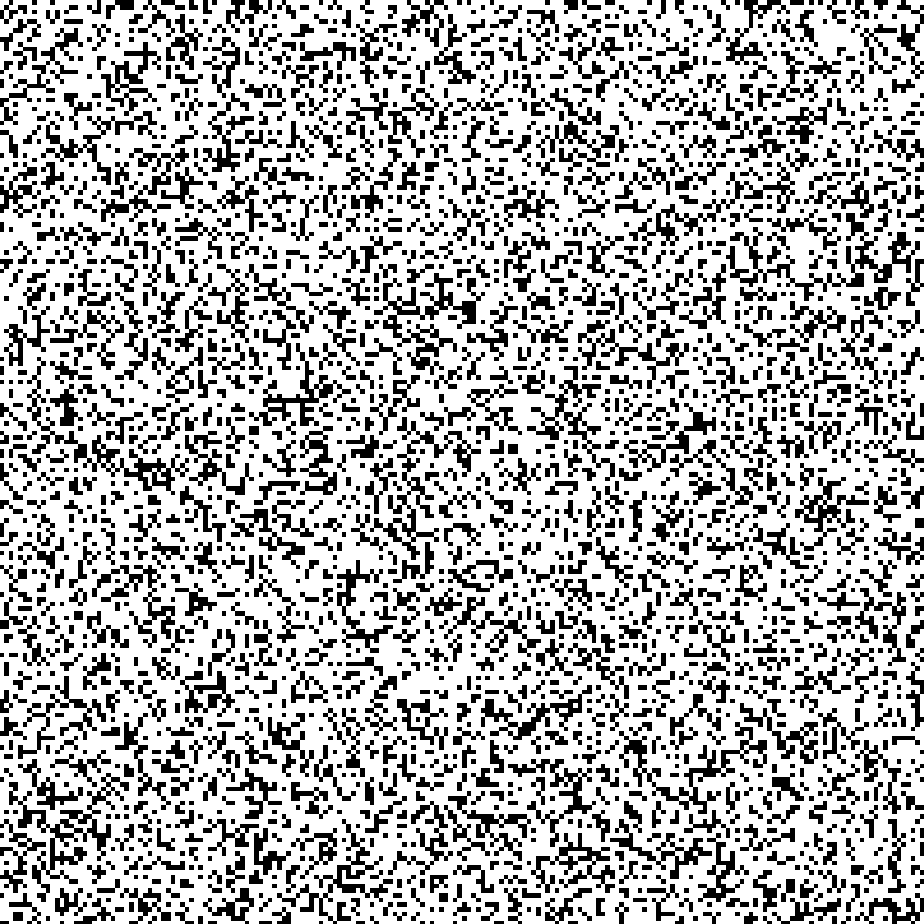}}\\
				\vspace{-2mm}
				{\footnotesize t=10}
			\end{minipage}
			\begin{minipage}{0.188\linewidth}
				\centering
				\fbox{\includegraphics[width=\linewidth]{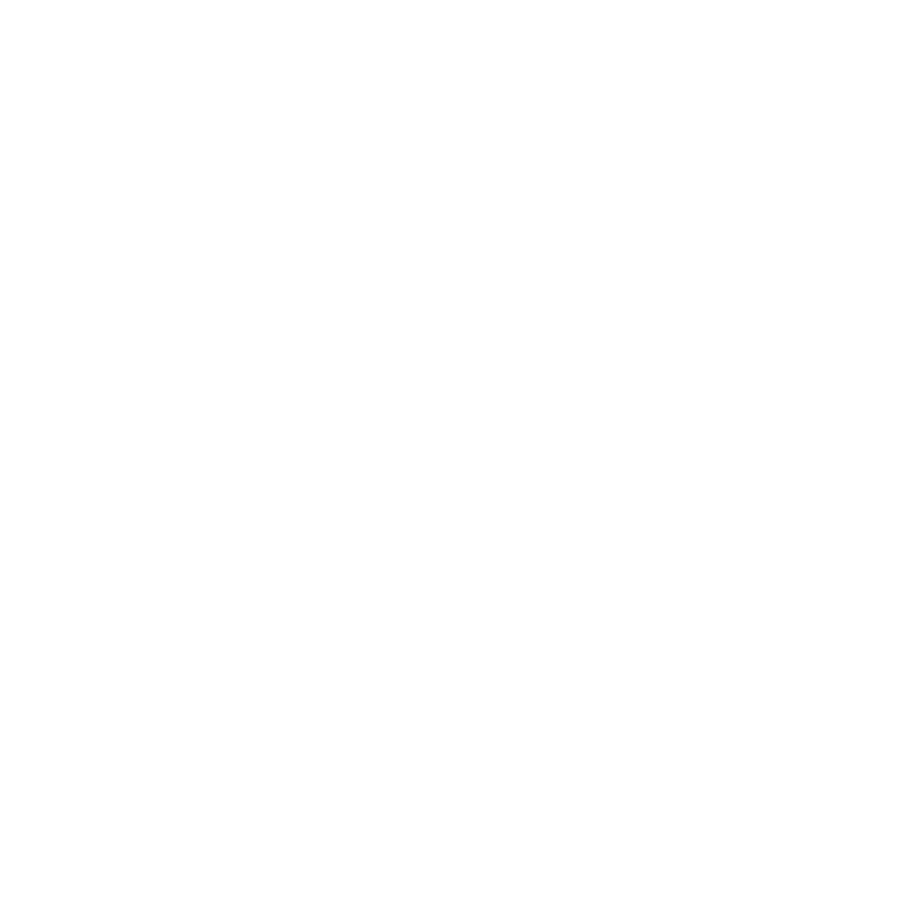}}\\
				\vspace{-2mm}
				{\footnotesize t=100}
			\end{minipage}
			\begin{minipage}{0.188\linewidth}
				\centering
				\fbox{\includegraphics[width=\linewidth]{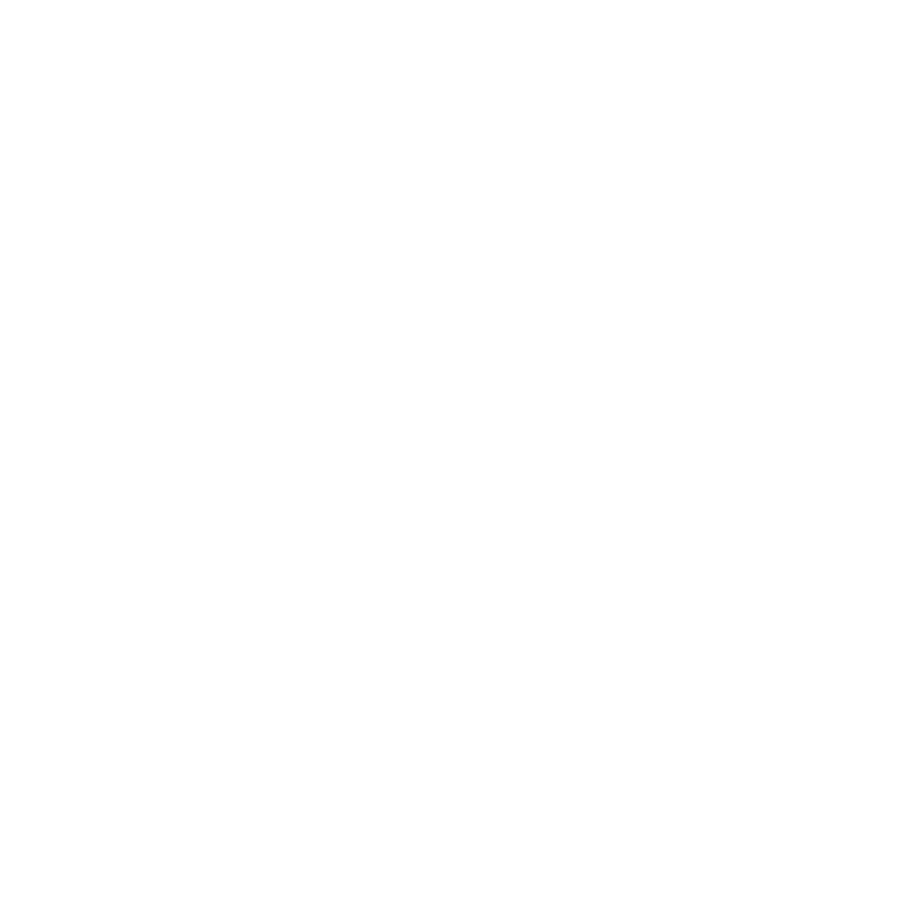}}\\
				\vspace{-2mm}
				{\footnotesize t=1000}
			\end{minipage}
			\vspace{-2mm}
			\caption*{\footnotesize (b) r=3.5}
		\end{minipage}
		\\
		[2mm]
		\begin{minipage}{\linewidth}
			\begin{minipage}{0.188\linewidth}
				\centering
				\includegraphics[width=\linewidth]{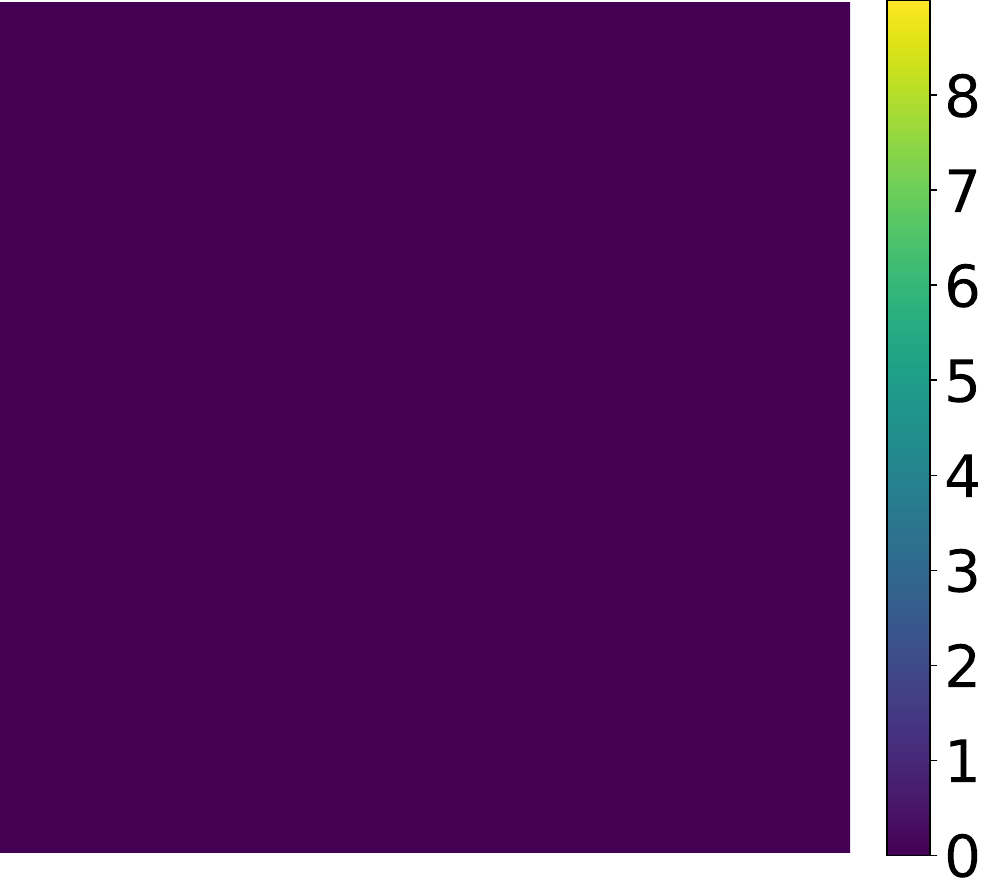}\\
				\vspace{-2mm}
				{\footnotesize t=0}
			\end{minipage}
			\hfill
			\begin{minipage}{0.188\linewidth}
				\centering
				\includegraphics[width=\linewidth]{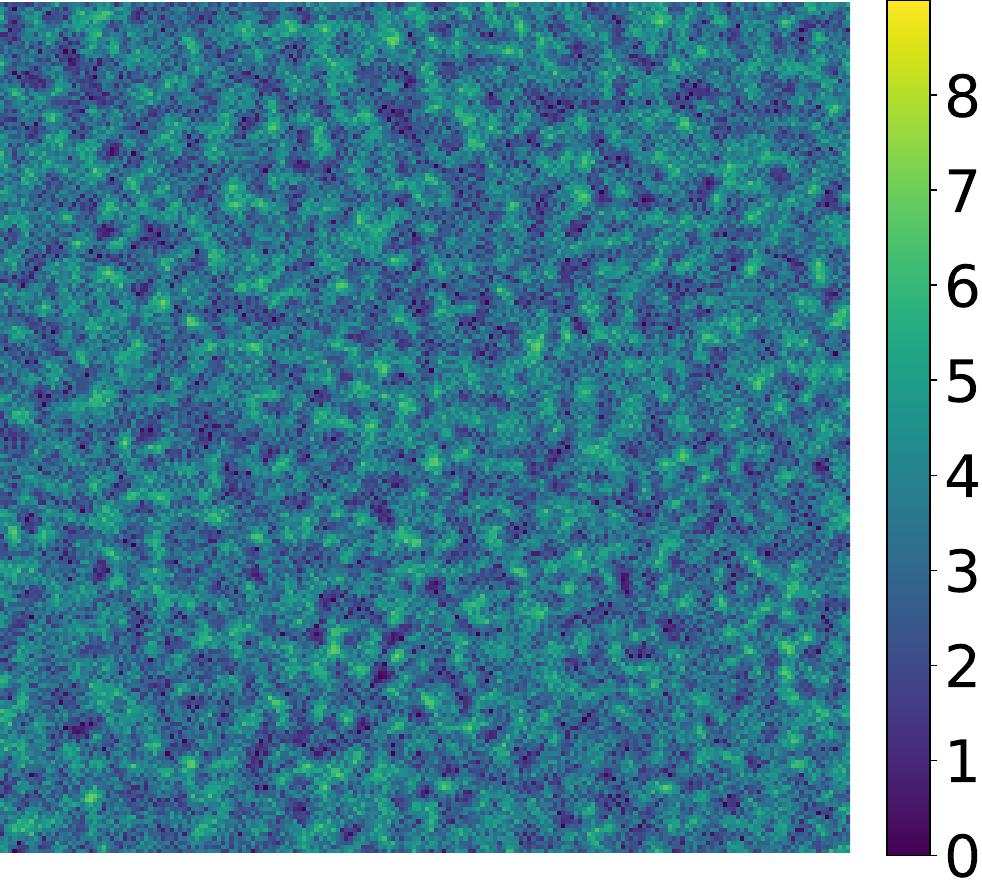}\\
				\vspace{-2mm}
				{\footnotesize t=1}
			\end{minipage}
			\hfill
			\begin{minipage}{0.188\linewidth}
				\centering
				\includegraphics[width=\linewidth]{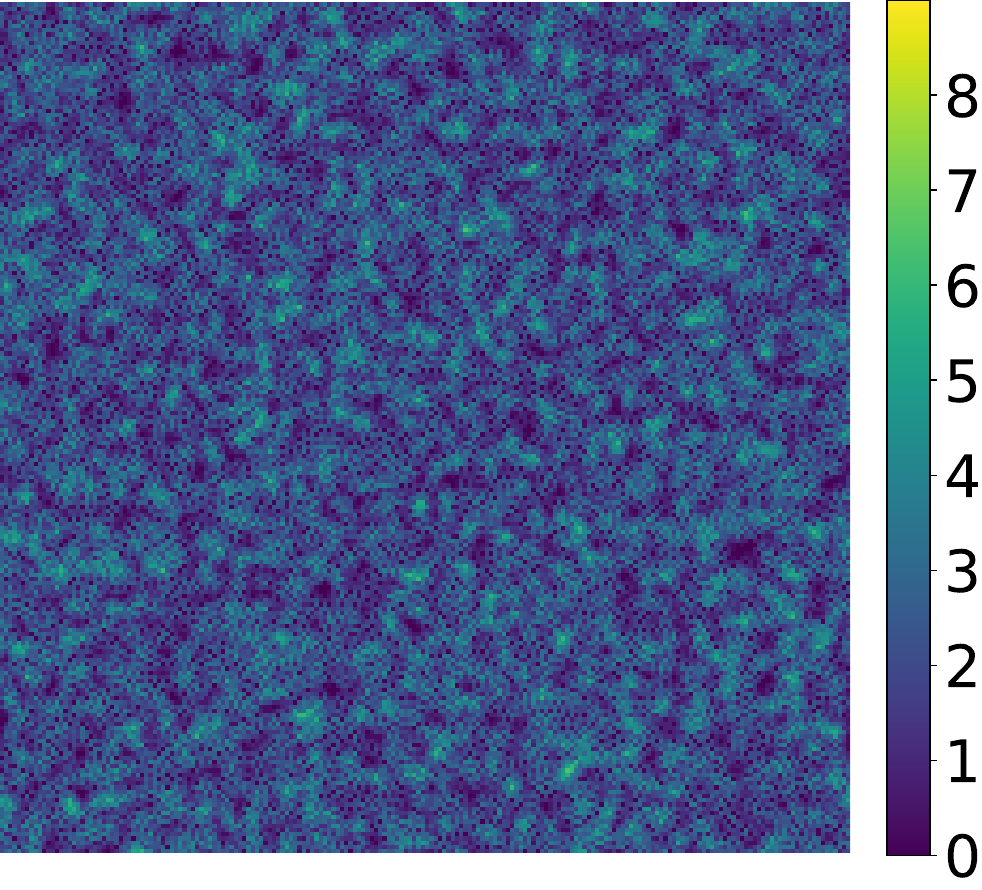}\\
				\vspace{-2mm}
				{\footnotesize t=10}
			\end{minipage}
			\hfill
			\begin{minipage}{0.188\linewidth}
				\centering
				\includegraphics[width=\linewidth]{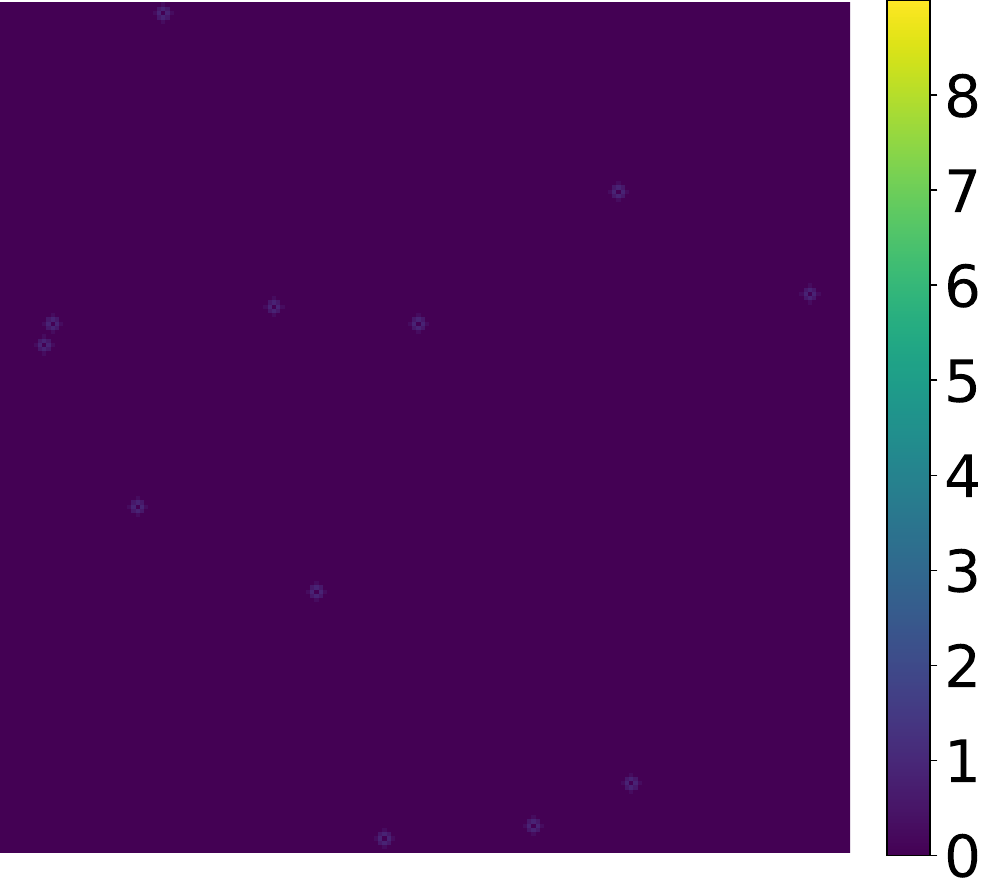}\\
				\vspace{-2mm}
				{\footnotesize t=100}
			\end{minipage}
			\hfill
			\begin{minipage}{0.188\linewidth}
				\centering
				\includegraphics[width=\linewidth]{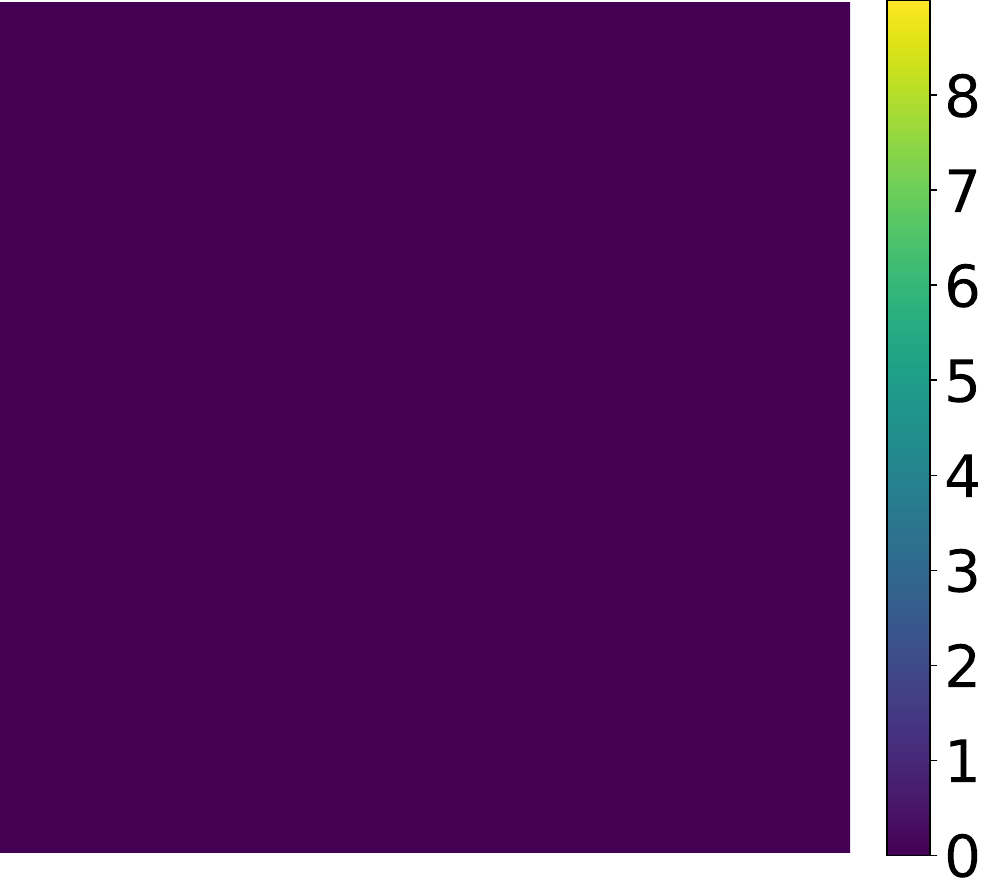}\\
				\vspace{-2mm}
				{\footnotesize t=1000}
			\end{minipage}
			\vspace{-2mm}
			\caption*{\footnotesize (c) r=3.0 (Payoff heatmaps)}
		\end{minipage}
		\\
		[2mm]
		\begin{minipage}{\linewidth}
			\begin{minipage}{0.188\linewidth}
				\centering
				\includegraphics[width=\linewidth]{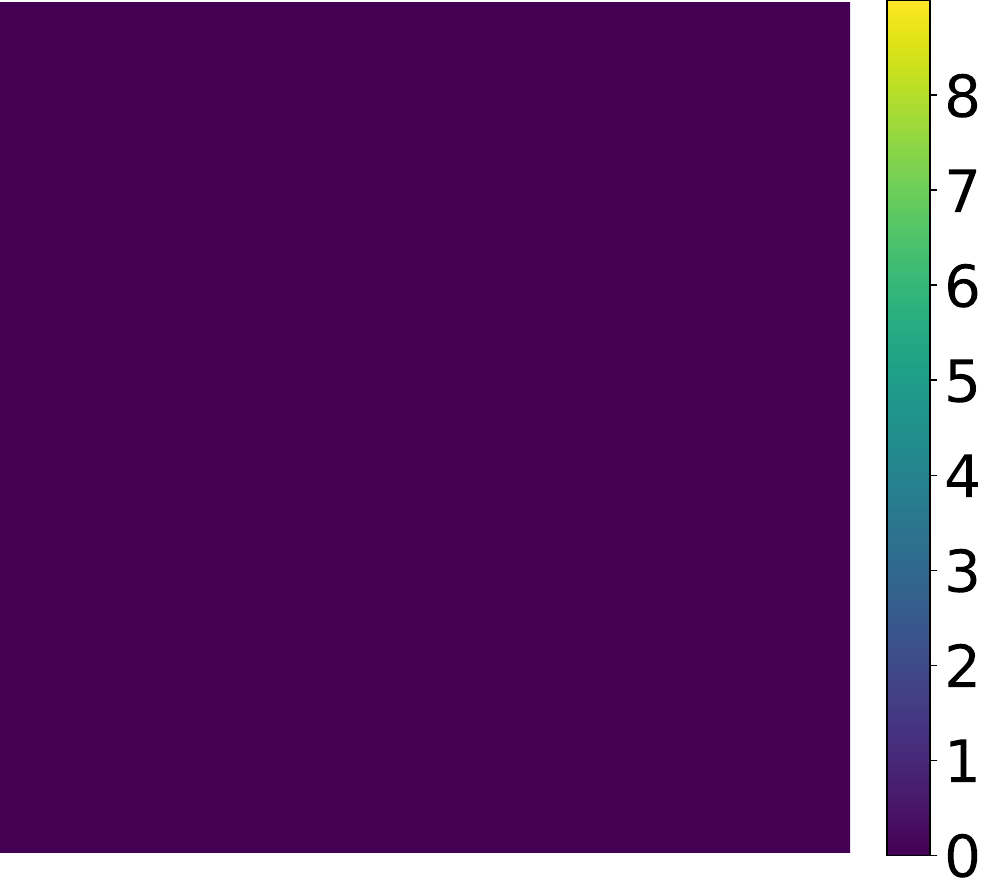}\\
				\vspace{-2mm}
				{\footnotesize t=0}
			\end{minipage}
			\hfill
			\begin{minipage}{0.188\linewidth}
				\centering
				\includegraphics[width=\linewidth]{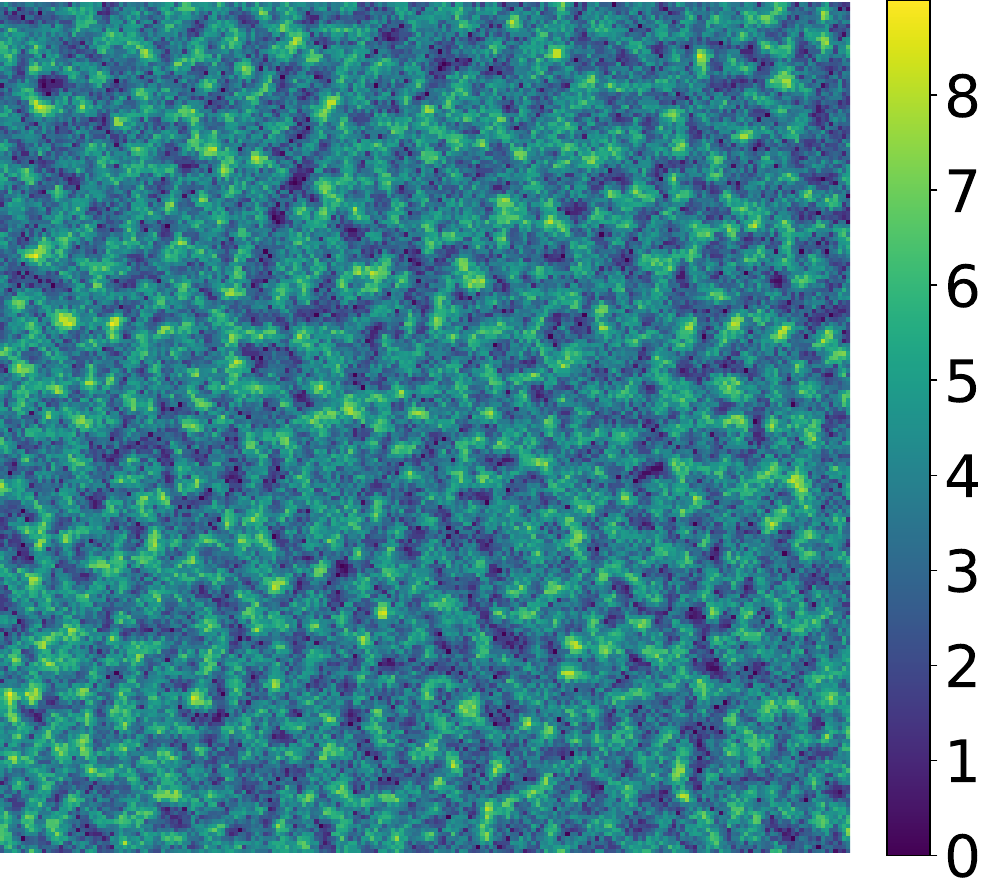}\\
				\vspace{-2mm}
				{\footnotesize t=1}
			\end{minipage}
			\hfill
			\begin{minipage}{0.188\linewidth}
				\centering
				\includegraphics[width=\linewidth]{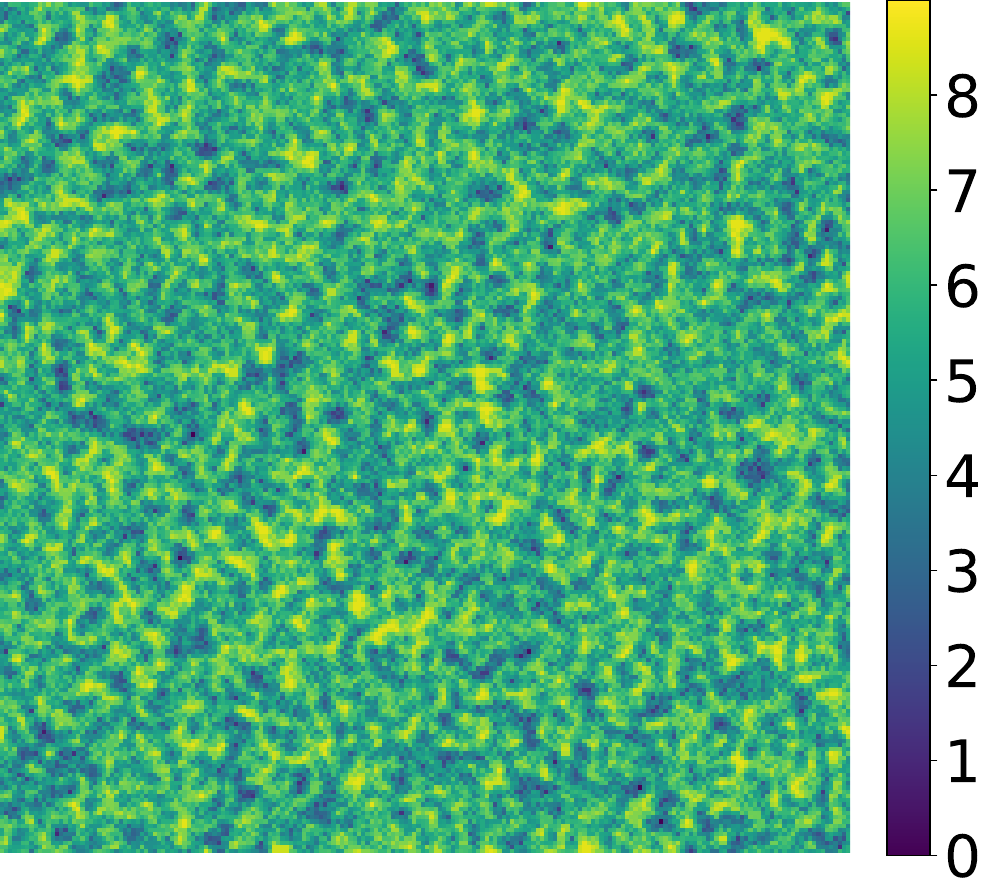}\\
				\vspace{-2mm}
				{\footnotesize t=10}
			\end{minipage}
			\hfill
			\begin{minipage}{0.188\linewidth}
				\centering
				\includegraphics[width=\linewidth]{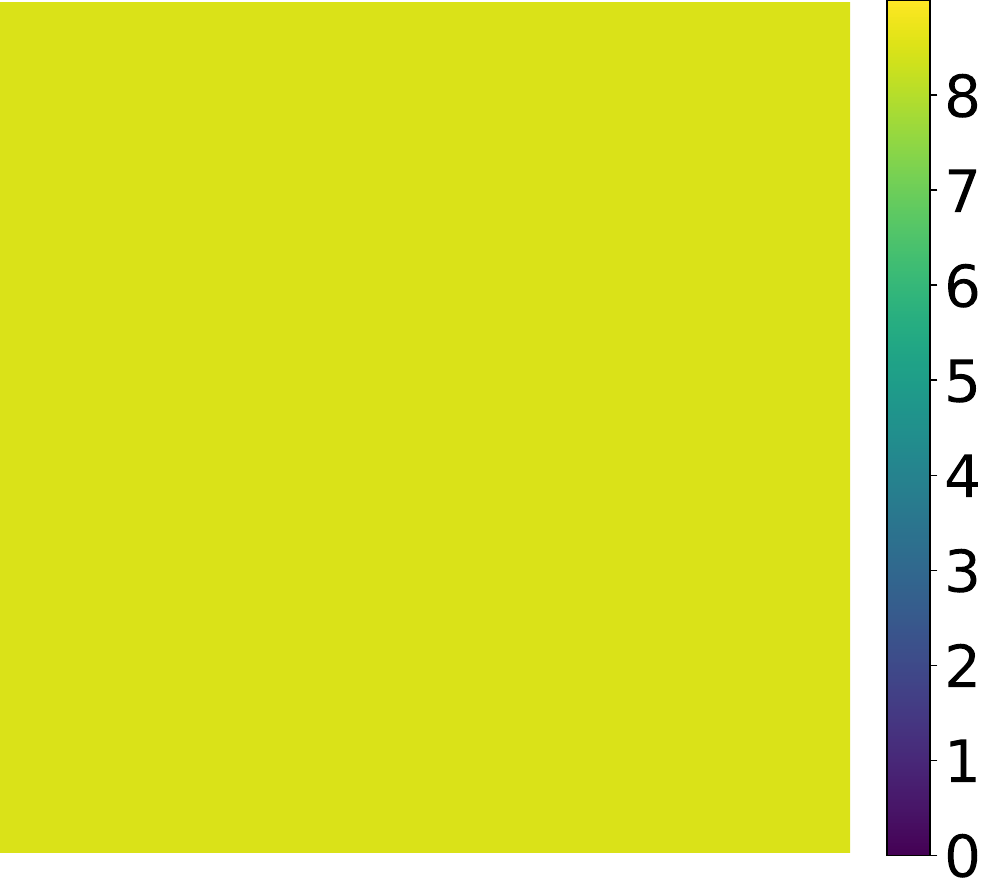}\\
				\vspace{-2mm}
				{\footnotesize t=100}
			\end{minipage}
			\hfill
			\begin{minipage}{0.188\linewidth}
				\centering
				\includegraphics[width=\linewidth]{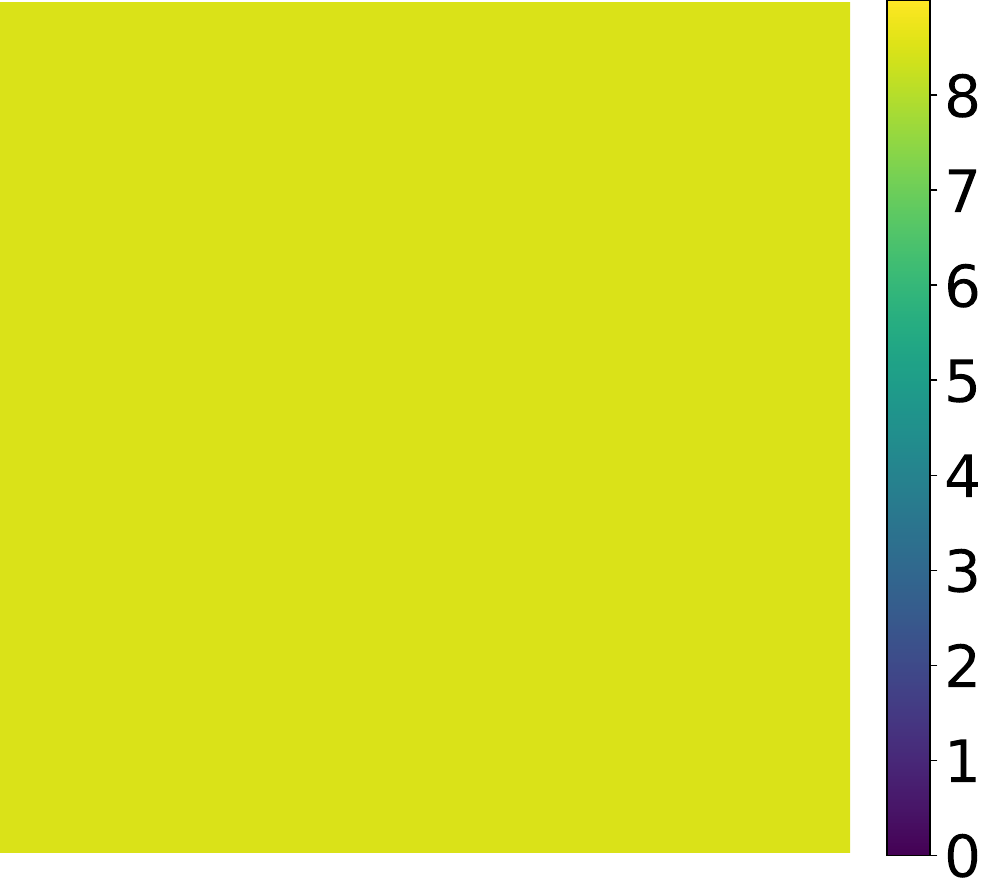}\\
				\vspace{-2mm}
				{\footnotesize t=1000}
			\end{minipage}
			\vspace{-2mm}
			\caption*{\footnotesize (d) r=3.5 (Payoff heatmaps)}
		\end{minipage}
		\caption{Experiment in SPGG using TUC-PPO with all-defectors initialization. \textbf{(a, b)} Temporal evolution curves show cooperators fractions in blue and defectors fractions in red, with corresponding strategy snapshots displaying defectors as black blocks and cooperators as white blocks. \textbf{(c, d)} Payoff heatmaps track individual reward changes during evolution. \textbf{(a, c)} Display results for enhancement factor $r = 3.0$ and \textbf{(b, d)} present $r = 3.5$ conditions. When all agents' strategies initialize as defectors, their transition to cooperation proves significantly more challenging.}
		\label{fig:TUC-PPO_unique}
	\end{figure*}

	This investigation examines the evolutionary dynamics of TUC-PPO under extreme initial conditions where all agents initially adopt defector strategies. As shown in Fig.~\ref{fig:TUC-PPO_unique}, this experimental design demonstrates the framework's capability to overcome fundamental limitations inherent in traditional evolutionary game theory methods. Conventional approaches like the Fermi update rule often fail to initiate cooperative emergence from homogeneous defector populations due to their reliance on neighborhood strategy diversity. The TUC-PPO framework addresses this challenge through its team-utility-constrained RL architecture, which enables the discovery of cooperative strategies even in these demanding initial conditions.
	
	Fig.~\ref{fig:TUC-PPO_unique} demonstrates TUC-PPO's robust performance when starting from all-defectors initialization. The random initialization of Actor-Critic network parameters leads to stochastic initial strategy selection. Under $r=3.0$ conditions, the cooperation rate first increases to around $50\%$ before gradually declining to zero. Spatial patterns show temporary cooperative clusters being replaced by expanding defector regions, while payoff maps reflect decreasing individual rewards during this process. For $r=3.5$, the system exhibits stable growth in cooperation, reaching full cooperation within 40 iterations. The visualizations demonstrate cooperators systematically replacing defectors, with payoff distributions showing continuous improvement across the population.
	﻿
	
	These results reveal two important findings. First, TUC-PPO maintains consistent convergence properties regardless of the initialization scheme, proving its algorithmic robustness. Second, the framework demonstrates significantly superior performance compared to rule-based approaches like the Fermi update rule. Crucially, the Fermi method shows high sensitivity to initial conditions and cannot escape all-defectors equilibria without external intervention. The payoff dynamics further demonstrate that TUC-PPO reliably guides the system toward economically superior cooperative equilibria when enhancement factors are sufficient.

	\subsection{Hyperparameter Sensitivity Analysis}
	\label{sec:hyperparam}
	
	The entropy regularization coefficient $\rho$ exhibits dominant control over cooperation emergence in the proposed framework. Conversely, other hyperparameters including learning rate $\alpha$, discount factor $\gamma$, and value loss weight $\delta$ demonstrate negligible sensitivity within standard operational ranges. Experimental results across enhancement factors $r \in [1,5]$ reveal distinct behavioral regimes governed by $\rho$ selection.

\begin{figure}[htbp!]
		\centering
		\includegraphics[width=0.7\linewidth]{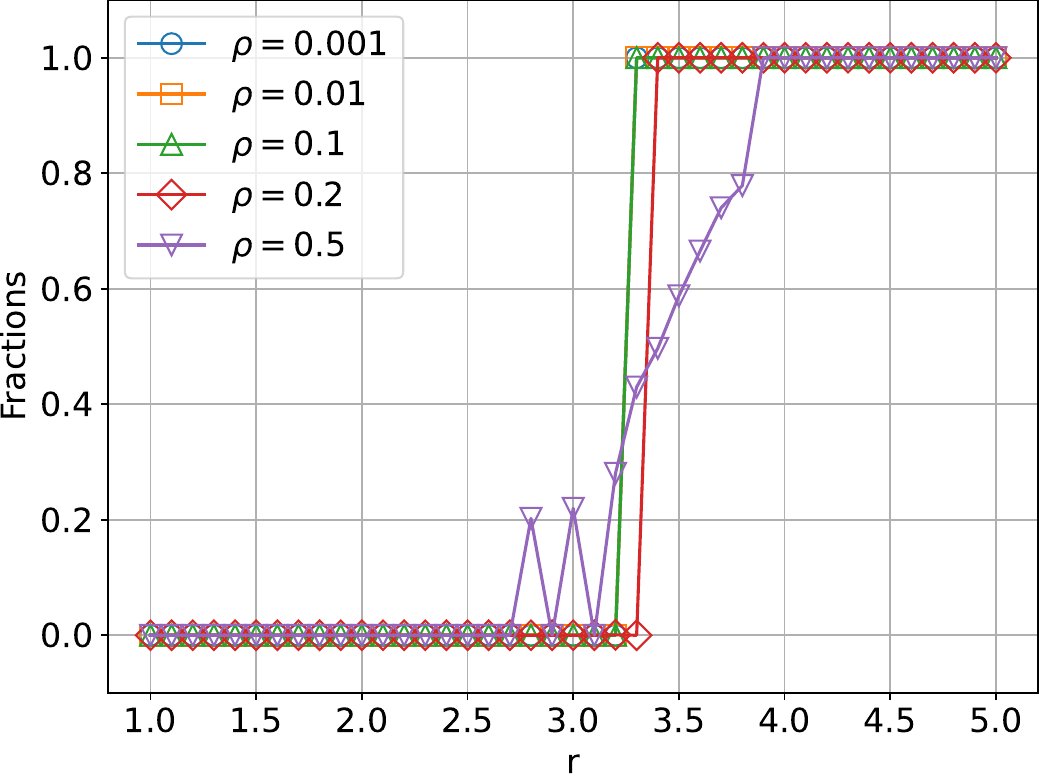}
		\caption{Impact of discount factor $\rho$ on TUC-PPO.}
		\label{fig:cdrcombinedrho}
	\end{figure}

	Fig.~\ref{fig:cdrcombinedrho} shows the sensitivity analysis of the entropy regularization coefficient $\rho$ in TUC-PPO. In PPO, the choice of $\rho$ critically influences the exploration-exploitation trade-off during policy optimization. As shown in Fig.~\ref{fig:cdrcombinedrho}, when $\rho$ exceeds 0.1, a higher reward $r$ is required to ensure all agents adopt cooperative strategies. This occurs because the variance of policy updates grows nonlinearly with increasing $\rho$, causing the policy gradient direction to deviate from the intended optimization path. Consequently, agents struggle to converge to the deterministic optimal policy within a reasonable number of iterations. Furthermore, excessive policy entropy disrupts the stability of value function estimation, leading to distorted advantage signals and undermining the theoretical foundation of importance sampling.
	
	TUC-PPO's entropy regularization term effectively encourages exploration when $\rho$ is appropriately set. However, surpassing a critical threshold introduces excessive stochasticity that violates the core assumptions of policy gradient optimization. In continuous action spaces, this manifests as non-convergent agent behaviors. For discrete decision-making scenarios, it causes the loss of deterministic policy choices at critical state nodes. Empirical studies indicate that optimal $\rho$ values must balance task-specific state-space complexity and exploration needs. These values typically correspond to the minimal stochasticity level required for maintaining policy diversity. Based on these findings, we set the default value of $\rho$ to 0.01. Other hyperparameters are set according to commonly used values.

	\section{Conclusions}
	\label{sec:con}
	
	The TUC-PPO framework establishes a novel paradigm for cooperation evolution in SPGG by integrating team utility constraints with proximal policy optimization. This synthesis creates a mathematically grounded approach where constrained policy updates explicitly balance individual rewards with collective welfare requirements. Through rigorous theoretical analysis and comprehensive experimental validation, we demonstrate that TUC-PPO's team-constrained optimization mechanism delivers significant advantages in cooperative system dynamics. Key findings reveal TUC-PPO's dual performance superiority over conventional methods. This method achieves stable cooperation at substantially lower enhancement factors ($r \geq 3.6$), significantly outperforming baseline PPO which requires $r > 5.0$. Additionally, TUC-PPO exhibits accelerated convergence rates compared to conventional approaches. Crucially, TUC-PPO maintains robust performance across diverse initialization schemes. This includes challenging all-defector scenarios where traditional methods fail, confirming the framework's adaptability to adverse initial conditions.

	Theoretical contributions encompass the first integration of team utility constraints into policy gradient optimization for evolutionary games. This integration enables precise control between individual rationality and collective welfare via Lagrangian dual-ascent. The framework employs a self-adjusting constraint mechanism that dynamically adapts penalty coefficients through batch-wise violation evaluation. This ensures team utility thresholds are met while maintaining policy update stability. This approach overcomes fundamental limitations in conventional multi-agent RL by demonstrating how explicit team welfare objectives foster cooperation emergence. 
	
	From an implementation perspective, TUC-PPO's spatial dynamics reveal emergent self-organization patterns. In these patterns, cooperators form stable clusters that minimize boundary exposure to defectors. This spatial configuration provides algorithmic-level validation of network reciprocity theory, demonstrating how localized team constraints generate system-wide cooperative equilibria. For practical applications, TUC-PPO provides novel design principles for multi-agent systems requiring robust cooperation. Potential implementations include resource distribution networks, community-based sustainability initiatives, and institutional frameworks that balance individual incentives with group welfare.
	
	While demonstrating significant advantages, TUC-PPO presents opportunities for further refinement. Future research should extend this framework to dynamic network topologies and investigate heterogeneous agent capabilities within team constraints. Additionally, exploration of asymmetric reward structures in collective action problems and optimization of computational efficiency for large-scale deployments are critical next steps. In summary, TUC-PPO advances evolutionary game theory by formalizing the relationship between local team constraints and global cooperation emergence. The framework combines mathematical rigor, empirical performance, and implementation flexibility. This triad establishes a new foundation for designing cooperative multi-agent systems in computational social science and distributed AI applications.
	﻿
	﻿
	
	\section*{CRediT authorship contribution statement}
	
	\textbf{Zhaoqilin Yang}: Writing – original draft, Writing – review and editing, Validation, Methodology, Conceptualization.
	\textbf{Xin Wang}: Conceptualization, Investigation, Writing – review and editing.
	\textbf{Ruichen Zhang} : Writing – review and editing, Validation, Supervision, 
	\textbf{Chanchan Li}: Writing – review and editing, Visualization, Software.
	\textbf{Youliang Tian}: Funding acquisition, Resources, Supervision.
	
	\section*{Declaration of competing interest }
	
	The authors declare that they have no known competing financial interests or personal relationships that could have appeared to influence the work reported in this paper.
	
	\section*{Data availability}
	
	No data was used for the research described in the article.
	
	\section*{Acknowledgments}
	This work was supported by the Natural Science Special Project (Special Post) Research Foundation of Guizhou University (No.[2024] 39); National Key Research and Development Program of China under Grant 2021YFB3101100; National Natural Science Foundation of China under Grant 62272123; Project of High-level Innovative Talents of Guizhou Province under Grant [2020]6008; Science and Technology Program of Guizhou Province under Grant [2020]5017, [2022]065; Science and Technology Program of Guiyang under Grant [2022]2-4.
	
	\bibliographystyle{elsarticle-num}
	\bibliography{cas-refs}
\end{document}